\documentstyle[sprocl]{article}

\def\be{\begin{equation}}
\def\ee{\end{equation}}
\def\bea{\begin{eqnarray}}
\def\eea{\end{eqnarray}}

\begin{document}

\begin{flushright}
\vspace{1mm}
hep-th/9910096\\
FIAN/TD/24--99\\
\end{flushright}

\vspace{1cm}

\title{HIGHER SPIN GAUGE THEORIES: STAR-PRODUCT AND ADS SPACE}

\author{M.A. VASILIEV}

\address{I.E.Tamm Department of Theoretical Physics,\\
 Lebedev Physical Institute,\\
 Leninsky prospect 53, 117924, Moscow, Russia}


\maketitle\abstracts{
We review  the theory of higher spin
gauge fields in 2+1 and 3+1 dimensional anti-de Sitter
space and present some new results on the structure of
higher spin currents and explicit solutions of the
massless equations.
A previously obtained d=3 integrating flow is
generalized to d=4 and is shown to
give rise to a
perturbative solution of the d=4 nonlinear higher spin equations.
A particular attention is paid to the
relationship between the star-product
origin of the higher spin symmetries, AdS geometry
and the concept of space-time locality.}

\newcommand{\ty}{\hat{y}}
\newcommand{\bee}{\begin{eqnarray}}
\newcommand{\eee}{\end{eqnarray}}
\newcommand{\nn}{\nonumber}
\newcommand{\hy}{\hat{y}}
\newcommand{\by}{\bar{y}}
\newcommand{\bz}{\bar{z}}
\newcommand{\go}{\omega}
\newcommand{\e}{\epsilon}
\newcommand{\half}{\frac{1}{2}}
\newcommand{\ga}{\alpha}
\newcommand{\gal}{\alpha}
\newcommand{\U}{\Upsilon}
\newcommand{\ups}{\upsilon}
\newcommand{\bu}{\bar{\upsilon}}
\newcommand{\dga}{{\dot{\alpha}}}
\newcommand{\dgb}{{\dot{\beta}}}
\newcommand{\gb}{\beta}
\newcommand{\gga}{\gamma}
\newcommand{\gd}{\delta}
\newcommand{\gl}{\lambda}
\newcommand{\gk}{\kappa}
\newcommand{\gep}{\epsilon}
\newcommand{\gvep}{\varepsilon}
\newcommand{\gs}{\sigma}
\newcommand{\V}{|0\rangle}
\newcommand{\ws}{\wedge\star\,}
\newcommand{\gee}{\epsilon}
\newcommand{\ggg}{\gamma}
\newcommand\ul{\underline}
\newcommand\un{{\underline{n}}}
\newcommand\ull{{\underline{l}}}
\newcommand\um{{\underline{m}}}
\newcommand\ur{{\underline{r}}}
\newcommand\us{{\underline{s}}}
\newcommand\up{{\underline{p}}}
\newcommand\uq{{\underline{q}}}
\newcommand\ri{{\cal R}}
\newcommand\punc{\multiput(134,25)(15,0){5}{\line(1,0){3}}}
\newcommand\runc{\multiput(149,40)(15,0){4}{\line(1,0){3}}}
\newcommand\tunc{\multiput(164,55)(15,0){3}{\line(1,0){3}}}
\newcommand\yunc{\multiput(179,70)(15,0){2}{\line(1,0){3}}}
\newcommand\uunc{\multiput(194,85)(15,0){1}{\line(1,0){3}}}
\newcommand\aunc{\multiput(-75,15)(0,15){1}{\line(0,1){3}}}
\newcommand\sunc{\multiput(-60,15)(0,15){2}{\line(0,1){3}}}
\newcommand\dunc{\multiput(-45,15)(0,15){3}{\line(0,1){3}}}
\newcommand\func{\multiput(-30,15)(0,15){4}{\line(0,1){3}}}
\newcommand\gunc{\multiput(-15,15)(0,15){5}{\line(0,1){3}}}
\newcommand\hunc{\multiput(0,15)(0,15){6}{\line(0,1){3}}}
\newcommand\ls{\!\!\!\!\!\!\!}

\section{Introduction}

The concept of supersymmetry as an extension
of the space--time symmetries
originally introduced by Yuri Golfand
and Evgeny Likhtman in 1971 \cite{GL}
plays tremendously important role in the modern
high energy physics. Nowadays it is a text-book example of
how important is to know a true symmetry
when investigating
a theory of fundamental interactions. In this contribution
we focus on some attempt to find a
larger (infinite-dimensional) symmetry that extends
ordinary supersymmetry discovered by Golfand and Likhtman
and underlies the theory of higher spin gauge fields.
The main motivation is that this further extension of
supersymmetry may lead to a most
symmetric phase of a theory of fundamental interactions.
In particular  we emphasize that quantum mechanical
nonlocality of star-product algebras of auxiliary spinor variables,
identified  with the higher spin symmetry algebras,
results in space-time nonlocality of higher spin interactions.

A theory of fundamental interactions is
presently identified with
still mysterious M-theory,\cite{M} which
should possess a number of properties such as:

\noindent
{\bf (i)} M-theory is some relativistic theory in $d = 11$.
d=11 SUGRA is a low-energy limit of  M-theory.

\noindent
{\bf (ii)} M-theory gives rise to superstring models in $d\leq 10$
providing a geometric explanation to dualities.

\noindent
{\bf (iii)}
Star-product (Moyal bracket) plays important role in a certain phase
of M-theory with nonvanishing vacuum expectation value of the
antisymmetric field $B_{\un\um}$.\cite{mmoyal,ch} In the limit
$\ga^\prime \to 0$, $B_{\un\um}= const$ string theory reduces to
noncommutative Yang-Mills Theory.\cite{SW}

\noindent
{\bf (iv)} A particularly interesting version of
the M-theory is expected
to have anti-de Sitter (AdS) geometry explaining duality between
AdS SUGRA and conformal models at the boundary of the AdS space.\cite{adsconf}

The most intriguing question is: ``what is M-theory?''.
It is instructive to analyze the situation
from the perspective of spectrum of elementary excitations.
Superstrings  describe massless modes of lower
spins $s\leq 2$ like graviton ($s=2$), gravitino ($s=3/2$),
vector bosons ($s=1$) and matter fields with spins 1 and 1/2,
as well as certain antisymmetric tensors.
On the top of that there is an infinite tower of
massive excitations of all spins. Since the corresponding massive
parameter is supposed to be large,
massive higher spin  excitations are not directly observed at
low energies. They are important however for the consistency of the
theory. Assuming that M-theory is some relativistic theory
admitting a covariant perturbative interpretation we conclude that
it should necessarily contain higher spin modes to
describe superstring models as its particular vacua.
There are two basic alternatives:
(i) $m\neq 0$: higher spin modes in M-theory are massive or
(ii) $m= 0$: higher spin modes in M-theory are massless.
Since M-theory is supposed to be formulated in eleven
(or may be higher) dimensions in the both cases  massive higher
spin modes in the compactified superstring
models may in principle result from compactification of extra dimensions.

Each  of these alternatives is not straightforward. In the massive case
it is generally believed that  no consistent superstring theory exists
beyond ten dimensions and therefore there is no good guiding principle
towards M-theory from that side. For the massless option the situation
is a sort of opposite: there is a very good guiding principle but it
looks like it might be too strong.
Indeed,  massless fields of high spins are
gauge fields. Therefore this type of theories should be based on some
higher spin gauge symmetry principle with the symmetry generators
corresponding to various representations of the
Lorentz group. It is very well known however that
it is  a hard problem to build a nontrivial theory
with higher spin gauge symmetries.
One argument is due to the Coleman-Mandula theorem and
its generalizations \cite{cm} which claim that symmetries of S-matrix in a
non-trivial (i.e., interacting) field theory in a flat space
can only have sufficiently low spins. Direct  arguments
come \cite{diff,WF} from the explicit attempts to construct
higher spin gauge interactions in physically interesting situations
(e.g. when the gravitational
interaction is included).  These arguments convinced  most of
experts that no consistent
nontrivial higher spin gauge theory can exist at all.

However, some positive results \cite{pos} were
obtained on the existence of consistent interactions
of higher spin gauge fields in the flat space
with the matter fields and with themselves but not with
gravity. Somewhat later it was realized  \cite{FV1} that the
situation changes drastically once, instead of the flat space,
the problem is analyzed in the AdS space
with nonzero curvature $\Lambda$.
This generalization led to the solution of the problem
of consistent higher spin gravitational interactions in the cubic
order at the action level \cite{FV1} and, later,  in all orders
in interactions at the level of equations of motion.\cite{con,Pr,more}

The role of AdS background in higher spin gauge theories
is very important. First it cancels the Coleman-Mandula
argument which is hard to implement in the AdS background.\cite{Witstr}
{}From the technical side the cosmological constant plays a crucial
role as well, allowing new types of interactions with higher derivatives
which have a structure
$
\Delta S^{int}_{p,n,m,k} \sim  \Lambda^p \partial^n \phi \partial^m \phi
\partial^k \phi \,,
$
where $\phi$ denotes any of the fields involved and $p$ can
take negative powers to
compensate extra dimension carried by
higher derivatives of fields in the
interactions (an order of derivatives which appear
in the cubic interactions increases linearly with spin \cite{pos,FV1}).
An important general conclusion is that $\Lambda$ should
necessarily be nonzero in the phase with unbroken higher spin
gauge symmetries. In that respect higher spin gauge theories are
analogous to gauged supergravities with charged gravitinos
which also require \cite{FVpr,DF} $\Lambda \neq 0$.

Higher spin gauge theories contain infinite sets
of spins $0\leq s < \infty$. This implies that higher spin
symmetries are infinite-dimensional.
Suppose now that higher spin gauge symmetries are spontaneously broken
by one or another mechanism.
Then, starting from the phase with massless higher spin gauge fields, one
will end up with a spontaneously broken phase with all  fields
massive except for a subset corresponding to an unbroken subalgebra.
The same time a value of the cosmological
constant will be redefined because fields acquiring a nonvanishing
vacuum expectation value may contribute to the vacuum energy. So,
there is in principle a possibility to have a
spontaneously broken phase with $m\neq 0$ for higher spins
and $\Lambda =0$ (or $\Lambda$ small). A most natural mechanism for
spontaneous breakdown of higher spin gauge symmetries is via
dimensional compactification. It is important that in the known
d=3 and d=4 examples the maximal finite-dimensional subalgebras
of the higher spin superalgebras coincide with the ordinary
AdS SUSY superalgebras giving rise to gauged SUGRA models.
Provided that the same happens in higher dimensional models,
this opens a natural way for obtaining superstring type theories
in $d\leq 10$ starting from some maximally symmetric
higher spin gauge theory in $d\geq 11$.

In this contribution  we would like to draw attention to a
deep parallelism of some of the properties of M-theory
with the theory of higher spin gauge fields, focusing
mainly on the higher spin
symmetries and the closed nonlinear higher spin equations
of motion. For more detail on the Lagrangian formulation
we refer the reader to \cite{rev} and original papers.\cite{FV1}

\section{Higher Spin Currents}
\label{Higher Spin Currents}

Usual inner symmetries are related via the Noether
theorem to the conserved spin 1 current that can be
constructed from different matter fields.
For example, a current  constructed  from scalar fields in
an appropriate representation of the gauge group
\be
\label{cur1} J^\un{}_i{}^j =  \bar{\phi}_i \partial^\un \phi^j -
\partial^\un \bar{\phi}_i \phi^j
\ee
is conserved on the solutions of the scalar field equations
\be
\partial_\un J^{\un}_i{}^j  =
\bar{\phi}_i (\Box +m^2) \phi^j - (\Box +m^2) \bar{\phi}_i \phi^j \,.
\ee
(Underlined indices are used for differential forms and vector fields
in d-dimen\-sional space-time, i.e. $\un = 0,\ldots ,d-1$ while $i$ and
$j$ are inner indices. Conventions
used throughout the paper are summarized in the Appendix).

Translational symmetry is associated with the spin 2 current called
stress tensor. For scalar matter it has the form
\be
\label{scalstress}
T^{\um\un} = \partial^\um \phi \partial^\un \phi -
\frac{1}{2} \eta^{\um\un}
\left(\partial_\ur \phi \partial^\ur \phi -m^2\phi^2\right)\,.
\ee

Supersymmetry is based on the conserved current
called supercurrent. It has fermionic statistics and
is constructed from bosons and fermions.
For massless scalar $\phi$ and massless spinor $\psi_\nu$
it has the form
\be
\label{cur3/2}
J^\un{}_\nu = \partial_\um \phi  (\gamma^\um \gamma^\un \psi )_\nu \,,
\ee
where $\gamma^n{}_{\mu}{}^{\nu}$ are Dirac matrices in
$d$ dimensions.

The conserved charges, associated with these conserved
currents, correspond, respectively,
to generators of inner symmetries $T^i{}_j$, space-time
translations $P^n$ and supertransformations $Q_\nu^i$. The
conserved current associated with Lorentz rotations can be
constructed from the symmetric stress tensor
\be
\label{ang2}
S^{\un}{}_{ ;}{}^{\um\ull} = T^{\un\um}\,x^\ull\, -\,
T^{\un\ull}\,x^\um\,,\qquad T^{\un\um}=T^{\um\un}\,.  \ee
These exhaust the standard lower spin conserved currents usually used in the
field theory.

The list of lower spin currents admits a natural extension
to higher spin currents containing higher derivatives of the
physical fields.
The higher spin currents associated with the integer spin $s$
\be
J^{\un}{}_{ ; m_1 \ldots m_t ,n_1 \ldots n_{s-1} }
\ee
are vector fields (index $\un$) taking values in all
representations of the Lorentz group described by the
traceless two-row Young diagrams
\bigskip
\be
\label{dia}
\begin{picture}(70,50)
\put(20,45){s-1} \put(33,35){\circle*{2}}
\put(25,35){\circle*{2}}
\put(17,35){\circle*{2}}
\put(25,25){\circle*{2}}
\put(17,25){\circle*{2}}
\put(33,25){\circle*{2}}
\put(00,40){\line(1,0){70}}
\put(00,30){\line(1,0){70}} \put(50,30){\line(0,1){10}}
\put(60,30){\line(0,1){10}} \put(70,30){\line(0,1){10}}
\put(00,20){\line(1,0){60}} \put(00,20){\line(0,1){20}}
\put(10,20){\line(0,1){20}} \put(40,20){\line(0,1){20}}
\put(60,20){\line(0,1){20}} \put(50,20){\line(0,1){20}} \put(20,10){t}
\end{picture}
\ee
with $0\leq t\leq s-1$. This means that the currents
$J^{\un ; m_1 \ldots m_t ,n_1 \ldots n_{s-1} }$ are symmetric in
the indices $n$ and $m$, satisfy the relations
\be \label{trind}
(s-1)(s-2)J^\un{}_{ ; m_1 \ldots m_t \,,r}{}^r {}_{n_3 \ldots n_{s-1} }
=0\,,\quad
\ee
\be
\label{trdep}
t(s-1)
J^\un{}_{ ;r m_2 \ldots m_t \,,}{}^r{}_{ n_2 \ldots n_{s-1}}
\,,\qquad
t(t-1)J^\un{}_{ ; m_3 \ldots m_t r }{}^r{}_{\,,n_1 \ldots n_{s-1}}
=0\,,\quad
\ee
and obey the antisymmetry property
\be
\label{asym}
J^\un{}_{ ; m_2 \ldots m_t \{n_s\,,  n_1 \ldots n_{s-1}\}_n} =0\,,
\ee
implying that symetrization over any $s$ indices $n$
and/or $m$ gives zero.

Let us now explain notation, which simplifies analysis of
complicated tensor structures and is useful  in the component
analysis. Following \cite{V2}
we combine the Einstein rule that upper and lower indices denoted by the
same letter are to be contracted with the convention that
upper (lower) indices denoted by the same letter imply
symmetrization which should be carried out prior contractions.
With this notation it is enough to put a number of symmetrized
indices in brackets writing e.g.  $X_{n(p)}$ instead of $X_{n_1 \ldots n_p}$.

Now, the higher spin currents are
$J^\un{}_{ ;m(t)\,,  n(s-1) }$
($1\leq t\leq s-1$) while the conditions (\ref{trind})-(\ref{asym})
take the form
\be
\label{hhtr}
J^\un{}_{ ;m(t)\,,n(s-2) }{}^n  =0\,,\quad
J^\un{}_{ ; m(t-1)}{}^n{}_{\,, n(s-1)}=0\,,\quad
J^\un{}_{ ;m(t) }{}^{m}{}_{\,,n(s-1)}=0\,,\quad
\ee
and
\be
\label{asymc}
J^\un{}_{ ; m(t-1)n\,, n(s-1)} =0\,.
\ee
The higher spin supercurrents associated with half-integer spins
\be
J^{\un}{}_{ ;m_1 \ldots m_t ,n_1 \ldots n_{s-3/2}}{}_{;\nu}
\ee
are vector fields (index $\un$) taking values in all
representations of the Lorentz group described by the
$\gamma$ -transversal  two-row Young diagrams
\bigskip
\be
\label{diaf}
\begin{picture}(70,50)
\put(20,45){s-3/2} \put(33,35){\circle*{2}}
\put(25,35){\circle*{2}}
\put(17,35){\circle*{2}}
\put(25,25){\circle*{2}}
\put(17,25){\circle*{2}}
\put(33,25){\circle*{2}}
\put(00,40){\line(1,0){70}}
\put(00,30){\line(1,0){70}} \put(50,30){\line(0,1){10}}
\put(60,30){\line(0,1){10}} \put(70,30){\line(0,1){10}}
\put(00,20){\line(1,0){60}} \put(00,20){\line(0,1){20}}
\put(10,20){\line(0,1){20}} \put(40,20){\line(0,1){20}}
\put(60,20){\line(0,1){20}} \put(50,20){\line(0,1){20}} \put(20,10){t}
\end{picture}
\ee
i.e., the irreducibility conditions for the higher spin
supercurrents \\ $J^\un{}_{ ;m(t)\,,  n(s-3/2) ;\nu }$ read
\be
\label{asymf}
tJ^\un{}_{;m(t-1)n,n(s-3/2);\nu} = 0
\ee
and
\be
\label{trf} (s-5/2 )\,
\gamma^n{}_{\mu}{}^{\nu} J^\un{}_{;m(t),n(s-3/2);\nu} = 0\,.
\ee
{}From these conditions it follows that
\be
\label{trbf} (s-5/2 )
\gamma^m{}_{\mu}{}^{\nu} J^\un{}_{;m(t),n(s-3/2);\nu} = 0
\ee
and all tracelessness conditions (\ref{trind}) and
(\ref{trdep}) are satisfied.

To avoid complications resulting from the projection
to the space of irreducible (i.e. traceless or
$\gamma-$ transversal) two--row Young diagrams
we study the currents
\be
J^\un ({\xi}) =
\!J^{\un}{}_{;}{}^{m(t)}{}_{,}{}^{n(s-1)}
\xi_{m(t),n(s-1)},\quad\!
J^\un ({\xi}) =
\xi_{m(t),n(s-3/2);}{}^{\nu}
J^{\un}{}_;{}^{m(t)}{}_,{}^{n(s-3/2)}{}_{;\nu},
\ee
where
$\xi_{m(t),n(s-1)} $ and
$\xi_{m(t),n(s-3/2)}{}_;{}^\nu$
 are some constant
parameters which themselves
satisfy analogous irreducibility conditions.
The conservation law then reads
\be
\label{cc}
\partial_\un
J^\un ({\xi}) =  0\,.
\ee

The currents corresponding to one-row Young diagrams
(i.e. with $t=0$) generalize the spin 1 current (\ref{cur1}),
supercurrent (\ref{cur3/2}) and stress tensor (\ref{scalstress}).
An important fact is that they can be chosen in the form
\be
\label{JT}
J^{m}{}_{ ;\,,}{}^{  n(s-1)}
\xi_{n(s-1)}= T^{m\,n(s-1)} \xi_{n(s-1)}
\ee
\be
\label{JTF}
J^{m}{}_{ ;\,,}{}^{n(s-3/2)}{}_{;\nu }
\xi_{n(s-3/2)}{}_;{}^\nu
= T^{m\,n(s-3/2)}{}_{;\nu}
\xi_{n(s-3/2)}{}_;{}^\nu \,,
\ee
with totally symmetric conserved currents
$T^{n(s)}$ or supercurrents $T^{n(s-1/2)}{}_{;\nu}$,
\be
\label{dT}
\xi_{n(s-1)}\partial_n T^{n(s)} =0 \,,\qquad
\xi_{n(s-3/2)}{}_;{}^\nu
\partial_n T^{n(s-1/2)}{}_{;\nu} =0\,
\ee
$(\xi^n{}_{n(s-2)}=0$, $(\xi_{n(s-3/2)}{} \gamma^n )^\nu =0$).

Analogously to the formula
(\ref{ang2}) for the angular momenta current, the symmetric
(super)currents $T$ allow one to construct explicitly
$x$-dependent higher spin ``angular" currents.
An observation is that the angular higher spin (super)currents
\be
\label{JTB}
\!J^\un ({\xi}) =
T^{\un n(s-1)}x^{m(t)}
\xi_{m(t),n(s-1)},\quad\!
J^\un ({\xi}) =
T^{\un\,n(s-3/2)}{}_{;\nu} x^{m(t)}
\xi_{m(t),n(s-3/2);}{}^\nu,
\ee
where we use the shorthand notation
\be
\label{xs}
x^{m(s)} =\underbrace{x^{m} \ldots x^m}_s\,,
\ee
also conserve as a consequence of (\ref{dT}) because when the
derivative in (\ref{cc}) hits a factor of $x^m$, the result
vanishes by symmetrization of too many indices in the parameters
${\xi}$  forming the two-row Young diagrams.

Since the parameters $\xi_{m(t),n(s-1)}$ and $\xi_{m(t),n(s-3/2);}{}^{\nu}$
are traceless and  $\gamma-$transversal, only the double traceless part of
$T^{\un n(s-1)}$
\be
T^{n(2)}{}_{n(s-2)} = 0 \,,\qquad s\geq 4
\ee
and triple $\gamma-$transver\-sal part of $T^{\un\,n(s-3/2)}{}_{;\nu}$
\be
\gamma^n T^{n}{}_{n(s-3/2)} = 0 \,,\qquad s\geq 7/2\,.
\ee
contribute to (\ref{JTB}).
These are the (super)currents of the formalism of symmetric
tensors (tensor-spinors).\cite{Fr,WF}
The currents with integer spins $T^{n(s)}$
were considered in \cite{cur,curan} for the particular case
of massless matter fields.

Integer spin currents built from scalars of equal masses
\be
\label{sfe}
(\Box+m^2 ) \phi^i =0\,
\ee
have the form
\bee
\label{t2k}
T^{n(2k)ij} =\!\!\!&{}&\!\!\!  ( \partial^{n(k) }\phi^i \partial^{n(k) }\phi^j
-\frac{k}{2} \eta^{nn}\partial^{n(k-1) }\partial_m \phi^i
\partial^{n(k-1) }\partial^m \phi^j\nn\\
\!\!\!&{}&\!\!\!
+\frac{k}{2} m^2 \eta^{nn}\partial^{n(k-1) } \phi^i
\partial^{n(k-1) }\phi^j )+i\leftrightarrow j
\eee for even spins and
\bee \label{t2k+1} T^{n(2k+1)ij} =\!\!\!&{}&\!\!\! (
\partial^{n(k+1) }\phi^i \partial^{n(k) }\phi^j
-\frac{k}{2} \eta^{nn}\partial^{n(k) }\partial_m \phi^i
\partial^{n(k-1) }\partial^m \phi^j\nn\\
\!\!\!&{}&\!\!\!
+\frac{k}{2} m^2 \eta^{nn}\partial^{n(k) } \phi^i
\partial^{n(k-1) }\phi^j )
-i\leftrightarrow
j \eee
for odd spins, where we ignore terms containing more than one flat
metric $\eta^{nn}$, which do not contribute to the charges (\ref{JTB}),
and use notation analogous to (\ref{xs})
\be \label{ds}
\partial^{n(s)} =\underbrace{\partial^{n} \ldots \partial^n}_s\,.\qquad
\ee

Higher spin supercurrents built from scalar and spinor
with equal masses
\be
\label{sbfe}
(\Box+m^2 ) \phi =0\,,\qquad
(i\partial_\un \gamma^\un + m )\psi_\nu =0
\ee
read
\be
\label{tfk+1}
\!T^{n(k+1)}{}_{;\nu} =
\partial^{n(k+1) }\phi \psi_\nu
-\frac{k+1}{2}
\Big (
(\gamma^n \gamma^m \psi )_\nu\partial^{n(k) }\partial_m \phi
+im (\gamma^n  \psi )_\nu\partial^{n(k)}\phi \Big )\,.
\ee

Inserting these expressions into
(\ref{JTB}) we obtain the set
of conserved ``angular" higher spin currents of even, odd
and half-integer spins.
The usual angular momentum current corresponds to the case
$s=2$, $t=1$.

Remarkably, the conserved higher spin currents listed above are
in one--to--one correspondence with the
higher spin gauge fields (1-forms) $\go_{\un ; m(t),n(s-1)}$ and
$\go_{\un ; m(t),n(s-3/2);\nu}$ introduced for the
 boson \cite{LV} and fermion \cite{V} cases in arbitrary $d$.
To the best of our knowledge, the fact that any of the
higher spin gauge fields has a dual conserved current
was never discussed before. Of course, such a
correspondence is expected because, like the
gauge fields of the supergravitational multiplets, the
higher spin gauge fields should take their values in a
(infinite-dimensional) higher spin algebra identified with
the global symmetry algebra in the corresponding
dynamical system (this fact is explicitly demonstrated
below for the cases of $d=3$ and $d=4$). The higher spin currents can
then be derived via the Noether theorem from the global
higher spin symmetry and give rise to the conserved charges
identified with the Hamiltonian generators of the same symmetries.

A few comments are now in order.

Higher spin currents contain higher derivatives. Therefore,
higher spin symmetries imply, via the Noether procedure, the
appearance of higher derivatives
in interactions. The immediate question is whether higher spin
gauge theories are local or not. As we shall see the answer is ``yes"
at the linearized level and ``probably not" at the interaction level.

It is well known that if the stress tensor is traceless this
indicates a larger  (conformal) symmetry. This property extends
to the higher spin currents provided that  the higher spin currents
are traceless.\cite{CHC}

Nontrivial (interacting) theories exhibiting
higher spin symmetries are formulated in AdS
background rather than in the flat space. Therefore
an important problem is to generalize the constructed currents
to the AdS geometry. This problem was solved recently \cite{PV2}
for the case d=3.

Explicit form of the higher spin algebras is known for
$d\leq 4$  \cite{} although a conjecture
was made in \cite{V} on the structure of higher spin
symmetries in any $d$. The knowledge of the structure of the
higher spin currents in arbitrary dimension may be very useful
for elucidating a structure of the higher spin symmetries
in any d.

\section{Higher-Spin Symmetries}
\label{Higher-Spin Symmetries}

The key element of
the theory of massless higher spin fields is
the higher spin gauge symmetry principle. Its role is as fundamental
as that of the Poincare superalgebra discovered by
Golfand and Likhtman for supersymmetric theories.
{}From the $d=4$ \cite{hsa4,OP1,KV1} and $d=3$
\cite{bl,BBS,Aq,Eq,PV} analysis it is known
that the relevant higher spin symmetry algebras $h$
are certain infinite-dimensional Lie superalgebras which give
rise to infinite chains of spins and contain $AdS_d$
algebras $o(d-1,2)$ and their superextensions as
(maximal) finite-dimensional subalgebras. To fix  conventions
let us note that the generators of the
$AdS_d$ algebra $o(d-1,2)$ can be identified with
the Lorentz generators $L^{mn}$ of $o(d-1,1)$  and the
generators of $AdS_d$ translations $P^m \in o(d-1,2)/o(d-1,1)$
with the commutation relations
\bea
&{}&[L^{mn},L^{kl}] = \eta^{nk}L^{ml}
 - \eta^{mk}L^{nl}
+\eta^{ml}L^{nk}-\eta^{nl}L^{mk}\,,\nn\\
&{}&[ L^{mn} , P^{k}]= \eta^{nk}P^{m} - \eta^{mk}P^{n}\,, \nn\\
&{}&[P^{m},P^{n}]=-\lambda^2 L^{mn}\,.
\eea
Here $\lambda^{-1}$ is identified with the $AdS$
radius. It serves as the In\"onu-Wigner contraction parameter:
 $\lambda \to 0$ in the flat limit.

A structure of higher spin algebras $h$
is such that no higher spin ($s > 2 $)  field
can remain massless unless it belongs to an infinite chain of
massless higher spins with infinitely increasing spins.
Unbroken higher spin symmetries require AdS background.
One can think however of some spontaneous
breakdown of the higher spin symmetries followed by a flat
contraction via a shift of the vacuum energy in the
broken phase. In a physical phase with
$\lambda =0$ and $m\gg m^{exp} $ for higher spin fields,
$h$ should break down to a finite-dimensional subalgebra
$
g=\{M^{mn} ,\, P^m ,\,Q^i_\nu \} \oplus T^i{}_j \,,
$
giving rise to usual lower spin gauge fields.
Here the first set of the generators corresponds to a
SUSY algebra while the second one describes some inner
(Yang-Mills) part. {}From this perspective the Coleman-Mandula
type theorems can be  re-interpreted as statements concerning a
possible structure of $g$ rather than the whole higher spin
algebra $h$ which requires AdS geometry.
These arguments are based on the $d\leq 4$ experience
but we expect them to large extend to be true
for higher dimensions. The origin of higher spin
symmetries can be traced back \cite{cur} to the
 higher spin conserved currents with higher derivatives
discussed in the Sec. \ref{Higher Spin Currents}.
Let us summarize the main results for $d=3$ and $d=4$ following
to \cite{hsa4,OP1,KV1,Aq,Eq,PV}.

\subsection{d=3 Higher Spin Symmetries and Deformed Oscillators}
\label{d=3 Higher Spin Symmetries and Deformed Oscillators}

The $AdS_3$ algebra is semisimple, $o(2,2)\sim sp(2;R)\oplus sp(2;R)$
with the diagonal subalgebra $sp(2;R)\sim o(2,1)$
identified with the Lorentz subalgebra. A particularly
useful realization of the $AdS_3$ generators is
\be
\label{al}
   L_{\gal\gb}=\frac1{4i}\{\hat{y}_\gal,
    \hat{y}_\gb\}\,,\qquad \,
   P_{\gal\gb}=\frac1{4i}\{\hat{y}_\gal,
    \hat{y}_\gb\}\psi  \,
\ee
with the generating elements $\hat{y}_\ga$ and $\psi$ obeying the relations
$[\hat{y}_\alpha , \hat{y}_\beta ] = 2i \epsilon_{\alpha\beta}$,
$\epsilon_{\alpha\beta}=-\epsilon_{\beta\alpha}$, $\epsilon_{12}=1$
and
\be
\psi^2=1\,,\qquad [\psi ,\hat{y}_\alpha ]=0\,.
\ee
The projectors to the simple components of $o(2,2)$ are identified with
$\Pi_\pm = \frac{1}{2} (1\pm \psi )$.

In \cite{BF,BBS,H,Aq,FLU} it was shown that there exists a
one-parametric class of infinite-dimensional algebras which we
denote $ hs(2;\nu )$ ($\nu$ is an arbitrary real parameter),
all containing $sp(2)$ as a subalgebra. This allows one to
define a class of higher spin algebras
$
g=hs(2;\nu )\oplus hs(2;\nu ).
$
The supertrace operation was defined in \cite{Aq} where also a
useful realization of the supersymmetric extension of
$ hs(2;\nu )$ was given, based on a certain deformed oscillator algebra.
Since this construction will play a key role below
let us explain its properties in somewhat more detail.

Consider an associative algebra
$Aq(2;\nu )$
with generic element of the form
\be
\label{sel}
f(\ty,k )=
\sum^\infty_{n=0}\sum_{A=0,1}
\!
\frac{1}{n!}
 f^{A\,\alpha_1\ldots\alpha_n}(k)^A
 \ty_{\alpha_1}\ldots \ty_{\alpha_n}\,,
\ee
under condition that the coefficients
$ f^{A\,\alpha_1\ldots\alpha_n}$ are symmetric with respect to
 the indices
$\ga_j =1,2$ and that the generating elements $\ty_\ga$ satisfy the
 relations
\bee
\label{modosc}
[\ty_\ga ,\ty_\gb ]=2i
\gee_{\ga\gb} (1+\nu k)\,,\quad k\ty_\ga
=-\ty_\ga k\,,
\quad k^2 =1\,,
\eee where $\nu $ is an arbitrary constant (central element).
In other words,
$Aq(2;\nu )$ is the enveloping algebra for the relations
(\ref{modosc}) often called deformed oscillator algebra.

An important property of this algebra is that, for all $\nu$,
the bilinears
\be
\label{lorq}
T_{\ga\gb}=\frac{1}{4i}\{\ty_\ga \,,\ty_\gb \}
\ee
have $sp(2)$ commutation relations and rotate $\ty_\ga$ as a $sp(2)$ vector
\bee
\label{q1com}
[T_{\ga \gb},
T_{\ggg \eta}]\!=\! \gee_{\ga\ggg} T_{\gb\eta}\!+\!
\gee_{\gb\ggg} T_{\ga\eta}\!+\!
\gee_{\ga\eta} T_{\gb\ggg}\!+\! \gee_{\gb\eta} T_{\ga\ggg}\,,
\eee
\bee
\label{q2com}
[T_{\ga\gb} , \ty_{\ggg }]
\!=\! \gee_{\ga\ggg}\ty_{\gb} \!+\! \gee_{\gb\ggg}\ty_{\ga}.
\eee

The deformed oscillators described above
have a long history and
were originally discovered by Wigner \cite{wig}
who addressed a question whether it is possible to modify
the commutation relations for the normal oscillators
$a^\pm$ in such a way
that the basic commutation relations $[H, a^\pm] =\pm a^\pm $,
$H=\frac{1}{2} \{a^+ ,a^- \}$ remain
valid. By analyzing this problem in the Fock-type space
Wigner found a one-parametric deformation of the standard commutation
relations which corresponds to a particular realization of the
commutation relations (\ref{modosc})  with the identification
$a^+ =\ty_1$,
$a^- =\frac{1}{2i} \ty_2$, $H=T_{12}$ and $k=(-1)^N$ where $N$ is the
particle number operator. These commutation relations were
discussed later by many authors in particular
in the context of parastatistics (see, e.g.,~\cite{des}).

According to (\ref{lorq}) and (\ref{q2com})
the $sp(2)$ symmetry generated by $T_{\ga\gb}$
extends to $osp(1,2)$ by identifying the supergenerators with
$\ty_\ga$. In fact, as shown in \cite{BWV}, one can start from the
$osp(1,2)$ algebra to derive the deformed oscillator commutation
relations. Since this construction is instructive in many respects
we reproduce it here.

One starts with the (super)generators
$T_{\alpha\beta}$ and $\ty_\alpha$
which, by definition of $osp(1,2)$, satisfy the commutation
relations (\ref{lorq})-(\ref{q2com}). Since
 $\alpha$ and $\beta$ take only two values one can write
\be
\label{co}
[\ty_\ga ,\ty_\gb ]=2i
\gee_{\ga\gb} (1+Q)\,,
\ee
where $Q$ is some new ``operator" while the unit term is singled out for
convenience.
Inserting this
back into (\ref{q2com}) with the substitution
of (\ref{lorq}) and completing the commutations
one observes that (\ref{q2com}) is
true if and only if $Q$ anticommutes with $\ty_\alpha$\,,
\be
\label{Q}
Q \ty_\alpha =-\ty_\ga Q.
\ee
The relation (\ref{q1com}) does not add anything new since it
is a consequence of (\ref{lorq}) and (\ref{q2com}).
As a result we arrive \cite{BWV} at the following
important fact:  the enveloping algebra of $osp(1,2)$,
$U(osp(1,2))$, is isomorphic to the enveloping algebra of the deformed
oscillator relations (\ref{co}) and (\ref{Q}).
In other words, the associative algebra with the generating elements
$\ty_\ga$ and $Q$ subject to the
relations (\ref{co}) and (\ref{Q})
is the same as the associative algebra with the generating elements
$\ty_\ga$ and $T_{\ga\gb}$ subject to the osp(1,2) commutation relations
(\ref{lorq})-(\ref{q2com}).

Computing the quadratic Casimir operator of $osp(1,2)$
\be
C_2 = -\frac{1}{2}T_{\ga\gb}T^{\ga\gb} -\frac{i}{4}\ty_\ga \ty^\ga\,,
\ee
one derives using (\ref{co}) that
\be
C_2 =-
\frac{1}{4} (1- Q^2)\,.
\ee

Let us now consider the factor algebra of $U(osp(1,2))$
with respect to its
ideal
$I_{(C_2 + \frac{1}{4} (1- \nu^2))}$
generated by the central element
$(C_2 + \frac{1}{4} (1- \nu^2) )$ where $\nu$ is an
arbitrary number. In other words, every element of
$U(osp(1,2))$ of the form $\left( C_2 + \frac{1}{4} (1- \nu^2)\right) a$,
$\forall a\in U(osp(1,2))$ is supposed to be equivalent to zero.
This factorization can be achieved in terms of the
deformed oscillators (\ref{co}), (\ref{Q}) by
setting~\footnote{The point $\nu =0$ is special since it may
happen that $Q^2 =0$, $Q\neq 0$.}
\be
Q=\nu k\,,\qquad k^2 =1\,, \qquad k\ty_\ga =-\ty_\ga k\,.
\ee

Thus, it is shown \cite{BWV} that the algebra $Aq(2,\nu )$ introduced in
\cite{Aq} is isomorphic to
$U(osp(1,2))/
I_{(C_2 + \frac{1}{4} (1- \nu^2))}$.
This fact has  a number of simple
but important consequences. For example, any representation of
$osp(1,2)$ with $C_2 =  -\frac{1}{4} (1- \nu^2)$
forms a representation of
$Aq(2,\nu )$ ($\nu \neq 0$) and vice versa (for any $\nu$).
In particular this is the case for finite-dimensional
representations corresponding to the values $\nu =2l+1$, $l\in
{\bf Z}$ with $C_2 = l(l+1)$.

Let us note that the even
subalgebra of $Aq(2;\nu )$
spanned by the elements of the form (\ref{sel}) with
$f(\ty,k)=f(-\ty,k)$
decomposes into a direct sum of two subalgebras
$Aq^E_\pm (2;\nu )$ spanned by the elements
$P_\pm f(\ty,k)$ with
$f(-\ty,k)=f(\ty,k)$, $P_\pm = \frac{1}{2} (1\pm k)$. These algebras can be
shown  to be  isomorphic to the factor algebras
$U(sp(2))/I_{(C_2 + \frac{3\pm 2\nu -\nu^2}{4})} $
where
$C_2 = -\frac{1}{2}T_{\ga\gb}T^{\ga\gb}$ is the quadratic
Casimir operator of $sp(2)$ and can be interpreted
as (infinite-dimensional) algebras interpolating between the ordinary
finite-dimensional matrix algebras. Such interpretation
of $U(sp(2))/I_{(C_2 -c)} $ was given by Feigin in \cite{BF}
and Fradkin and Linetsky in \cite{FLU}.

An important property of
$Aq(2;\nu )$
is that it admits \cite{Aq} a uniquely defined supertrace operation
\be
\label{str}
str (f) =f^0 -\nu f^1\,,
\ee
such that
$str(fg) = (-1)^{\pi_f \pi_g} str(gf)
$, $\forall f,g $ having a definite parity,
$f(-\ty,k)
\!\!=\!\!(-1)^{\pi_f}\!f(\ty,k)
$
(i.e. $str (1)=1\,,\quad str(k) =-\nu$ while all higher
 monomials of $\ty_\ga$ in (\ref{sel}) do not contribute under the supertrace).
This supertrace reduces
to the ordinary supertrace of finite-dimensional algebras
for the special values of the parameter $\nu=2l+1$ corresponding
to the values of the Casimir operator related to the finite-dimensional
representations of $osp(1,2)$ ($sp(2)$ in the bosonic case).
This property allows one to handle the algebras
$Aq(2;\nu )$ very much the same way as ordinary finite-dimensional
(super)matrix algebras. What happens for special values of
$\nu=2l+1$ is that $Aq(2;\nu )$
acquires ideals $I_l$ such that $Aq(2;\nu )/I_l$
amounts to appropriate (super)matrix algebras. These ideals
were described  in \cite{Aq} as null vectors of the invariant
bilinear form $str (ab)$, $a,b \in Aq(2;\nu )$.
Note that the limit $\nu\to\infty$ corresponds to the algebra of
area preserving diffeomorphisms in accordance with the original
matrix analysis \cite{HOPPE}.

By construction, $Aq(2;\nu )$
possesses $N=1$ supersymmetry as inner $osp(1,2)$
automorphisms. A more interesting property \cite{BWV} is that
it admits $N=2$ supersymmetry $osp(2,2)$ with the generators
\be
\label{N2}
T_{\ga\gb}=\frac{1}{4i}\{\ty_\ga \,,\ty_\gb \}\,,\quad
Q_\ga =\ty_\ga \,,\quad
S_\ga =\ty_\ga k\,,\quad J=k+\nu\,.
\ee

The deformed oscillator algebras are important
for the description of the higher spin dynamics
not only in d=3 but also in d=4. The reason is
that, as shown in section \ref{Interaction Ambiguity},
they allow one to formulate a non-linear dynamics
with explicit local Lorentz symmetry as a consequence of
(\ref{q1com}). In its turn, the analysis of the higher spin
dynamics in the section \ref{Nonlinear Equations in d=3}
is interesting in the context of the deformed oscillator algebra
itself because algebraically it reduces to the construction of
some its embedding  into a direct product of two Weyl
(i.e. oscillator) algebras equipped with certain twist operators.

Thus, the $AdS_3$ algebra is the algebra of bilinears in the
oscillators $\hat{y}$. Extension to higher spin
 algebras consists of allowing
arbitrary powers in the oscillators. Namely the
 higher spin gauge fields
$w(\ty,\psi,k|x)$=$dx^\nu w_\nu (\ty,\psi,k|x)$
have a form
\be
\label{ghs}
w(\ty,\psi,k|x)
  \!\! =  \!\!\!\! \sum^\infty_
    {B=0,1\,n=0 }
    \frac{1}{n!}  w^{A\,B\gal_1\ldots\gal_n}(x) k^A \psi^B
    \ty_{\gal_1}\ldots \ty_{\gal_n}\,.
\ee
The components $w^{A\,B\gal_1\ldots\gal_n}(x)$
are identified with the higher spin gauge fields in the d=3
space-time with the coordinates $x^\un$.
The higher spin field strengths have a standard form
\be
\label{cur}
   R(\ty,\psi,k|x) = dw(\ty,\psi,k|x) -
   w(\ty,\psi,k|x) \wedge w(\ty,\psi,k|x) \,,
\ee
where $d=dx^\un \frac{\partial}{\partial x^\un} $.
This construction for the ordinary oscillators was suggested
in \cite{bl}.

The structure of the higher spin gauge fields
(\ref{ghs}) suggests that the d3 higher spin currents
should form a similar set
\be
\label{d3c}
  J^{\un ; A\,B\gal_1\ldots\gal_n}(x) \,.
\ee
Here, the two indices $A=0,1$ and $B=0,1$ play
different roles. The label $A$ describes the doubling of all fields
as a consequence of $N=2$ supersymmetry in the theory.
This doubling can in principle be avoided in an appropriately
truncated theory \cite{PV}. The label $B$ distinguishes
between the Lorentz--type fields $(B=0)$ and frame--type
fields $(B=1)$ and therefore leads to two different
types of currents of any spin $s=\half n$.

To see that such a structure of currents fits the
systematics of the section
\ref{Higher Spin Currents} one observes that in three dimensions all
traceless two-row Young diagrams
(\ref{dia}) with $t>1$ vanish. Indeed, consider the quantity
\be
\label{YX}
 Y^{cc}{}_{b(t),a(p)} =
\gep^{abc} \gep^{abc} X_{b(t+2),a(p+2)}
\ee
with some traceless $X_{b(t+2),a(p+2)}$ satisfying the antisymmetry condition\\
$X_{b(t+1)a,a(p+2)}=0$. On the one hand, one gets
\be
X_{b(t+2),a(p+2)} =\frac{p+1}{p+3}
\gep_{abc} \gep_{abc}
 Y^{cc}{}_{b(t),a(p)} \,.
\ee
But on the other hand, from (\ref{YX}) it follows that
$ Y^{cc}{}_{b(t),a(p)} =0$ since at least one pair of indices
of $X_{b(t+2),a(p+2)} $ will be contracted after the two epsilon symbols
are replaced by the combinations of the metric tensors.

As a result, only the currents corresponding to the Young diagrams
(\ref{dia}) with $t=0$ and $t=1$ survive in the d3 case.
The one-row currents ($t=0$) describe
stress tensor-type higher spin currents and correspond
to the currents (\ref{d3c}) with $B=1$ while
those with a single box in the second row describe higher spin
angular momentum--type currents and correspond to
the currents (\ref{d3c}) with $B=0$.

As shown in \cite{BPV}, the ambiguity in $\nu$
identifies with
the ambiguity in the parameter of mass in the d=3 matter systems.
Thus, a structure of global higher spin algebra depends on a
particular type of matter systems. This property manifests
the fact \cite{PV} that
 the parameter of mass in the matter field sector of
the d3 higher spin system arises from a vacuum expectation
value of a certain scalar field in the model, while the
global symmetry algebra identifies with the centralizer of
the vacuum solution in the larger gauge algebra having a
universal structure (see section \ref{Nonlinear Equations in d=3}).

A  special property of the d3 higher spin systems is
that higher spin gauge fields do not propagate in d=3 in analogy
with the usual Chern-Simons gravitational \cite{d3gr}
and Yang-Mills fields
although the higher spin gauge symmetries remain
nontrivial, like the gravitational (spin 2) and inner (spin 1)
symmetries.  The matter fields do propagate.
In other words, higher spin gauge fields and matter fields belong
to different higher spin multiplets and the parameter of
mass of matter fields remains arbitrary. In higher dimensions
the situation changes significantly  because the
higher spin gauge fields are propagating massless fields and
 matter fields may belong to the same multiplet of higher spin
symmetry with the higher spin gauge fields thus necessarily
being massless just as in ordinary SUSY supermultiplets.

\subsection{d=4 Higher Spin Symmetries}
\label{d=4 Higher Spin Symmetries}

Algebraically, the situation in d=4 is analogous. The isomorphism
$o(3,2)\sim sp(4|R)$ allows one to realize the generators of
Lorentz transformations
$L_{\gal\gb}$,
$\bar{L}_{\dot{\gal}\dot{\gb}}$ and  AdS translations $P_{\gal\dgb}$
as bilinears
\be
\label{al4}
   L_{\gal\gb}=\frac1{4i}\{\hat{y}_\gal,
    \hat{y}_\gb\}\,,\qquad \,
   \bar{L}_{\dot{\gal}\dot{\gb}}
=\frac1{4i}\{\hat{\bar{y}}_{\dot{\gal}},
    \hat{\bar{y}}_{\dot{\gb}}\}\,,\qquad \,
   P_{\gal\dot{\gb}}=\frac1{2i}\hat{y}_\gal
    \hat{\bar{y}}_{\dot{\gb}}\,
\ee
in the oscillators obeying the relations
\be
\label{d4osc}
[\hat{y}_\alpha , \hat{y}_\beta ] = 2i \epsilon_{\alpha\beta}\,,\qquad
[\hat{\bar{y}}_{\dot{\alpha}} , \hat{\bar{y}}_{\dot{\beta}} ] = 2i
\epsilon_{\dot{\alpha}\dot{\beta}}\,,\qquad
[\hat{y}_\alpha , \hat{\bar{y}}_{\dot{\beta}} ] =0\,.
\ee

The simplest version of the d=4 higher spin algebra \cite{FVA}
 is identified with
the algebra of all polynomials of the oscillators
and Klein operators
 $k$ and
$\bar{k}$ having the properties
\be
k^2 = \bar{k}^2 =1\,,\qquad  k\hat{y}_\gal = - \hat{y}_\gal k\,,\qquad
\bar{k}\hat{\bar{y}}_{\dot{\gal}}= - \hat{\bar{y}}_{\dot{\gal}}\bar{ k}
\ee
(all barred operators commute to all unbarred ones),
i.e. the corresponding gauge fields have a form \cite{FVA}
\be
W=\!\!\!\sum_{A,B=0,1}\! \sum^\infty_ {n,m=0}
   \! \frac{1}{2i m!n!} dx^\un w_\un{}^{A\,B\gal_1\ldots\gal_n\,
{\dga}_1 \ldots {\dga_m}} (x) k^A \bar{k}^B
    \ty_{\gal_1}\ldots \ty_{\gal_n}\,
    \hat{\bar{y}}_{\dga_1}\ldots \ty_{\dga_m}.
\ee
In accordance with (\ref{al4}),
the gauge fields bilinear in the oscillators are identified with
the gravitational fields.

The operators $k$ and $\bar{k}$ again play a role in the
construction and lead to N=2 supersymmetry \cite{FVA} $osp(2,4)$
with the supergenerators
\be
\label{n2su}
Q^1_\gal = \hat{y}_\gal\,,\qquad
Q^2_\gal = ik\bar{k}\hat{y}_\gal\,,\qquad
\bar{Q}^1_\dga = \hat{\bar{y}}_\dga\,,\qquad
\bar{Q}^2_\dga = ik\bar{k}\hat{\bar{y}}_\dga\,,\qquad
\ee
and the $o(2)$ generator
\be
\label{U1}
J= ik\bar{k}\,.
\ee
Factoring out the trivial $u(1)$ subalgebra associated with
the unit element (as explained in the section \ref{Star Product},
the resulting subalgebra is spanned by traceless elements) this
$osp(2,4)$ becomes a maximal finite-dimensional subalgebra
of the infinite-dimensional d=4 higher spin algebra.

According to \cite{V2,VA,FVA} the fields
\be
W (y,\hat{\bar{y}}, k,\bar{k};x)=
W (y,\hat{\bar{y}}, -k,-\bar{k};x)
\ee
describe the higher spin fields while the fields
\be
\label{af}
W (y,\hat{\bar{y}}, k,\bar{k};x)= -
W (y,\hat{\bar{y}}, -k,-\bar{k};x)
\ee
are auxiliary, i.e. do not describe nontrivial degrees of freedom.
(For that reason this version of the higher spin algebra was called
in \cite{FVA} algebra of higher spins and auxiliary fields $shsa(1)$.)
Therefore we have two sets of higher spin potentials
\be
    w^{A\,A\gal_1\ldots\gal_n\,
{\dga}_1 \ldots {\dga}_m}(x) \,,\qquad \mbox{A=0 or 1}\,.
\ee
The subsets associated with spin $s$ are fixed by the condition
\cite{V2} $s=1+\half (n+m)$. The complex
conjugation transforms dotted indices into undotted and vice versa, thus
mapping $n$ to $m$.

For a fixed value of $A$ we therefore expect a set of currents
\be
    J^{\un}{}_{; \gal_1\ldots\gal_n\,,
{\dga}_1 \ldots {\dga}_m}(x) \,.
\ee
In accordance with the results of the section
\ref{Higher Spin Currents}, it indeed
 describes in terms of two-component spinors the
set of all two-row Young diagrams (\ref{dia}) both in the integer spin
case ($n+m$ is even) and in the half-integer spin case ($n+m$
is odd), with the identification
\be
s = \half (n+m) +1 \,,\qquad t = \left[\half |n-m|\right]\,,
\ee
where $[a]$ denotes integer part of $a$.
The fact that d3 and d4 higher spin algebras give rise
to the sets of gauge fields which exactly match the
sets of conserved higher spin currents of the
section \ref{Higher Spin Currents} is very significant.

\subsection{Extended Higher Spin Algebras}
\label{Extended Higher Spin Algebras}

Higher spin dynamical systems
admit a natural extension to the case with non-Abelian
internal (Yang-Mills) symmetries, as was discovered
for the $d4$ case in \cite{Ann,KV1} and then confirmed
for the $d3$ case in \cite{Eq,PV}. The key observation is that
higher spin dynamics remains consistent if components of all
fields take their values in an arbitrary associative algebra $M$
with unity $I_M$. For example
$
W(y,\bar{y}, k, \bar{k}|x )\rightarrow
W^i{}_j (y,\bar{y}, k, \bar{k}|x )$,
$ i,j =1\ldots n$
for $M=Mat_n$.
The gravitational sector is associated with the fields proportional
to $I_M$. Therefore, $M$ describes internal symmetries in the model.
For the case of semisimple finite-dimensional inner symmetries,
$M$ has to be identified with some matrix algebra.
This observation leads to a class of higher spin systems with
different semisimple Yang-Mills algebras classified
for d=4 in \cite{KV1} and for  $d3$  in \cite{PV}.
Note that, the Yang-Mills coupling constant $g^2$ is the
only dimensionless constant in the theory. It identifies
with the dimensionless combination of the cosmological constant
and the gravitation constant $g^2 \sim \Lambda \kappa^2$
analogously to the case of gauged supergravity \cite{FVpr,DF}.

In this section, we list following to \cite{KV1}
all global extended d=4 higher spin
symmetries with finite-dimensional internal symmetries.
The case d=3 can be considered analogously \cite{PV}.

To solve the problem it is useful to investigate
most general truncations
of the extended higher spin theories with arbitrary matrix algebras
$M$, which lead to consistent higher spin dynamics.
To this end one finds such
automorphisms $\tau$ of the higher spin algebras which
leave invariant the $AdS_4$ subalgebra associated with
the gravitational sector and the nonlinear equations
discussed in the section
\ref{Nonlinear Equations in d=4}. The latter condition reduces to
the simple additional requirement that, in presence of fermions,
 the allowed automorphisms should leave invariant the operator
\be
\label{KK}
K=k\bar{k}\,,
\ee
which enters explicitly the nonlinear higher spin equations
as explained in the section
\ref{Nonlinear Equations in d=4}. (In the purely bosonic case,
$K$ becomes a central element and can be truncated away.)
Truncating away all degrees of freedom except for singlets with respect to
the discrete symmetries associated with $\tau$
one is left with some consistent truncation of the full system
being itself a consistent dynamical system. In particular, in the
sector of gauge potentials the truncation conditions read
$\tau (w )= w$. The requirement that the AdS algebra associated
with the gravitational sector is $\tau-$invariant is imposed to
guarantee that the vacuum gravitational fields
necessary for any relativistic perturbative
interpretation of the model survive in the truncated system.

For example the automorphism $\tau$ defined according to
\be
\label{taukk}
\tau (k) = -k\,,\qquad \tau (\bar{k}) = -\bar{k}
\ee
can be used to
truncate away the auxiliary fields (\ref{af}). In the rest of this
section we will be only concerned with the sector of dynamical fields.
As a result, the generating elements $k$ and
$\bar{k}$ only appear in this sector in the combination $K$.

Using a combination of an appropriate involutive inner automorphism $t$
of the matrix algebra $M$ with the boson-fermion parity automorphism $f$
of the algebra of oscillators
 one arrives at the class of higher spin algebras
realized as $(n+m)\times (n+m)$ matrices with the elements
depending on the
operators $\hat y_\alpha $ and $\hat{\bar{y}}_{\dot{\alpha}}$
\begin{equation}
\label{P}
P_i{}^j(\hat{y},\hat{\bar{y}})={\kern 2em\raise
2.5ex\hbox{$n$}}{\kern -0.7em\raise -2.5ex\hbox{$m$}}\phantom{n}
\begin{tabular}{|l|l|} \multicolumn{2}{l}{\kern 2.5em\hbox{$n$}\kern
6em\hbox{$m$}} \\ \hline $P^E{}_{i^\prime}{}^{j^\prime}
(\hat{y},\,\hat{\bar{y}})\phantom{\biggl(}$  &
$P^O{}_{i^\prime}{}^{j^{\prime\prime}}(\hat{y},\,\hat{\bar{y}})$ \\
\hline
$P^O{}_{i^{\prime\prime}}{}^{j^\prime}(\hat{y},\,\hat{\bar{y}})
\phantom{\biggl(}$
& $P^E{}_{i^{\prime\prime}}{}^{j^{\prime\prime}}(\hat{y},\,\hat{\bar{y}})$ \\
\hline
\multicolumn{2}{c}{}
\end{tabular}
\end{equation}
 for the conditions that
\be
\label{odda}
P^E{}_i{}^j (\hat{y},\hat{\bar{y}})=P^E{}_i{}^j(-\hat{y},-\hat{\bar{y}})\,,
\qquad
P^O{}_i{}^j (\hat{y},\hat{\bar{y}})=-P^O{}_i{}^j(-\hat{y},-\hat{\bar{y}})\,,
\ee
$i.e.$ the diagonal blocks $P^E$ are bosonic
(the power series coefficients carry even numbers of spinor
indices) while the off-diagonal blocks $P^O$ are fermionic
(the power series coefficients carry odd numbers
of spinor indices).
The operator $K$ that anticommutes to fermions is realized as
\begin{equation}
\label{K}
K={\kern 2em\raise 2.5ex\hbox{${}$}}{\kern -0.7em\raise
-2.5ex\hbox{${}$}}\phantom{n} \begin{tabular}{|l|l|} \multicolumn{2}{l}{\kern
1.5em\hbox{${}$}\kern 1em\hbox{${}$}} \\
\hline
$\phantom{)}I$  &
$\phantom{)}0$ \\ \hline
$\phantom{)}0$  &
$-I$ \\
\hline \multicolumn{2}{c}{} \end{tabular} \quad\,.
\end{equation}

By definition (see e.g. \cite{kirill}), Weyl algebra $A_{n}$
is the associative algebra with unity $I$ and
the generating elements $y_\nu$ ($\nu ,\mu = 1 \ldots 2n$)
satisfying $[ y_\nu ,y_\mu ]= C_{\nu\mu}I$
for some nondegenerate skewsymmetric matrix $C_{\nu\mu}=-C_{\mu\nu}$,
i.e.$A_{n}$ is the algebra of $n$ oscillators,
$A_{n}=(\otimes A_1 )^n $. We denote the associative algebra
spanned by the elements of the form (\ref{P}), (\ref{odda})
$A_{l}^{n,m}$. Obviously, $A_{l}^{n,m}\sim A_{l}^{m,n}$.
The complex higher spin Lie superalgebra $hgl(n,m|2l\mid{\bf C})$
is defined via supercommutators
in $A_{l}^{n,m}$ with respect to the grading (\ref{odda}) (of course,
any other field instead of ${\bf C}$ can be formally used).
The algebra $hgl(n;m|2l)$ is not simple, containing a trivial central
element associated with the unit element of $A\otimes M$, i.e.
\be
\label{hsl}
hgl(n;m|2l) = u(1)\oplus hsl(n;m|2l)\,,
\ee
where $hsl(n;m|2l)$ is  the traceless part
of $hgl(n;m|2l)$ according to the definition of the trace \cite{OP1}
explained in the section \ref{Star Product}.

The complex algebra $hgl(n,m|2l)$ admits a number of real forms.
Of the most interest are those
which lead to unitary dynamics and, in particular, to compact inner
symmetries. The corresponding real higher spin algebras $hu(n;m|4)$
are spanned by elements  obeying the reality conditions
\begin{equation}
[P_i{}^j(\hat{y},\hat{\bar{y}})]^\dagger
=-(i)^{\pi (P_i{}^j)}\,P_i{}^j(\hat{y},\hat{\bar{y}})
\end{equation}
 with
\begin{equation}
(\hat{y}_\alpha )^\dagger =\hat{\bar{y}}_{\dot{\alpha}},\qquad \pi (P^E
)=0,\qquad \pi (P^O )=1\,
\end{equation}
(the algebras $hu(n,m|2l)$ are defined analogously for $l=2k$).

Although being non-simple due to (\ref{hsl})
the algebra $hsu(n;m|4)$ plays a
fundamental role in the higher spin theory because all its
gauge fields appear in the spectrum including the spin 1 field
associated with the $u(1)$ central subalgebra. This is a manifestation
of the general fact that the higher spin gauge theories are based
on the  associative algebra structure rather than on the Lie
(super)algebra structure as lower spin theories.

It turns out that the algebras $hu(n;m|2l)$
exhaust all elementary higher spin superalgebras
which can be obtained with the aid of automorphisms
of the underlying associative algebra $A_2^{n,m}$.
Further truncations of $hu(n;m|4)$ are based on the antiautomorphisms of
$A_2^{n,m}$ which induce automorphisms of $hu(n;m|4)$
(for more detail on the relationship between antiautomorphisms of
associative algebras and automorphisms of the related Lie superalgebras
see for example \cite{OP1}). These give rise to
the higher spin algebras of orthogonal and
symplectic types denoted $ho(n;m|4)$ and $husp(n;m|4)$,
 respectively, which also lead to consistent equations
of motion for infinite sets of massless fields of different spins
via appropriate truncations of the higher spin equations
corresponding to $hu(n;m|4)$ \cite{{Ann},{KV1}}. These
subalgebras are extracted from $hu(n;m|4)$ by the conditions
\begin{equation}
P_k{}^l(\hat{y},\hat{\bar{y}}) =
-(i)^{\pi (P_k{}^l )}\,
\eta^{lu} P_u{}^v(i\hat{y},i\hat{\bar{y}})\eta^{-1}{}_{vk}
\end{equation}
 with some nondegenerate bilinear form $\eta _{kl}$. If $\eta _{kl}$ is
symmetric, $\eta _{kl}=\eta _{lk}$, this leads to the orthogonal algebras
$ho(n;m|4)$. The skewsymmetric form, $\eta _{kl}=-\eta _{lk}$,
gives rise to the symplectic algebras $husp(n;m|4)$ ($n$ and $m$
should be even in the latter case). The isomorphisms
$hu(n;m|4)\sim hu(m;n|4)$, $ho(n;m|4)\sim ho(m;n|4)$ and
$husp(n;m|4)\sim husp(m;n|4)$ take place as a consequence of the
isomorphism $A_{l}^{n,m}\sim A_{l}^{m,n}$.

Spin$-1$ Yang-Mills subalgebras of the higher spin algebras defined this way
are spanned by the matrices independent of the operators $\hat y_\alpha $ and
$\hat{\bar y}_{\dot{\alpha}}$
As a result, the Yang-Mills subalgebras coincide with $u(n)\oplus u(m)$,
$o(n)\oplus o(m)$ and $usp(n)\oplus usp(m)$ for $hu(n;m|4)$, $ho(n;m|4)$ and
$husp(n;m|4)$, respectively. Thus, all types of compact
Lie algebras which belong to the
classical series $a_n$, $b_n$, $c_n$ and $d_n$
can be realized as spin-1 Yang-Mills
symmetries in appropriate higher spin theories.

The multiplicities of massless particles of different spins
in higher spin theories
based on extended superalgebras are \cite{KV1}
\begin{equation}
\label{multi}
\begin{tabular}{|l|c|c|c|} \hline \hfill spin                &
       $ \mbox{ odd }$            &   $\mbox{ even }  $                &
       $\mbox{ half-}$         \\ algebra\hfill                 &
&               &    $\!\!\!\mbox{- integer }\!\!\! $  \\ \hline

$hu(n;m|4)\phantom{\kern -1em\biggl(}$ &    $n^2 +m^2$ &    $n^2+m^2$ &
$2n m$ \\
\hline
$ho(n;m|4)\phantom{\kern -1em\biggl(}$ &
$\!\!\frac{1}{2}(n(n-1)+m(m-1))\!\!$ &
$\!\!\frac{1}{2}(n(n+1)+m(m+1))\!\!$ &   $nm$     \\
\hline
     $husp(n;m|4)\phantom{\kern -1em\biggl(}$ &
$\!\!\frac{1}{2}(n(n+1)+m(m+1))\!\!$       &
$\!\!\frac{1}{2}(n(n-1)+m(m-1))\!\!$ &   $n m$     \\
\hline
\end{tabular}
\end{equation}
Let us note that the fields of all odd spins belong to the
adjoint representations of the corresponding Yang-Mills algebras
while even spins always belong to a reducible representation
which contains a singlet component. This property is
important since this singlet component corresponds to the spin$-2$
colorless field to be identified with graviton. In other words,
the finite-dimensional algebra spanned by the elements
$(I\otimes$  $bilinears$ $ in$ $\hat y$ $, $ $\hat{\bar y})$
is a proper subalgebra of all higher spin algebras.
This is not surprising of course, because only those truncations
have been considered that leave invariant the gravitational subalgebra
$sp(4)$.

The original higher spin algebra constructed in the section
\ref{d=4 Higher Spin Symmetries} (with truncated auxiliary
sector by virtue of (\ref{taukk})) is isomorphic to
$hu(1;1|4)$. To see this one identifies $K$ according to
(\ref{K}) and
\begin{equation}
\label{Y}
(\hy_\ga ,\hat{\bar{y}}_\dga) \rightarrow
{\kern 2em\raise 2.5ex\hbox{${}$}}{\kern -0.7em\raise
-2.5ex\hbox{${}$}}\phantom{n} \begin{tabular}{|l|l|} \multicolumn{2}{l}{\kern
2.5em\hbox{${}$}\kern 3em\hbox{${}$}} \\
\hline
$\phantom{M}0\phantom{\biggl(}$  &
$(\hy_\ga ,\hat{\bar{y}}_\dga)$ \\ \hline
$(\hy_\ga ,\hat{\bar{y}}_\dga)\phantom{\biggl(}$  &
$\phantom{M}0$ \\\hline \multicolumn{2}{c}{} \end{tabular} \,\qquad.
\end{equation}
The case with $n=0$ or $m=0$ corresponds to
the purely bosonic higher spin theories.

{}From the structure of the higher spin
superalgebras it is clear why consistent interaction for a spin $s\geq 2$
field in d=3+1 is only possible in presence of infinite sets of massless
fields of infinitely increasing spins. The reason is that
any field of spin $s\geq 2$ corresponds to
generators which are some
 $deg >2$ polynomials of $\hat{y}$ and $\hat{\bar{y}}$
so that their  commutators lead to higher
and higher polynomials.
The same happens when one replaces the unit matrix $I$ by some
non-Abelian matrix algebra for bilinear polynomials in $\hat y$ and
$\hat{\bar y}$, $i.e$. spin$-2$ particles possessing a non-Abelian
structure require higher spin fields.

Although no direct proof that the list of higher spin
algebras is complete is known,
we conjecture that the three two-parametric
classes of elementary algebras described above and their direct sums
exhaust all d=4 higher spin superalgebras with
finite-dimensional Yang-Mills symmetries. This conjecture gets a
nontrivial support from the analysis of the unitary representations
of the higher spin algebras in \cite{KV,KV1} where it was shown that
the algebras listed above admit unitary representations with
the spectra of massless states having exactly the same spin
multiplicities as it follows from the field theoretical
analysis of the higher spin equations of motion based on these
superalgebras. Moreover, the algebras which
can be obtained from the original higher spin algebras by
truncations incompatible with the structure
of the higher spin equations
do not possess unitary representations containing enough states
for all massless fields associated with
the gauge fields of these algebras.

A simplest example of a truncation that is well defined at the
algebra level but is incompatible with field equations can be
obtained by truncating away the dependence on the Klein operators
$k$ and $\bar{k}$. The resulting algebra gives rise to the
set of gauge fields $w (\hy,\hat{\bar{y}})$
corresponding to massless fields of all spins $s \geq 1$, every
spin appears once. However it neither admits a unitary massless
representation with this spin spectrum \cite{KV},
nor corresponds to any consistent truncation of the higher spin
equations \cite{Ann}. One can however get rid of the operators
$k$ and $\bar{k}$ in the purely bosonic case with
$w(\hy,\hat{\bar{y}})=w(-\hy,-\hat{\bar{y}})$. The resulting
bosonic algebra is $hu(1;0|4)$.

The higher spin superalgebras $hu(n;m|4)$, $ho(n;m|4)$ and $husp(n;m|4)$
are supersymmetric in the standard sense only if $n=m$. Indeed, one
observes that (averaged) numbers of bosons and fermions
in (\ref{multi}) coincide
only for this case. All higher spin superalgebras with
$n=m$ contain the AdS superalgebra $osp(1;4)$
subalgebra. The algebras
$hsu(n;n|4)$ contain the N=2 superalgebra $osp(2;4)$.
Direct identification of the generators is
according to (\ref{n2su}) and (\ref{U1})
with the  realization (\ref{K}) and (\ref{Y}).

For the special values of $n=m=2^{[N/2]}$, the higher spin algebras
admit larger finite-dimensional supersymmetry subalgebras.
This is a particularly appealing special case with $M$ identified
with the Clifford algebra $C_N$
($C_N\sim Mat_{\frac{N}{2}}$ for even $N$ and
$C_N\sim Mat_{\frac{N-1}{2}} \oplus Mat_{\frac{N-1}{2}}$
for odd $N$).

The corresponding higher spin algebras can be described by
``functions"
of additional operators $\phi_i$, the generating elements
of the Clifford algebra
\be
\{\phi_i ,\phi_j \}=2\delta_{ij}\,,
\ee
$ i,j =1\ldots N$.
A nice feature of these algebras is that they naturally
contain extended AdS supersymmetry $osp(N,4)$ with the
supercharges
\be
Q^i_\gal = \hat{y}_\gal \phi^i
\,,\qquad
\bar{Q}^i_{\dot{\gb}}=\hat{\bar{y}}_{\dgb}\phi^i
\ee
 and $o(N)$
generators
\be
T^{ij} = \frac{1}{4}[\phi^i ,\phi^j ]\,.
\ee
For odd $N$, the Clifford algebras are semisimple,
thus leading to semisimple higher spin superalgebras.
We therefore will only consider the case of even N.

These higher spin algebras were originally
introduced in \cite{OP1}, where we used notation $shs^E (N,4)$
for the algebras spanned by the even ``functions"
\be
f(-\phi , -y, -\bar{y}) = f(\phi , y, \bar{y})\,.
\ee
Further truncation can be achieved by virtue of the
natural  antiautomorphism $\rho$
of the Clifford algebra defined by $\rho (\phi^i )= \phi^i$.
As argued in \cite{KV1},
this truncation is compatible with the higher spin dynamics for $N=4p$
when the operator $K$ identified with $\phi_1 \ldots \phi_N$
is invariant.
The resulting algebras denoted $shs^E(N,4|0)$ in \cite{OP1}
 form minimal higher spin algebras  possessing $N$-extended
supersymmetry and containing the  gauge
fields of the N-extended SUGRA supermultiplet $e_{\nu\gal\dot{\gb}}$,
$\go_{\nu\gal\gb}$, $\bar{\go}_{\nu\dot{\gal}\dot{\gb}}$ (spin 2),
$\psi^i_{\nu\gal}$, $\bar{\psi}^i_{\nu\dot{\gal}}$ (spin 3/2)
and $A_\nu^{ij}$ (spin 1) within the set of higher spin gauge fields
$W(y,\bar{y}, k, \bar{k},\phi |x )$.
The spin 1/2 and spin 0
fields from the SUGRA multiplets are contained in the sector of 0-forms
\cite{Ann,KV1} discussed in the section \ref{Nonlinear Equations in d=4}.

Let us stress that, in the framework of the higher spin gauge
theories, there is no barrier $N\leq 8$ which was a consequence of
the restriction $s\leq 2$ in the framework of supergravity.
The models with $N > 8$ can be considered equally well. The
restriction $N\leq 8$ should be reinterpreted as a restriction
on a number of unbroken supersymmetries in the phase with
broken higher spin symmetries.

Although the higher spin algebras based on the Clifford
algebra $M=C_N$ are distinguished by a larger finite-dimensional
supersymmetry, they are still particular cases of the algebras
described in the beginning of this section. The identification
is as follows \cite{KV1}:
\be
 shs^E (2l,4) =hu(2^{l-1};2^{l-1}|4)\,,
\ee
\be
 shs^E (8p,4|0) =ho(2^{4p-1};2^{4p-1}|4)\,,\quad
 shs^E (8p+4,4|0) =husp(2^{4p+1};2^{4p+1}|4)\,.
\ee

The particular case of the higher spin theory with N=8
that corresponds to a higher spin extension of N=8 SUGRA was
considered recently in great detail in \cite{SS}.
Within  the classification above this is
the case of~\footnote{To avoid misunderstandings
let us note that unfortunately the authors of \cite{SS} used
the notation $shs^E (8,4)$ for the algebra called $shs^E (8,4|0)$
in \cite{OP1,KV1} and in this paper. }
\be
 shs^E (8,4|0) =ho(8;8|4).
\ee
The appearance
of 8 on the left and right hand sides of this isomorphism is a
manifestation of triality of $so(8)$: in
$ shs^E (8,4|0)$ 8 stands for the vector representation of
$so(8)$ while in
$ ho(8;8|4)$ two its spinor representations are referred to.

Two comments are now in order.

The construction can be generalized to
infinite-dimensional algebras $M$. In particular, one can consider the
higher spin superalgebras $h\ldots(n;m|4)$ with $n \rightarrow  \infty $\,\,
or/and $\ m \rightarrow  \infty $,\, that will lead to theories with infinite
numbers of massless particles of every spin. Such theories may be of interest
in the context of spontaneous breakdown of higher spin gauge symmetries
and a relationship with string theory.

The analysis of extended
symmetries in the higher spin theories elucidates
 their deep parallelism with the string
theory and, in particular, with the Chan-Paton structure
of inner symmetries. The lesson is that higher spin
gauge theories are based on the associative
structure rather than on the Lie structure as lower
spin theories like Yang-Mills and (super)gravity. In this respect,
the $hsu(n,m|4)$ higher spin theories are similar to the oriented
open superstring theories while the $ho(n,m|4)$ and $husp(n,m|4)$
higher spin theories singled out with the help of antiautomorphisms
are analogous to non-oriented
open string theories. Remarkably, all higher spin theories contain
even spins and, in particular, gravity. From that perspective
higher spin theories necessarily
contain sectors to be associated with
closed superstring.

\subsection{Towards $d > 4$}

The challenging problem
 is to find  higher spin algebras for higher dimensions $d>4$.
At the moment we neither know a complete structure of
the higher spin
conserved currents  nor a structure of the
higher spin symmetry superalgebras in arbitrary $d$.

Algebraically, a straightforward generalization of the definition
of $d3$ and $d4$ higher spin algebras first conjectured in \cite{V}
seems to be a most natural candidate for higher spin algebras
in higher dimensions. It consists of introducing spinors in
$d$ dimensions, $\hy_\mu$ and $\hat{\bar{y}}^\nu$
($\mu , \nu = 1\ldots \left [ \frac{d}{2} \right ]$)
satisfying the commutation relations
\be
[\hy_\mu \,,\hat{\bar{y}}^\nu ] = \delta_\mu^\nu
\ee
with possible identification
\be
\hy_\mu = \hat{\bar{y}}^\nu C_{\nu\mu}
\ee
if antisymmetric charge conjugation matrix
$ C_{\nu\mu}$
exists (otherwise one has to require
$[\hy_\mu \,,\hy_\nu ] =0$ and
$[\hat{\bar{y}}^\mu \,,\hat{\bar{y}}^\nu ] = 0$)
and/or imposing
appropriate chirality conditions.
The basis of the infinite-dimensional Weyl algebra
is formed by monomials
\be
T_{\mu_1 \ldots \mu_n }{}^{\nu_1 \ldots \nu_m }
= \hy_{\mu_1}\ldots \hy_{\mu_n}\hat{\bar{y}}^{\nu_1} \ldots
\hat{\bar{y}}^{\nu_m}\,.
\ee
The generators
\be
T_\mu{}^\nu = \{ \hy_\mu \,,\hat{\bar{y}}^\nu \}
\ee
form a finite-dimensional subalgebra $u(2^{[d/2]})$
(or $sp(2^{[d/2]})$ in the Majorana case) which contains
standard AdS algebra $o(d-1,2)$ as
the subalgebra spanned by the
elements
\be
P_n =  \hy_\mu \gamma_n{}^\mu{}_\nu \hat{\bar{y}}^\nu  \,,\qquad
M_{mn} =  \hy_\mu [\gamma_m\,,\gamma_n ]{}^\mu{}_\nu \hat{\bar{y}}^\nu \,,
\ee
where $\gamma_n$ denotes Dirac matrices in $d$ dimensions.

It is  an interesting algebraic problem
to analyze the field content of the higher spin gauge fields
originating from this construction in higher dimensions.
To the best of our knowledge a solution of
the related problem of decomposition
of arbitrary symmetric tensor products of spinor representation
into irreducible representations of the Lorentz group has not
been yet elaborated for arbitrary $d$.
The situation in lower dimensions is simplified by the
isomorphisms of the AdS algebras with the algebras of
bilinears of oscillators,
$sp(4)\sim o(3,2)$ and $sp(2)\oplus sp(2) \sim o(2,2)$.

\section{Star-Product}
\label{Star Product}

In practice, instead of working with the algebras of
operators as discussed in the section \ref{Higher-Spin Symmetries},
it is convenient to use usual functions endowed with
the product law
\be
\label{wprod}
 (f*g)(y)
   =\frac{1}{(2\pi)^{2p}}\int d^{2p}u d^{2p} v\exp(iu_\mu v^\mu)
   f(y+u)g(y+v)\,.
\ee
Here $f(y)$ and $g(y)$ are functions (polynomials or formal
power series) of commuting variables
$y_\mu$ where $\mu = 1 \ldots 2p$.
This formula defines the associative algebra with the
defining relation
\be
y_\mu *y_\nu - y_\nu * y_\mu = 2iC_{\mu\nu}\,,
\ee
where $C_{\mu\nu}$ is the symplectic form used to raise and
lower indices,
\be
u^\mu = C^{\mu\nu}u_\nu\,,\qquad
u_\mu =u^\nu C_{\nu\mu}     \,.
\ee
The star-product defined this way describes
the product of Weyl ordered (i.e. totally symmetric)
polynomials of oscillators in terms of symbols of operators \cite{sym,BS}.
Thus, this construction gives a particular realization of the
Weyl algebra $A_p$.

Usual differential versions \cite{moyal,sym,BS} can be derived from
(\ref{wprod}) by elementary Gaussian integration of the Taylor
expansion $f(y) = \exp y^\mu \frac{\partial}{\partial{z^\mu}} f(z)|_{z=0}$,
\be
\label{moy}
A(y)* B(y) =
e^{i\frac{\partial}{\partial y^1_\mu}
\frac{\partial}{\partial y^{2\mu}}} A(y+y^1 ) B(y+y^2 ) |_{y^1 = y^2 =0}\,.
\ee
This Weyl product law (often called Moyal bracket
 for commutators constructed from (\ref{moy})) is obviously
nonlocal. This is of course the ordinary
quantum-mechanical nonlocality. Note that the integral
formula (\ref{wprod}) is sometimes called triangle formula
\cite{BS} because, for two-component spinors,
 the term $u_\mu v^\mu$ in the exponential
is equal to the (oriented) area of the triangle with vertices
at $y$, $y+u$ and $y+v$. In most cases the triangle formula
(\ref{wprod}) is  more convenient
than (\ref{moy}) for practical computations and has a broader
area of applicability beyond the class of polynomial functions.

An important property of the star-product
is that it admits a uniquely
defined supertrace operation \cite{OP1}
\be
\label{STR}
str (f(y)) = f(0)
\ee
possessing the standard property
\be
\label{strpr}
str (f* g) = (-1)^{\pi (f) \pi (g)} str (g* f)
\ee
with the  parity definition
\be
\label{pari}
f (-y) = (-1)^{\pi (f)} f (y)\,.
\ee
This is in accordance with the normal spin-statistics
relation once $\mu$ and $ \nu$ are interpreted as
spinor indices. Note that the additional sign factor originates
from antisymmetry of the matrix $C_{\mu\nu}$ leading to
\be
u_\mu v^\mu = -v_\mu u^\mu \,.
\ee
(To prove (\ref{strpr}) one takes into account
that $str (f* g) = 0$ if $\pi (f)\neq  \pi (g)$ because
 $str (h) = 0$ if $\pi (h)=1$.)

The fact of existence of the supertrace operation is very important
for the theory of higher spin gauge fields because it allows one to
build invariants of the higher spin transformations
\be
\delta A = A *\epsilon - \epsilon * A
\ee
as supertraces of products
$str (A*B\ldots )$  provided that
 the ``fields" $A$, $B \ldots$ have appropriate
Grassmann grading for fermions.

The star-product algebras (\ref{wprod})
can be interpreted as algebras of differential
operators with polynomial coefficients (often identified with
$W_{1+\infty }$ algebras) by setting
\be
y_1 = 2i \frac{\partial}{\partial z}\,,\qquad y_2 = z
\ee
(for the case $p=1$). From this perspective it may look
surprising that the algebra of  differential
operators with polynomial coefficients
admits the
uniquely defined supertrace and no usual trace operation.

The star-product formulae (\ref{wprod}) and (\ref{moy})
are very handy for practical computations with
oscillators. Unfortunately, no useful analog of these
formulae is known for the deformed oscillators (\ref{modosc}).

For the description of the nonlinear higher spin
dynamics we will also need another associative
star-product defined on the space of functions of two
symplectic (spinor) variables
\be
\label{star2}
(f*g)(z;y)=\frac{1}{(2\pi)^{2p}}
\int d^{2p} u\,d^{2p} v \exp{[iu^\mu v^\nu C_{\mu\nu}]}\, f(z+u;y+u)
g(z-v;y+v) \,,
\ee
where $ u^\mu $ and $ v^\mu $ are real integration variables.
It is a simple exercise with Gaussian integrals
to see that this star-product is associative
\be
(f*(g*h))=((f*g)*h)
\ee
and is
normalized such that 1 is a unit element of the star-product
algebra, i.e. $f*1 = 1*f =f\,.$

The star-product (\ref{star2})  again yields a particular
realization of the Weyl algebra
\be
[y_\mu,y_\nu]_*=-[z_\mu,z_\nu ]_*=2iC_{\mu\nu},,\qquad
[y_\mu,z_\nu]_*=0
\ee
($[a,b]_*=a*b-b*a$).
These commutation relations are particular
cases of the following simple formulae
\be
\label{y,f}
    [y_{\mu}, f]_*=2i{\partial f\over \partial y^\mu}\,,
\ee
\be
\label{z,f}
    [z_{\mu}, f]_*=-2i{\partial f\over \partial z^\mu} \,,
\ee
which are true for an arbitrary $f(z,y)$.

The star-product (\ref{star2}) corresponds
to the normal ordering of the Weyl
algebra with respect to the generating elements
\be
   a^+_\mu = \frac{1}{2} (y_\mu - z_\mu )\,,\qquad
   a_\mu = \frac{1}{2} (y_\mu + z_\mu )\,,\qquad
\ee
which satisfy the commutation relations
\be
\label{com a}
   [a_\mu, a_\nu]_*=[a^+_\mu, a^+_\nu]_* =0 \,,\quad
   [a_\mu, a^+_\nu]_* =iC_{\mu\nu} \,
\ee
and can be interpreted as creation and annihilation operators. This is
most evident from the relations
\be
\label{pr a}
   a^+_\mu * f(a^+, a) =
      a^+_\mu f(a^+, a) \,,\qquad
   f(a^+, a) * a_\mu =
      f(a^+, a)  a_\mu \,.
\ee

Star-product (\ref{star2}) admits the
supertrace operation
\be
str(f (z,y)) =
\frac{1}{(2\pi)^{2p}}\int d^{2p} ud^{2p} v\exp(-iu_\mu v^\mu)
   f(u,v)\,
\ee
which can be obtained from the Weyl
supertrace operation
(\ref{STR}) (for the doubled number of variables)
by changing the ordering prescription.
Obviously,
\be
str({f(z,y)})=str({f(-z,-y)})\,,
\ee
i.e. fermions have zero supertrace once the boson-fermion grading
is defined in accordance with the standard relationship between
spin and statistics
\be
f(-z,-y)=(-1)^{\pi(f)}f(z,y)\,
\ee
(for $y_\mu$ and $z_\mu$ interpreted as spinors).
Then it is elementary to see that (\ref{strpr}) holds.

An important property of the star-product (\ref{star2}) is that
it admits the inner Klein operator
\be
\Upsilon = \exp i z_\mu y^\mu \,,
\ee
which behaves as $(-1)^{N_f} ,$ where $N_f$ is the fermion number
operator. It is a simple exercise to show that
\be
\U *\U =1,
\ee
\be
\label{[UF]}
\U *f(z;y)=f(-z;-y)*\U\,,\quad
\ee
and
\be
\label{Uf}
(\U *f)(z;y)=\exp{i z_\mu y^\mu }\, f(y;z) \,.
\ee
{}From (\ref{Uf}) we see that, up to the exponential factor,
$ \U $ interchanges the arguments $ z $ and $ y $.
For the d=4 problem we will need the left and right
inner Klein operators
\be
\label{kk4}
\ups =\exp i z_\ga y^\ga\,,\qquad
\bu =\exp i \bar{z}_\dga \bar{y}^\dga\,,
\ee
 which act analogously
on the undotted and dotted spinors, respectively:
\be
\label{uf}
\!(\ups *f)(z,\!\bar{z};y,\!\bar{y})\!=\!\exp{i z_\ga y^\ga }\,\!
f(y,\!\bar{z};z,\!\bar{y}) ,\quad\!
(\bu *f)(z,\!\bar{z};y,\!\bar{y})\!=\!\exp{i \bar{z}_\dga \bar{y}^\dga }\,\!
f(z,\!\bar{y};y,\!\bar{z}) ,
\ee
\be
\label{[uf]}
\ups *f(z,\bar{z};y,\bar{y})=f(-z,\bar{z};-y,\bar{y})*\ups\,,\quad
\bu *f(z,\bar{z};y,\bar{y})=f(z,-\bar{z};y,-\bar{y})*\bu\,,
\ee
\be
\ups *\ups =\bu *\bu =1\,.
\ee

The star-product (\ref{star2}) is regular:
given two polynomials $f$ and $g$, $f*g$ is also some polynomial.
Particular star-products corresponding to one or another ordering
prescription are of course equivalent in the class of polynomials
but may be inequivalent beyond this class. The reason is that
reordering of infinite series may lead to divergent coefficients
(e.g., to an infinite ``vacuum energy" constant term). Moreover, it
is not {\it a priori} guaranteed that star-product of non-polynomial
functions is well defined. The reader can easily construct
examples of functions $f$ and $g$ such that $f*g$ will
diverge because the bilinear form in the Gaussian integral will
degenerate. The special property of the star-product (\ref{star2})
is that it is defined in a certain class of nonpolynomial functions
\cite{Pr,PV} containing  the nonpolynomial Klein operators $\U$, $\ups$
and $\bu$ in the sense that the product is still associative and
no infinities appear in this class. This is not the case
for the Weyl ordered star-product in the doubled spinor space
where it is not clear how to define the Klein operators $\U$, $\ups$
and $\bu$ (at least analogous exponentials are ill-defined
in the algebra). Since the Klein operators play important role in the
construction, the star-product (\ref{star2}) acquires a distinguished
role. It is of course not surprising that the star-product associated
with the certain normal ordering (\ref{pr a})
has better properties with respect to potential divergencies.
In fact, as shown in \cite{Pr,PV}, the star-product (\ref{star2})
is well defined for a class of regular functions
which appears in the process of solution of the nonlinear
constraints in the section \ref{Nonlinear Higher-Spin Equations}.
This guarantees that the non-linear higher spin
equations are well-defined.

{}From (\ref{star2}) it follows that functions $f(y)$ independent
of $z$ form a proper subalgebra.
Due to (\ref{z,f})  this subalgebra
identifies with the centralizer of the elements $z_\nu$.
Note that for $z-$independent functions
the star-product (\ref{star2}) reduces to the Weyl star-product
(\ref{wprod}). This is why we use the same symbol $*$ for the
both product laws.

\section{AdS Vacua}
\label{AdS Vacua}

 AdS background plays distinguished  role in the higher spin theories.
It appears naturally in the framework of the
higher spin algebras as a particular solution of the
zero-curvature equation
\be
\label{vacu}
dw = w* \wedge w\,.
\ee
(From now on we only consider the cases of $d=3$ and $d=4$.)

Any vacuum solution $w_0$ of the equation (\ref{vacu})
breaks the local higher spin symmetry to
its stability  subalgebra
with the infinitesimal parameters $\epsilon_0 (y|x)$
satisfying the equation
\be
\label{D0}
D_0 \epsilon_0\equiv  d\epsilon_0 -w_0 * \epsilon_0
+\epsilon_0 * w_0 =0 \,.
\ee
The consistency of the equation (\ref{D0}) is guaranteed
by the vacuum equation (\ref{vacu}). As a result, (\ref{D0})
admits a unique solution in some neighborhood of an arbitrary
point $x_0$ with the initial data
\be
\label{gex0}
\epsilon_0 (y|x_0 ) = \epsilon_0 (y)\,,
\ee
where $\epsilon_0 (y)$ is an arbitrary $x$- independent element
of the Weyl algebra.

In the higher spin theories no further symmetry breaking
is induced by other field equations. Therefore,
$\epsilon_0 (y)$ parametrizes the global symmetry
of the theory, the higher spin global symmetry. As a result,
we conclude that global symmetry higher spin algebras
identify with the Lie superalgebras constructed from the
(anti)commutators of the elements of the Weyl algebra and their
extensions with the Klein operators and matrix indices. Note that
fields carrying odd numbers of spinor fields are
anticommuting thus inducing a structure of superalgebra into
(\ref{vacu}).

Since functions bilinear in $y_\ga$ form a
closed  subalgebra with respect to commutators it is
a consistent ansatz to look for a solution of the vacuum
equation (\ref{vacu}) in the form

\be
\label{w0}
w_0 =
\frac{1}{8i}
\left ( \go_0^{\ga\gb} (x)\{y_\ga ,y_\gb \}_*
+\lambda h_0^{\ga\gb} (x)\psi \{y_\ga ,y_\gb \}_* \right )\,
\ee
for $d=3$ and
\be
\label{anz0}
w_0 =
\frac{1}{8i}
\left ( \go_0^{\ga\gb} (x)\{y_\ga ,y_\gb \}_*
+ \bar{\go}_0^{\dot{\ga}\dot{\gb}} (x)\{\bar{y}_{\dot{\ga}} ,
\bar{y}_{\dot{\gb}} \}_*
+2\lambda h_0^{\ga\dot{\gb}} (x) \{y_\ga ,
\bar{y}_{\dot{\gb}} \}_* \right )
\ee
for d=4, respectively. Here $\{a,b\}_* = a*b + b*a$. For the
star-product (\ref{wprod}) we have
\be
\{y_\ga ,y_\gb \}_* =2 y_\ga y_\gb\,,\quad
\{\bar{y}_{\dot{\ga}} \bar{y}_{\dot{\gb}} \}_* =
2\bar{y}_{\dot{\ga}} \bar{y}_{\dot{\gb}}\,,\quad
 \{y_\ga ,\bar{y}_{\dot{\gb}} \}_*  =2 y_\ga \bar{y}_\dgb\,.
\ee

Inserting these formulae into
(\ref{vacu}) one finds that the fields $\go_0 (\bar{\go_0 })$ and
$h_0$ identify with the Lorentz connection and the frame field
of $AdS_3$ or $AdS_4$, respectively, provided that the
1-form $h_0$ is
invertible. The parameter $\lambda= r^{-1}$ is identified with
the inverse AdS radius. Thus,  the fact that the higher spin algebras
are star-product (oscillator) algebras leads to the AdS geometry as
a natural vacuum solution.

The vacuum equation (\ref{vacu}) has a form of
zero-curvature equation and therefore admits a pure gauge
solution
\be
\label{pg}
w_0 = -g^{-1}(y|x)* d g(y|x)
\ee
with some invertible element $g(y|x)$ of the
Weyl algebra $g*g^{-1} = g^{-1} *g = I$. The equation
(\ref{D0}) then solves as
\be
\label{gs}
\epsilon_0 (y|x) = g^{-1}(y|x)* \epsilon_0 (y)* g(y|x) \,.
\ee
(Clearly, (\ref{gex0}) is true for a point $x_0$ such that
$g(y| x_0 ) = I$.)
This fact was known long ago \cite{} but unless recently it was not
used for practical computation. In this presentation we
use it for the analysis of the higher spin dynamics in $AdS_4$
following the recent work of Kirill Bolotin and the
author \cite{BV}.

A particular solution of the
vacuum equation (\ref{vacu}) corresponding to the stereographic
coordinates has a form
\be
\label{h0S}
h_\un {}^{\ga\dgb} =-z^{-1} \sigma _\un {}^{\ga\dgb}\,,
\ee
\be
\label{w0S}
\go_\un {}^{\ga\ga} =-\lambda^2 z^{-1} \sigma _\un {}^{\ga\dgb}x^\ga{}_\dgb\,,
\ee
\be
\label{bw0S}
\bar{\go}_\un {}^{\dgb\dgb}
=-\lambda^2 z^{-1} \sigma _\un {}^{\ga\dgb}x_\ga{}^\dgb\,,
\ee
where we use notation
\be
x^{\ga\dgb}=x^\un \sigma _\un {}^{\ga\dgb}\,,\quad
x^2 = \half x^{\ga\dgb}x_{\ga\dgb}\,,\quad z= 1+\lambda^2 x^2 \,.
\ee
Let us note that $z\to 1$ in the flat limit and $z\to 0$ at the
boundary of AdS.

The form of the gauge function $g$ reproducing these
vacuum background fields (with all $s\neq 2$  fields
vanishing) turns out to be remarkably simple \cite{BV}
\be
\label{gz}
g(y,\bar{y}| x) = 2\frac{\sqrt{z}}{1+\sqrt{z}}
\exp[\frac{i\lambda}{1+\sqrt{z}}x^{\ga\dgb}y_\ga \bar{y}_\dgb ]
\ee
with the inverse
\be
\label{g-1}
g^{-1}(y,\bar{y}| x) = 2\frac{\sqrt{z}}{1+\sqrt{z}}
\exp[\frac{-i\lambda}{1+\sqrt{z}}x^{\ga\dgb}y_\ga \bar{y}_\dgb ]\,.
\ee

As shown in the section \ref{$AdS_4$},
when solving relativistic field equations,
$g$ plays a role of a sort of evolution operator. From this
perspective $\lambda$ is
analogous to the inverse of the Planck constant $\hbar$,
\be
\label{lh}
\lambda \sim \hbar^{-1}\,.
\ee
This parallelism indicates that the flat limit
$\lambda \to 0$ may be essentially singular.

Let us draw attention to
an important difference between the d=3 and d=4 cases.
The ansatz (\ref{w0}) in the $d3$ case solves the
vacuum equation not only for the usual oscillators $y_\ga$
but also for the deformed oscillators
(\ref{modosc})
\be
\label{w}
w_0 =
\frac{1}{8i}
\left ( \go^{\ga\gb} (x)\{\ty_\ga (\nu ),\ty_\gb (\nu )\}
+\lambda h^{\ga\gb} (x)\psi \{\ty_\ga (\nu ),\ty_\gb (\nu )\} \right )\,.
\ee
The zero-curvature vacuum equation still implies that the fields
$\go^{\ga\gb} (x)=dx^\un \go_\un{}^{\ga\gb} (x)$ and
$h^{\ga\gb} (x)=dx^\un h_\un{}^{\ga\gb} (x)$
 identify with the Lorentz connection
and dreibein of $AdS_3$ (under condition that
$h_\un{}^{\ga\gb} (x)$ is  non-degenerate).
The properties of the deformed oscillator algebra guarantee that
this is true for any value of the parameter $\nu$, i.e. the
differential equations for $\go^{\ga\gb}_0$ and $h^{\ga\gb}_0$
which follow from (\ref{vacu}) are $\nu$ - independent.

If we try to proceed similarly in the d=4 case with
\be
\label{anz}
\!\!\!w_0\! =\!
\frac{1}{8i}\!
\left (\! \go^{\ga\gb}_0 \! (x)\{\ty_\ga (\nu ),\ty_\gb (\nu )\}
\!+\! \bar{\go}^{\dot{\ga}\dot{\gb}}_0 \!(x)
\{\hat{\bar{y}}_{\dot{\ga}} (\bar{\nu} ),
\hat{\bar{y}}_{\dot{\gb}} (\bar{\nu} )\}
\!+\!2\lambda h^{\ga\dot{\gb}}_0 \! (x) \{\ty_\ga (\nu ),
\hat{\bar{y}}_{\dot{\gb}} (\bar{\nu} )\}\!\right )\!,
\ee
where $\ty_\ga (\nu )$ and $\hat{\bar{y}}_{\dot{\gb}} (\bar{\nu}) $ are
two mutually commuting sets of deformed oscillators,
the result would be that for $\nu=\bar{\nu}= 0$ the
zero-curvature vacuum equation
indeed describes $AdS_4$ while for $\nu \neq 0$ and/or $\bar{\nu}\neq 0$
(\ref{vacu}) becomes inconsistent, i.e. it admits no solution
with the ansatz (\ref{anz}).
The replacement of the ordinary oscillators
(\ref{d4osc}) by the deformed oscillators (\ref{modosc})
breaks $sp(4)$ down to its Lorentz subalgebra
$sl_2 (C)$ spanned by the generators $L_{\gal\gb}$ and
$\bar{L}_{\dot{\gal}\dot{\gb}}$.
Commutators of  AdS translations $P_{\gal\dgb}$ for non-zero
parameters $\nu$ and $\bar{\nu}$ give rise to higher and higher
polynomials in the deformed oscillator
algebra. To construct some consistent
vacuum solution for that case one has to introduce infinitely
many nonvanishing vacuum higher spin field components in the
d=4 analog of the expansion (\ref{ghs}).
This fact will have important consequences for the
analysis of the section \ref{Integrating Flow}.

\section{Free Equations}

The structure of higher spin currents suggests
that higher spin symmetries relate higher derivatives
of relativistic fields. Once spin $s>2$ symmetries
are present, the algebra becomes infinite-dimensional.
Therefore, higher spin symmetries relate derivatives of
physical fields of all orders. To have a
natural linear realization of the higher spin symmetries,
it is useful to introduce infinite multiplets reach enough
to contain dynamical fields along with
all their higher derivatives. Such multiplets admit
a natural realization in terms of the Weyl algebra.

Namely, in d=3 \cite{un} and d=4 \cite{Ann}, 0-forms
\be
\label{CY}
C(Y|x)=
  \sum^\infty_{ n=0 }
    \frac{1}{n!}  C^{\nu_1\ldots\nu_n\,}
    Y_{\nu_1}\ldots Y_{\nu_n}\,
\ee
taking their values in the Weyl algebra form
 multiplets of the higher spin symmetries for lower
spin matter fields and Weyl-type higher spin
curvature tensors (in $d=4$).
The free equations of motion have a form \cite{un,Ann}
\be
\label{cald0}
{\cal D}_0 C \equiv (d C -{w}_0 *C +C* \tilde{w}_0 ) = 0 \,,
\ee
where
\be
\tilde{f}(y,\bar{y} ) =f(y,-\bar{y} )
\ee
for d=4 and
\be
\tilde{f}(y,\psi ) =f(y,-\psi )
\ee
for d=3.

In the both cases,
tilde denotes an
involutive automorphism of the higher spin algebra
which changes a sign of the AdS translations
(\ref{al}) and (\ref{al4}).
As a result, the covariant derivative ${\cal D}_0$
corresponds to some representation of the higher spin
algebra which we call twisted representation.
The consistency of the equation (\ref{cald0}) is
guaranteed by the vacuum equation (\ref{vacu}).

In this formulation, called in \cite{un} ``unfolded formulation",
dynamical field equations have a form of covariant constancy conditions.
The fact that the equation (\ref{cald0}) is invariant under the
global higher spin symmetries
\be
\label{tw}
\delta C =  {\epsilon}_0 * C - C*\tilde{\epsilon}_0
\ee
with the  parameters satisfying (\ref{D0}) is obvious.
Moreover,
one can write down a general solution of the free field
equations (\ref{cald0}) in the pure gauge form
\be
\label{gC}
C(Y|x) = {g}^{-1}(Y|x)* C_0 (Y)* \tilde{g}(Y|x) \,
\ee
analogous to the form of the global symmetry
parameters (\ref{gs}), where $C_0(Y)$ is an arbitrary
$x-$independent element of the twisted representation
with $Y=(y,\bar{y} )$ for  d=4 and $Y= (y, \psi )$ for d=3.
Here $C_0(Y)$ plays a role of initial data.
Let us now explain in more detail a physical content of the
equations (\ref{cald0}) starting with the d4 case.

\subsection{$AdS_4$}
\label{$AdS_4$}

In d=4, by virtue of the star-product
(\ref{wprod}) the system (\ref{cald0}) reduces to
\be
\label{opcald}
{\cal D}_0 C (y,{\bar{y}}|x) \equiv
D^L C (y,{\bar{y}}|x) +i\lambda h^{\ga\dgb}
\Big (y_\ga \bar{y}_\dgb -\frac{\partial}{\partial y^\ga}
\frac{\partial}{\partial \bar{y}^\dgb}\Big ) C (y,{\bar{y}}|x) =0\,,
\ee
where
\be
\label{dlor}
D^L_0 C (y,{\bar{y}}|x) =
d C (y,{\bar{y}}|x) -
\Big (\go^{\ga\gb}y_\ga \frac{\partial}{\partial {y}^\gb} +
\bar{\go}^{\dga\dgb}\bar{y}_\dga \frac{\partial}{\partial \bar{y}^\dgb} \Big )
C (y,{\bar{y}}|x)\,.
\ee
Rewriting (\ref{CY}) as
\be
\label{Cyby}
C(Y|x)=C (y,{\bar{y}}|x)=
  \sum^\infty_{n,m=0}
    \frac{1}{m!n!}  C^{\gal_1\ldots\gal_n}{}_{,}{}^
{{\dga}_1 \ldots {\dga}_m}(x)
    y_{\gal_1}\ldots y_{\gal_n}\,
    {\bar{y}}_{\dga_1}\ldots {\bar{y}}_{\dga_m}\,,
\ee
one arrives at the following infinite chain of equations
\begin{equation}
\label{DCC}
D^LC_{\alpha (m),\,\dot{\beta} (n)}=
i\lambda h^{\gamma\dot{\delta}}C_{\alpha (m)\gamma,\,
\dot{\beta} (n) \dot{\delta}}
-i nm\lambda h_{\alpha\dot{\beta}}
C_{\alpha (m-1),\,\dot{\beta} (n-1)}\,,
\end{equation}
where $D^L$ is the Lorentz-covariant differential
\begin{equation}
D^L A_{\alpha\dot{\beta}}=dA_{\alpha\dot{\beta}}+\omega_\alpha{}^\gamma
\wedge
A_{\gamma\dot{\beta}}+\bar{\omega}_{\dot{\beta}}{}^{\dot{\delta}}
\wedge A_{\alpha\dot{\delta}}\,.
\end{equation}
Here we skip the subscript $0$ referring to the vacuum AdS solution
and use again the convention introduced in the section
\ref{Higher Spin Currents} with symmetrized indices denoted by
the same letter and a number of symmetrized indices indicated in
brackets.

The system  (\ref{DCC}) decomposes into a set of independent
subsystems with $n-m$ fixed. It turns out \cite{Ann}
that the subsystem
with $|n-m| = 2s$ describes a massless field of spin $s$
(note that the fields
$C_{\alpha (m),\,\dot{\beta}(n)}$
and
$C_{\beta (n),\,\dot{\ga} (m)}$ are
complex  conjugated).

It is instructive to consider the example of $s=0$ associated with
the  fields $C_{\alpha (n),\,\dot{\beta}(n)}$.
The equation (\ref{DCC}) at $n=m=0$ expresses the field
$C_{\ga,\dgb}$ via the first derivative of  $C$,
\be
\label{C00}
C_{\ga ,\dgb}=\frac{1}{2i\lambda} h^\un_{\ga\dgb} D^L_\un C\,,
\ee
where $h^\un_{\ga\dgb}$ is the inverse frame field
\be
\label{inv}
h^\un_{\ga\dgb} h_\un^{\gamma\dot{\delta}} =2
\delta_\ga^\gamma
\delta_\dgb^{\dot{\delta}}\,,\qquad g^{\un\um} = \half
h^\un_{\ga\dgb} h^\um{}^{\ga\dgb}\,
\ee
with the normalization chosen in such a way
that it is true for
$h^\un_{\ga\dgb} = \sigma^\un_{\ga\dgb}$ and
$h_\un^{\ga\dgb} = \sigma_\un^{\ga\dgb}$.

The second equation with $n=m=1$ contains more information.
First, one obtains by contracting indices with the
frame field and using (\ref{inv}) that
\be
h^\un_{\ga\dgb} (D^L_\un C^{\ga}{}_,{}^{\dgb} + 8i\lambda C )=0\,.
\ee
With the aid of (\ref{C00}) this reduces to the Klein-Gordon
equation in $AdS_4$
\be
\label{KG}
\Box C -8\lambda^2 C =0\,.
\ee
The rest part of the equation (\ref{DCC}) with $n=m=1$ expresses the
field $C_{\ga\ga ,\dgb\dgb }$ via second derivatives of $C$
\be
\label{C11}
C_{\ga\ga\,,\dgb\dgb}=\frac{1}{(2i\lambda)^2}
h^\un_{\ga\dgb} D^L_\un
h^\um_{\ga\dgb}D^L_\um C\,.
\ee

All other equations with $ n=m >1$ either reduce to identities
by virtue of the spin 0 dynamical equation (\ref{KG}) or
express higher components in the chain of fields
$C_{\alpha_1\ldots\alpha_n ,\,\dot{\beta}_1 \ldots {\dot{\beta}_n}}$
via higher derivatives in the space-time coordinates as
\be
\label{Cnn}
C_{\ga(n)\,,\dgb (n)}= \frac{1}{(2i\lambda)^n}
h^{\um_1}_{\ga\dgb} D^L_{\um_1}
\ldots h^{\um_n}_{\ga\dgb} D^L_{\um_n} C\,.
\ee
This completes the proof of the fact that the system
(\ref{DCC})  with $n=m$ describes a scalar field.
The value of the mass parameter in (\ref{KG}) is such that
$C$ describes massless scalar in $AdS_4$.

Spin 1/2 is described by the mutually conjugated chains of
fields
$C_{\alpha (m),\,\dot{\beta} (n)}$
with $|n-m| =1$. In this case the first equation with $n=0$ and
$m=1$ has a form
\be
\label{C10}
D_\un^LC_{\alpha}=
i\lambda h_\un^{\gamma\dot{\delta}}C_{\alpha\gamma,\,
\dot{\delta}}\,.
\end{equation}
Dirac equation is a simple consequence of this equation,
\be
\label{dirac}
h^\un_{\ga\dgb} D^L_\un C^{\ga}  =0\,.
\ee
All the rest equations again do not impose any further restriction
on the dynamical field $C_\ga$ just expressing higher members of the
chain via higher space-time derivatives of $C_\ga$.
The fact that, although overdetermined, the system
(\ref{DCC}) is consistent takes place
because in (\ref{cald0}) $({\cal D}_0)^2 = 0$
as a consequence of (\ref{vacu}).

Analogously, the equations (\ref{DCC}) with other values of
$n$ and $m$ describe free field equations for
spin $s=|n-m|$ massless fields. However, for spins $s\geq 1$
it is more useful to treat these equations not as fundamental ones
but as consequences of the higher spin equations
formulated in terms of gauge fields (potentials).
To illustrate this point let us first consider the example of gravity.

As argued in section \ref{Higher-Spin Symmetries},
Lorentz connection $1-$forms $\omega _{\alpha \beta }$,
$\bar \omega_{\dot \alpha\dot \beta}$
and vierbein $1-$form $h_{\alpha\dot \beta}$
can be identified with the $sp(4)-$gauge fields. The corresponding
$sp(4)-$curvatures read in terms of two-component spinors
\be
\label{nR}
R_{\alpha_1 \alpha_2}=d\omega_{\alpha_1 \alpha_2} +\omega_{\alpha_1}{}^\gamma
\wedge \omega_{\alpha_2 \gamma} +\lambda^2\, h_{\alpha_1}{}^{\dot{\delta}}
\wedge
h_{\alpha_2 \dot{\delta}}\,,
\ee
\be
\label{nbR}
\bar{R}_{\dot{\alpha}_1 \dot{\alpha}_2}=d\bar{\omega}_{\dot{\alpha}_1
\dot{\alpha}_2} +\bar{\omega}_{\dot{\alpha}_1}{}^{\dot{\gamma}}
\wedge \bar{\omega}_{\dot{\alpha}_2 \dot{\gamma}} +\lambda^2\,
h^\gamma{}_{\dot{\alpha}_1} \wedge h_{\gamma \dot{\alpha_2}}\,,
\ee
\begin{equation}
\label{nr}
r_{\alpha \dot{\beta}} =dh_{\alpha\dot{\beta}} +\omega_\alpha{}^\gamma \wedge
h_{\gamma\dot{\beta}} +\bar{\omega}_{\dot{\beta}}{}^{\dot{\delta}}
\wedge h_{\alpha\dot{\delta}}\,.
\end{equation}
The zero-torsion condition $r_{\alpha\dot \beta}=0$
expresses the Lorentz connection $\omega $ and $\bar \omega $ via
derivatives of $h$. After that, the $\lambda -$independent part of the
curvature $2-$forms $R$ (\ref{nR}) and $\bar R $ (\ref{nbR})
coincides with the Riemann tensor.
Einstein equations imply that the Ricci tensor vanishes
up to a constant
trace part proportional to the  cosmological constant.
This is equivalent to saying that only those
components of the tensors (\ref{nR}) and (\ref{nbR}) are allowed to be
non-vanishing which belong to the Weyl tensor.  As is well-known \cite{PR},
Weyl tensor is described by the fourth-rank mutually conjugated totally
symmetric multispinors $C_{\alpha _1\alpha _2\alpha _3\alpha _4}$ and $\bar
C_{\dot \alpha_1\dot \alpha_2\dot \alpha_3\dot \alpha_4}$.  Therefore,
Einstein equations with the cosmological term can be cast into the form
\begin{equation}
\label{e1} r_{\alpha \dot{\beta}} =0\,,
\end{equation}
\be
\label{e2}
R_{\alpha_1\alpha_2}=h^{\gamma_1 \dot{\delta}} \wedge h^{\gamma_2}
{}_{\dot{\delta}}
C_{\alpha_1 \alpha_2 \gamma_1 \gamma_2}\,,\qquad
\bar{R}_{\dot{\beta}_1 \dot{\beta}_2}=h^{\eta\dot{\delta}_1}\wedge
h_\eta{}^{\dot{\delta}_2}
\bar{C}_{\dot{\beta}_1 \dot{\beta}_2 \dot{\delta}_1 \dot{\delta}_2}\,.
\ee

It is useful to treat the
$0-$forms $C_{\ga (4)}$ and $\bar C_{\dga (4)}$
on the right hand sides of (\ref{e2})
as  independent field variables which identify with the Weyl
tensor by virtue of the equations (\ref{e2}). {}From (\ref{e2})
it follows that the $0-$forms $C_{\ga (4)}$ and
$\bar C_{\dga (4)}$  should obey certain
differential restrictions as a consequence of the Bianchi
identities for the curvatures $R$ and $\bar R$. It is not
difficult to make sure that these differential
restrictions can be equivalently rewritten in the form
\be
\label{b1}
D^L C_{\alpha (4)}=i\lambda h^{\gamma\dot{\delta}}
C_{\alpha (4) \gamma ,\dot{\delta}}\,,\qquad
D^L \bar{C}_{\dot{\beta} (4)}=i\lambda h^{\gamma\dot{\delta}}
\bar{C}_{\gamma ,\,\dot{\beta} (4)
\dot{\delta}}\,,
 \ee
where $C_{\alpha (5),\dot\delta}$ and
$\bar C_{\gamma,\dgb (5)}$ are new multispinor field variables
totally symmetric in the spinor indices of each type.
(The factor of $i\lambda$ is introduced for future convenience.)

Once again, Bianchi identities for the
left hand sides of (\ref{b1}) impose certain
differential restrictions on
$C_{\alpha (5),\dot\delta}$
and
$\bar C_{\gamma,\dgb (5)}$
which can be cast into the form analogous to (\ref{b1})
 by virtue of introducing new field variables
$C_{\alpha (6),\dot\delta (2)}$ and
$\bar C_{\alpha (2),\dot\delta (6)}$.
Continuation of this process leads to the infinite chains of
differential relations (\ref{DCC}) with $|n-m|=4$
\be
\label{c2}
D^L C_{\alpha ({n+4}),\,\dot{\beta} (n)}=i\lambda
( h^{\gamma\dot{\delta}} C_{\alpha
({n+4})\gamma,\,\dot{\beta} (n)\dot{\delta}} - n(n+4)
h_{\alpha\dot{\beta}}
C_{\alpha (n+3),\,\dot{\beta} (n-1)} )\! +\!
O(C^2 ),
\ee
\be
\label{bc2}
D^L\bar{C}_{\alpha (n),\,\dot{\beta} (n+4)}= i\lambda
(h^{\gamma\dot{\delta}}\bar{C}_{\alpha (n)\gamma,\,
\dot{\beta} (n+4) \dot{\delta}}
-n(n+4) h_{\alpha\dot{\beta}} \bar{C}_{\alpha
(n-1),\,\dot{\beta} (n+3)})\! +\! O(C^2 ),
\ee
where $O(C^2 )$ denotes nonlinear terms to be discarded in
the linearized approximation we are interested in.
All these relations contain no new dynamical information in addition to that
contained in the original Einstein equations in the form
(\ref{e1}), (\ref{e2}). Analogously
to the spin 0 case, (\ref{c2}) and (\ref{bc2}) merely express
highest $0-$forms $C_{\alpha(n+4),\dot\beta (n)}$
and $\bar C_{\alpha (n),\dot\beta (n+4)}$
via derivatives of the lowest $0-$forms
$ C_{\alpha (4)}$ and $\bar C_{\dot \beta (4)}$
containing at the same time all consistency conditions for (\ref{e2})
and the equations (\ref{c2}), (\ref{bc2}) themselves. Thus, the system of
equations (\ref{e1}), (\ref{e2}), (\ref{c2}) and (\ref{bc2})
turns out to be dynamically
equivalent to the Einstein equations with the cosmological term.

As shown in \cite{V2,Ann} this construction extends to all
spins $s\geq 1$. The linearized higher spin equations read
\be
\label{s}
R_{1\,\alpha (n) \,,\dgb (m)}=
\delta (m)h^{\gamma \dot{\delta}}
\wedge h^{\gamma}{}_{\dot{\delta}} C_{\alpha (n) \gamma (2)} +
\delta(n)h^{\eta\dot{\delta}}\wedge h_\eta{}^{\dot{\delta}}
\bar{C}_{\dot{\beta} (m) \dot{\delta}(2)}\,,
\ee
($\delta (n)= \delta^0_n$) plus the equations (\ref{DCC}).
Here the curvatures $R_1{}^{\alpha (n)}{}_,{}^{\dot{\beta}(m)}$
are the components of the linearized higher spin curvature tensor
\bee
\label{cur31}
R_1 (y,\bar{y}\mid x)\!\!\!\! &\equiv&\!\!\!\! d w(y,\bar{y}\mid x)\! -\!
w_0 (y,\bar{y}\mid x) * w (y,\bar{y}\mid x)\!+\!
w (y,\bar{y}\mid x) * w_0 (y,\bar{y}\mid x)\nn\\
\!\!\!\!&=&\!\!\!\!\sum_{n,m=0}^{\infty}
\frac{1}{2i\,n!m!}
{y}_{\alpha_1}\ldots {y}_{\alpha_n}{\bar{y}}_{\dot{\beta}_1}\ldots
{\bar{y}}_{\dot{\beta}_m
} R_1{}^{\alpha_1\ldots\alpha_n}{}_,{}^{\dot{\beta}_1
\ldots\dot{\beta}_m}(x)\,,
\eee
where $w_0$ denotes  the $AdS_4$ background fields (\ref{anz0}).
One obtains
\be
\label{RRR}
R_1 (y,\bar{y}\mid x) =D^L w (y,\bar{y} \mid x) -
\lambda h^{\ga\dgb}\Big (y_\ga \frac{\partial}{\partial \bar{y}^\dgb}
+ \frac{\partial}{\partial {y}^\ga}\bar{y}_\dgb\Big )
w (y,\bar{y} \mid x) \,,
\ee
where the Lorentz covariant derivative is defined in (\ref{dlor}).
The component expression for the linearized curvatures following from this
definition is
\be
\label{R1}
R_{1\,\alpha (n)\,,\dgb (m)}=D^L w_{\alpha (n)\dgb (m)}
+n\lambda h_{\alpha}{}^{\dot{\delta}}\wedge
w_{\alpha (n-1)\,,\dgb (m) \dot{\delta}}
+m\lambda h^{\gga}{}_{\dgb}\wedge
w_{\gga \alpha (n)\,,\dgb (m-1) }\,.
\ee

For spins $s\geq 3/2$ the equations
(\ref{DCC}), like in the case of gravity, do not contain any independent
dynamical information just expressing the highest multispinors
$ C_{\alpha (n),\,\dot{\beta} (m)}$
via derivatives of the generalized higher spin Weyl tensors defined
through (\ref{s}),
\be
\label{Cn>m}
\!\!C_{\ga (n)\,,\dgb (m) }\!=\!\frac{1}{(2i\lambda)^{\half (n+m -2s)}}
h^{\un_1}_{\ga\dgb}  D^L_{\un_1}
\!\ldots h^{\un_{\half (n+m -2s)}}_{\ga\dgb}\!
D^L_{\un_{\half (n+m -2s)}}\! C_{\ga (2s)}\,\quad n \geq m ,
\ee
or
\be
\label{Cn<m}
\!\!C_{\ga (n)\,,\dgb (m) }\!=\!\frac{1}{(2i\lambda)^{\half (n+m -2s)}}
h^{\un_1}_{\ga\dgb} D^L_{\un_1}
\!\ldots  h^{\un_{\half (n+m -2s)}}_{\ga\dgb} \!
D^L_{\un_{\half (n+m -2s)}} C_{\dgb (2s)}\,\quad n \leq m .
\ee

For $s=1$ the equation (\ref{s}) is just the definition of
the field strengths $C_{\ga (2)}$ and $C_{\dgb (2)}$ while the
equations (\ref{DCC}) contain Maxwell equations
in the bottom part of the chain \cite{Ann}. For spins 0 and 1/2,
the system (\ref{DCC}) is not linked to the equations for the gauge
potentials (\ref{s}).

Thus, it is shown \cite{} that the free equations of motion for
all massless fields in $AdS_4$ can be cast into the form
\be
\label{R_1}
R_1 (y,\bar{y}|x) = h^{\gga\dgb} \wedge h_{\gga}{}^{\dga}
\frac{\partial}{\partial \bar{y}^\dga}
\frac{\partial}{\partial \bar{y}^\dgb}
C(0 ,\bar{y} |x) +
 h^{\ga\dot{\gamma}} \wedge h^\gb{}_{\dot{\gamma}}
\frac{\partial}{\partial {y}^\ga}
\frac{\partial}{\partial {y}^\gb}
C({y},0 |x) \,,
\ee
\be
\label{DC}
{\cal D}_0 C(Y |x) =0\,.
\ee
This statement, which plays a key role from various
points of view, will be referred to as
Central On-Mass-Shell Theorem.

The infinite set of the $0-$forms $C$
forms a basis in the space of all on-mass-shell
nontrivial combinations of covariant derivatives
of matter fields and (lower and higher spin) curvatures.

A spin $s\geq 1$ dynamical massless field is identified with the 1-form
(potential)
\be
\label{physi}
w_{\ga (n),\dgb (n)}\phantom{MMMMMMMM}  \qquad n = (s-1)\,,
\qquad s\geq 1 \quad \mbox{integer}\,,
\ee
\be
\label{physo}
w_{\ga (n),\dgb (m)}
  \qquad n+m = 2(s-1)\,, \quad |n-m|
=1 \qquad s\geq 3/2 \quad \mbox{half-integer}.
\ee
The matter fields are described by the 0-forms
\be
\label{phys0} C_{\ga (0)\,,\dga (0)}\qquad \qquad\qquad\qquad\qquad\qquad\qquad
s=0\,,
\ee
\be
\label{phys1/2}
C_{\ga (1)\,,\dga (0)}     \oplus C_{\ga (0)\,,\dga (1)} \qquad
\qquad\qquad\qquad\qquad s=1/2\,.
\ee
Eqs. (\ref{R_1}) and (\ref{DC}) contain the free dynamical equations
for all massless fields and express all auxiliary components via higher
derivatives of the dynamical fields by virtue of  (\ref{Cn>m}) and
(\ref{Cn<m}) for the 0-forms and by analogous formulae
having a structure
\be
w_{\ga (n)\,,\dgb (m)} \sim \left ( \lambda^{-1}
\frac{\partial}{\partial x}
\right )^{[\frac{|n-m|}{2}]} w^{phys} +w^{gauge}
\ee
for the gauge 1-forms \cite{V2},
where $w^{phys}$ denotes some field from the list
(\ref{physi}), (\ref{physo}) while $w^{gauge}$
is a pure gauge part.

Let us stress that the equations (\ref{R_1}) and (\ref{DC})
are equivalent \cite{V2}
to the usual free higher spin equations in the AdS space
which follow from the standard actions proposed in \cite{Fr}.
In addition they link together derivatives in the
space-time coordinates $x^\un$ and in the
auxiliary spinor variables $y_\ga$ and $\by_\dga$.
In accordance with (\ref{Cn>m}) and (\ref{Cn<m}), in the sector of
0-forms the derivatives in the auxiliary spinor variables
can be viewed as a square root of the space-time derivatives,
\be
\label{dxdydy}
\frac{\partial}{\partial x^\un } C(y,\by |x)
\sim \lambda  h_\un{}^{\ga\dgb}
\frac{\partial}{\partial y^\ga}
\frac{\partial}{\partial \by^\dgb} C(y,\by |x)\,.
\ee
(This is most obvious from (\ref{opcald}).)
Analogous formula in the sector of higher spin gauge potentials
reads in accordance with (\ref{RRR}) and (\ref{R_1})
\be
\label{dxdyy}
\frac{\partial}{\partial x^\un } w(y,\by |x)
\sim \lambda  h_\un{}^{\ga\dgb}
\Big (\frac{\partial}{\partial y^\ga}
 \by_\dgb w(y,\by |x) +
 y_\ga
\frac{\partial}{\partial \by^\dgb} w(y,\by |x) \Big )\,.
\ee
As a result, any nonlocality in the auxiliary
variables $y$  may imply a space-time nonlocality.

As argued in the section \ref{Star Product}
the associative star-product acting
on the auxiliary spinor variables is nonlocal,
 thus indicating a potential nonlocality in the space-time sense.
The higher spin equations contain star-products via
terms  $C(Y|x) * X(Y|x)$ with some operators
$X$ constructed from the gauge and mater fields.
Once $X(Y|x)$ is at most quadratic in the auxiliary variables $Y^\nu$,
the resulting expressions are local,
containing at most two derivatives in $Y^\nu$.  This is the case
for the $AdS$ background gravitational fields
and therefore, in agreement with the analysis of this section,
the higher spin dynamics is local at the linearized level.
But this may easily be not the case beyond the linearized approximation.
We will illustrate this issue in terms of the integration flow defined in the
section \ref{Integrating Flow}.

Another important consequence of the formulae
(\ref{dxdydy}) and
(\ref{dxdyy}) is that they contain explicitly the inverse
AdS radius $\lambda$ and become meaningless in the flat limit
$\lambda \to 0$. This happens because, when resolving these equations
for the derivatives in the auxiliary variables
$y$ and $\by$, the space-time derivatives appear in the
 combination
\be
\lambda^{-1}
\frac{\partial}{\partial x^\un }
\ee
that leads to the inverse powers of $\lambda$
in front of the terms with higher derivatives in the higher
spin gauge interactions.  This is the main reason why higher
spin interactions require the cosmological constant to be nonzero
as was first concluded
in \cite{FV1}.

To summarize, the following facts are strongly correlated:

\noindent
(i) higher spin algebras are described by the (Moyal)
star-product in the auxiliary spinor space

\noindent
(ii)
relevance of the AdS background

\noindent
(iii) potential space-time nonlocality of the higher spin interactions
     due to the appearance of higher derivatives at the nonlinear level.

These properties are in many respects analogous to
 the superstring picture with the
deep parallelism between the cosmological constant and the
string tension parameter. The  fact that unbroken
higher spin symmetries require AdS geometry may provide an explanation
why the most symmetric higher spin phase is not seen in the
usual superstring picture with the flat background space-time.
Interestingly, recent insight into the structure of
superstring theory in~\cite{SW}
proves that star-product plays a key role in certain regimes.

\subsection{``Plane Waves" in $AdS_4$}
\label{``Plane Waves" in $AdS_4$}

The fact that $C(Y|x )$ describes all
derivatives of the physical fields compatible with the field
equations allows us to solve the dynamical
equations in the form (\ref{gC}). The arbitrary
parameters $C_0 (Y)$  in (\ref{gC}) describe all higher derivatives
of the field $C(Y|x_0 )$ at the point $x_0$ with $g(Y| x_0 ) =I$.
In other words, (\ref{gC}) describes a covariantized Taylor
expansion in some neighborhood of $x_0$. For the gauge function
(\ref{gz}), $x_0 =0$. Let us now illustrate how the formula
(\ref{gC}) can be used to produce explicit solutions
of the higher spin equations in $AdS_4$.

Let us set
\be
\label{CE}
C_0 (Y) =\exp i(y^\ga \eta_\ga + \by^\dga \bar{\eta}_\dga ) \,,
\ee
where $\eta_\ga$ is an arbitrary  commuting complex spinor and
$\bar{\eta}_\dga$ is its complex conjugate. Taking into account that
\be
\tilde{g}^{-1} (Y|x) = g (Y|x)\,,
\ee
inserting $g(Y|x)$ into (\ref{gC}) and using the product law
(\ref{wprod}) one performs elementary Gaussian integrations to
obtain \cite{BV}
\be
C(Y|x) = z^2 \exp i\left
[ -\lambda (y_\ga \by_\dgb+\eta_\ga\bar{\eta}_\dgb )x^{\ga\dgb}
+z(y^\ga \eta_\ga +\by^\dga \bar{\eta}_\dga )\right ]\,,
\ee
where
$z= 1+\lambda^2 \half x^{\ga\dgb}x_{\ga\dgb}\,.$
Using
\be
C_{\ga_1 \ldots \ga_n} (x) =  \frac{\partial}{\partial y^{\ga_1}}
\ldots \frac{\partial}{\partial y^{\ga_n}}
C(y,\by |x) |_{y=\by =0}\,,
\ee
one obtains for the matter fields
and higher spin Weyl tensors
\be
C_{\ga \ldots \ga_{2s}} (x) = z^{2(s+1)}
\eta_{\ga_1} \ldots \eta_{\ga_{2s}}
\exp i k_{\gamma\dgb} x^{\gamma\dgb} \,,
\ee
where
\be
k_{\ga\dgb} =- \lambda \eta_\ga \bar{\eta}_\dgb
\ee
is a null vector expressed in the standard way in terms of spinors.
(Expressions  for the conjugated Weyl tensors carrying dotted indices
are analogous).

Since $z\to 1$ in the flat limit, the
obtained solution indeed describes plane waves in the flat space
limit $\lambda \to 0$ provided that the parameters $\eta_\ga$ and
$\bar{\eta}_\dga$ are rescaled according to
\be
\eta_\ga \to \lambda^{-1/2} \tilde{\eta}_\ga\,,\qquad
\bar{\eta}_\dga \to \lambda^{-1/2} \tilde{\bar{\eta}}_\dga \,.
\ee
On the other hand, $z\to 0$ at the boundary of $AdS_4$ and therefore
the constructed AdS plane waves tend to zero at the boundary.

This approach is very efficient and can be applied
to  produce explicit solutions of the field equations in
many cases as we hope to demonstrate elsewhere.
So far we only focused on the equation (\ref{DC}) for 0-forms
$C$ which has a form of the  covariant constancy condition and
therefore admits an explicit solution (\ref{gC}).
Interestingly enough, although the equation (\ref{R1})
does not have a form of a zero-curvature equation, it also
can be solved in a rather explicit algebraic way
\cite{BV} using a more sophisticated technics explained in
the section \ref{Integrating Flow} and inspired by
the analysis of the nonlinear higher spin dynamics.

\subsection{$AdS_3$}
\label{$AdS_3$}

The d=3 linearized system is much simpler then the d=4
one because d3 ``higher spin" fields are of Chern-Simons type and
do not propagate analogously to the case of d3 gravity \cite{d3gr}.
Equivalent statement is that d3 higher spin fields
do not admit nonzero Weyl tensors. In fact, the name
``higher spin gauge fields" is  misleading  for d=3
because these gauge fields do not carry any degrees of freedom
and therefore do not describe any spin. Higher spin gauge symmetries
are however nontrivial.

Consequently, the d3 Central On-Mass-Shell
Theorem has a form
\be
\label{d3l}
R_1(\hat{y} ,\psi ,k|x) =0 \,,\qquad {\cal D}_0 C(\hat{y}, \psi ,k |x) =0\,,
\ee
where $R_1$ is the linearized part of the d3 curvature tensor
(\ref{cur}) and ${\cal D}_0$ is the covariant derivative
(\ref{cald0}).
As shown in  \cite{BPV}, in the sector of 0-forms,
(\ref{d3l}) describes four massive scalars
$C(\hat{y},\psi ,k|x)=C(-\hat{y},\psi ,k |x)$ and four massive spinors
$C(\hat{y},\psi ,k|x)=-C(-\hat{y},\psi ,k |x)$
arranged into N=2 d3 hypermultiplets.
The values of mass $M$
are expressed in terms of $\lambda$ and $\nu$ as follows \cite{BPV}
\be
\label{M}
  M^2_\pm =\gl^2\frac{\nu(\nu\mp 2)}2 \,
\ee
for bosons, and
\be
\label{M f}
  M^2_\pm =\gl^2\frac{\nu^2}2 \,
\ee
for fermions. The signs ``$\pm$'' correspond to the projections
\be
\label{pm}
   C^\pm=P_\pm C\, ,\qquad P_\pm=\frac{1\pm k}2\, .
\ee
One doubling of a number of fields of the same mass is due to
the dependence on $\psi$ ($\psi^2 =1$) while another one, with
the mass splitting in the bosonic sector, is due to $k$.
As expected, the flat ``higher spin" connections do not
describe any local degrees of freedom. The property that
the values of masses depend on a free parameter
$\nu$ in the deformed oscillator algebra (\ref{modosc})
is quite different from what happens in d=4 where
only massless matter fields appear because they  all
belong to the same  multiplet with massless higher spin gauge fields.

The component form of the covariant constancy conditions
(\ref{d3l}) with appropriately rescaled component fields
amounts to \cite{BPV}
\be
\label{chainbos}
   D^LC_{\ga(n)} = \lambda \Big (h^{\gb\gga}C_{\gb\gga\ga(n)}+
        n(n-1)\left( \frac{1}{4}-{M^2\over2 \lambda^2 (n^2-1)} \right)
        h_{\ga\ga}C_{\ga(n-2)} \Big )
\ee
for a boson ($n$ is even), and
\bee
\label{chainferm}
  D^LC_{\ga(n)} =\lambda\Big ( h^{\gb\gga}C_{\gb\gga\ga(n)}\!\!\! &-&\!\!\!
     \frac {\sqrt{2} M}{\lambda (n+2)}\; h_{\ga}{}^{\gb}C_{\gb\ga(n-1)}\nn\\
 \!\!\!&+&\!\!\! n(n-1)\left( {1\over 4}-{M^2\over 2\lambda^2 n^2} \right)
          h_{\ga\ga}C_{\ga(n-2)} \Big )
\eee
for a fermion ($n$ is odd).

Note that analogously to the d4 formulae (\ref{Cn>m}) and
(\ref{Cn<m}) one gets
\be
\label{Cn}
C_{\ga (n)}=\frac{1}{(\lambda)^{\half (n -2s)}}
h^{\um_1}_{\ga\ga} D^L_{\um_1}
\ldots h^{\um_{\half (n -2s)}}_{\ga\ga}
D^L_{\um_{\half (n -2s)}} C_{\ga (2s)}\,,\quad  s=0\,\,\,\mbox{or}\,\,\, 1/2\, .
\ee
As in d=4, this means that
\be
\label{3dxdydy}
\frac{\partial}{\partial x^\un } C(y |x)
\sim \lambda  h_\un{}^{\ga\gb}
\frac{\partial}{\partial y^\ga}
\frac{\partial}{\partial y^\gb} C(y |x)\,.
\ee

The formula (\ref{gC}) is true  for any value of $\nu$ in the d3
case. However it is not straightforward to apply it in the massive d3 case
because no practically useful formula is known
generalizing (\ref{wprod}) to the case of arbitrary $\nu$. The problem of
developing an efficient machinery of the symbols of
operators for the case of general $\nu$ is therefore quite interesting.

\section{Free Differential Algebras and Unfolded Formulation}
\label{Free Differential Algebras and Unfolded Formulation}

Let us consider\footnote{In this section, the identification of
indices may be different from the conventions summarized in  the
Appendix.} an arbitrary set of differential $p-$forms $W^A (x)$
 with $p\geq 0$ (0-forms are included). Let the generalized
curvatures $R^A$ be defined by the relations
\be
R^A= dW^A +F^A (W)\,,
\ee
where $d=dx^\un \frac{\partial}{\partial x^\un}$
and $F^A$ are some functions of $W^B$ built with the aid of the
exterior product of differential forms.  Given function
$F^A (W)$ satisfying the generalized Jacobi identity
\be
\label{BI}
F^B \frac{\delta F^A }{\delta W^B} \equiv 0\,
\ee
(the derivative with respect to $W$ is left), we say
following to \cite{FDA} that
it defines a free differential algebra. This property
 guarantees the generalized Bianchi identity
\be
dR^A = R^B
\frac{\delta F^A }{\delta W^B}\,,
\ee
which tells us that the differential equations on $W^A$
\be
\label{eq}
R^A =0
\ee
are consistent. Clearly, the requirement that the equation
(\ref{eq}) is consistent for generic fields $W^A$ is
equivalent to (\ref{BI}).

The property (\ref{BI})
 allows one to define the gauge transformations
\be
\label{delw}
\delta W^A = d \e^A -\e^B
\frac{\delta F^A }{\delta W^B}\,,
\ee
where $\epsilon^A (x) $ is a $(deg (W^A )-1)- $form (0-forms do not
give rise to any gauge parameters). With respect to these gauge
transformations the generalized curvatures transform as
\be
\delta R^A =-R^C
\frac{\delta }{\delta W^C} \left (\e^B
\frac{\delta F^A }{\delta W^B} \right )\,.
\ee
This implies gauge invariance of the equations (\ref{eq}).
Also, since the equations (\ref{eq}) are formulated
entirely in terms of differential forms, they are explicitly
general coordinate invariant.

For the particular case when the set $W^A$ consists of only
1-forms $w^i$, the function $F^i (w)$ is bilinear
\be
F^i = f^i_{jk} w^j \wedge w^k
\ee
and the relation (\ref{BI}) amounts to the usual Jacobi identity
for a Lie algebra $g$ with the structure coefficients
$ f^i_{jk} $ (or superalgebra if some of $w^i$ carry an
additional Grassmann grading). The equation (\ref{eq}) is then
the zero-curvature equation for $g$.

If the set $W^A$ also contains some  $p$-forms $C^\ga$
(e.g. 0-forms) and the functions $F^\ga$ are linear in $C$
\be
F^\ga = t_i{}^\ga {}_\gb w^i \wedge C^\gb\,,
\ee
the relation (\ref{BI}) implies that the matrices
$t_i{}^\ga {}_\gb$ form some representation $t$ of $g$
while the equations (\ref{eq}) contain zero-curvature
equations of $g$ along with the covariant constancy
equation $DC=0$ for the representation $t$.

We see that the vacuum equations (\ref{vacu}) $R = 0$
and free equations (\ref{cald0}) ${\cal D}_0 C=0$ are just of this
form. Therefore, the fields $C$ span some representation of
the AdS algebra. Moreover, since the equations (\ref{vacu})
and (\ref{cald0}) are formally consistent independently of a
particular solution $w_0$, the 0-forms $C$ form some
representation of the whole infinite-dimensional higher spin
algebra. This is the twisted representation defined by
(\ref{tw}).

This simple observation suggests the following
strategy for the analysis of the higher spin theories
(in fact, any dynamical system). Starting from a space-time
with some symmetry algebra $s$ and vacuum gravitational
gauge fields (1-forms) $w_0$ taking values in $s$
and satisfying the zero curvature equations $dw_0=w_0 \wedge w_0$,
one reformulates  field  equations of a given free dynamical system
in the ``unfolded form" ${\cal D}_0 C=0$. This can always
be done in principle and the only question is how simple is
the explicit expression for ${\cal D}_0 C$. Indeed, one starts writing
\be
D_\un^L C^i_0 =  h_\un{}_n C^{i,n}_{1}
\ee
for a given dynamical field $C^i_0 (x)$ with some set of
spinor and/or vector indices $i$. Next, one checks
whether the original
field equations impose any restrictions on the first
derivatives of $C_0^i$. If they do, as it happens for  fermions,
one expresses the corresponding components of
$ C^{i,n}_{1} $ in terms of $C^i_0$ treating the unrestricted part
$\tilde{C}^{i,n}_{1}$ of $ C^{i,n}_{1} $ as new independent fields
parametrizing  on-mass-shell nonvanishing components of first derivatives.
This  leads to the equation
\be
\label{D0C}
D_\un^L C^i_0 =  h_\un{}_n \tilde{C}^{i,n}_{1} +A_{1\un}^i (C_0 )\, .
\ee
Then one writes
\be
D_\un^L \tilde{C}^{i,n}_{1} =  h_\un{}_m {C}^{i,n,m}_{2}\,,
\ee
where ${C}^{i,n,m}_{2}$ parametrizes the second derivatives.
Once again one checks, taking into account the Bianchi identities
for  (\ref{D0C}), which components
$\tilde{C}^{i,n,m}_{2}$ of the second level fields
remain independent provided that the original
equations of motion are true, expressing the rest of the components
of ${C}^{i,n,m}_{2}$
in terms of the lower derivatives $C_0^i$ and $\tilde{C}^{i,n}_{1}$.
This process continues infinitely leading to a chain
of equations having a form of some covariant constancy conditions.
By construction, the resulting set of fields
$\tilde{C}^{i,n_1 \ldots n_m}_m$ realizes some representation
of the space-time symmetry algebra $s$ (e.g. Poincare' or AdS).
The interactions are then described by nonlinear deformations of
the resulting free differential algebra.

It is useful to address the question which infinite-dimensional
extension of $s$ can act on thus derived representation $t$ of $s$.
A natural candidate is a Lie superalgebra $g$ constructed via
(anti)commutators from the associative algebra
$H$ \be
H=U(s)/I(t)\,,
\ee
where $U(s)$ is the universal enveloping of $s$ while $I(t)$ is the
ideal of  $U(s)$ spanned by the elements which trivialize on
the representation $t$. Of course, this strategy is too naive
in general because a set of fields $C$ mixed by a higher spin
algebra and compatible with the nonlinear higher spin
dynamics may take values in a larger representation. In any case,
$U(s)$ is the reasonable starting point to look for a higher spin algebra.
Based on somewhat different arguments, this idea was put
forward by Fradkin and Linetsky in \cite{FLU}.
As shown in the section
\ref{d=3 Higher Spin Symmetries and Deformed Oscillators}
it works in d=3 at least for the case without inner symmetries.
Similar interpretation of the Weyl algebra (endowed with Klein
operators) underlying the d=4 higher spin superalgebra
can be given in terms of $U(sp(4))$ in the
bosonic case and $U(osp(1,4))$ in the supersymmetric
case\footnote{This is a
consequence of the fact that Weyl algebra acts on the
Fock space defined with respect to  creation and annihilation operators
built from $y_\ga$ and $\bar{y}_\dgb$ and identified
with the metaplectic (singleton) representation of $sp(4|R)$
(for more detail see \cite{KV1} and references therein).}.

The language of free differential algebras is perfectly
adequate for the study of interactions of higher spin theories.
The Central On-Mass-Shell Theorem is just a right starting
point to attack this problem which reduces to searching for
such a deformation of the equations
(\ref{R_1}) and (\ref{DC})
\be
\label{HSFDA}
R (w) = F_2 (w,C)\,,\qquad {\cal D} C = F_1 (w,C)
\ee
with a 2-form $F_2$ and 1-form $F_1$,
that is consistent in the sense of (\ref{BI}) and reproduces
the linearized equations in the lowest order of the expansion
in powers of $C$ and $w_1$ identified with the fluctuational
part  in $w=w_0 +w_1 $. {}From (\ref{R_1}) it
is clear that the nonlinear deformation is inevitable
in $d\geq 4$ because
\be
{\cal D}^2 (C) \sim R C \sim C^2
\ee
and therefore $F_1$  necessarily
starts from some terms bilinear in $C$. In its turn,
this may induce some further nonlinear correction
to $F_2$ as a result of the differentiation of $C$ in (\ref{R_1}).

It is not {\it a priori} guaranteed that some deformation
$F_{1,2}$ exists at all. If not, this would mean that no
consistent nonlinear higher spin equations exist.
(Note that dynamics of any consistent system
can in principle be rewritten in the
unfolded form.)
Once the complete form of $F_1$ and $F_2$ is found, the problem is
solved because the resulting equations are formally consistent,
gauge invariant and generally coordinate invariant as a
consequence of the general properties of free differential algebras.
By construction, it describes the correct dynamics at the
free field level.

Let us stress that one can proceed analogously for
other dynamical systems containing gravity either at
the dynamical level (i.e with the zero-curvature equation
for the gravitational fields deformed by the Weyl tensor)
or at the background level (i.e. with the background
gravitational field satisfying the vacuum zero-curvature
equation). The  ``unfolded formulation" has much in common with
the Penrose ``exact sets of fields" formulation
\cite{PR}. In this approach some infinite systems of equations also
appear, containing both dynamical equations and the relations
expressing higher components via higher derivatives of the dynamical
fields. The important difference is that the procedure we use
(``unfolding") is formulated in terms of differential forms
(gauge potentials) containing, in
particular, the gravitational field. As a result,  our formulation
brings together such seemingly different issues as general coordinate
invariance, gauge invariance and formal compatibility of the dynamical
equations. The common feature of the two approaches
is that all higher derivatives compatible with the field
equations at any point $x_0$ of the space-time are identified with
values of certain fields at $x_0$, the fields $C(Y|x_0 )$ in
the higher spin case.

Coming back to the higher spin problem, let us note that
there was one more requirement imposed
on the form of the higher spin free differential algebra.
It was assumed that the gravitational Lorentz
connection $\go_{\ga\gb}$ and $\bar{\go}_{\dga\dgb}$ only
appear via Lorentz covariant derivatives (and their commutator
in the Riemann tensor). Equivalently, the deformations
$F_{1,2}(w,C)$ were required to be independent of the Lorentz
connection. {}From (\ref{delw}) it follows that this requirement
means that the transformation law for local Lorentz rotations remains
undeformed. This requirement is natural and restrictive as
 explained in the section \ref{Interaction Ambiguity}.
The only other symmetry that remains undeformed
in the extended higher spin systems is the spin-1
Yang-Mills symmetry \cite{Pr}.
All other symmetries deform, acquiring some $C$-dependent
corrections to their transformation laws. In particular, this
happens with the AdS translations. It is not surprising,
since it is known from the SUGRA example \cite{PVN} that local
AdS translations acquire some curvature-dependent corrections
which deform them to diffeomorphisms.

This general problem setting was studied for d4 higher spin
theories  in a series of papers (see e.g. \cite{Ann,V6}) within
 the order by order analysis in powers of the
0-forms $C$. This perturbation expansion is natural
 because the 0-forms $C$ describe deviations
 of matter fields and higher spin Weyl tensors from
their vacuum values $C=0$, i.e. this is just a weak coupling
expansion. In every order, a contribution of the 1-forms $w$ was
taken into account completely. Of course, the important
difference between 0-forms $C$ and 1-forms $w$ is that the
latter can appear at most quadratically in this approach
while the former can appear in arbitrary power so that the
functions $F_{1,2}$ may be nonpolynomial in $C$. The concrete
form of the functions $F_{1,2}$ is quite cumbersome
and can hardly be written down explicitly in all orders.
However, in \cite{Pr,more} (and references therein)
it was suggested that these functions
can be described as solutions of certain
rather simple nonlinear equations in additional
spinor variables which can be proved to have
a unique solution modulo field redefinitions, thus,
among other things, proving the fact of existence of the
higher spin free differential algebra in all orders in
interactions. Skipping technical details of the derivation
of these equations we instead formulate in the
section \ref{Nonlinear Higher-Spin Equations}
the final result and show how it reproduces the higher spin free
differential algebra.

Let us note that although we know the closed equations for higher
spins, for the case of pure gravity or Yang-Mills theory an
explicit form of all terms nonlinear in $C$ on the
right hand sides of (\ref{c2}) and (\ref{bc2}) is still unknown.
The form of $C^2-$ type terms was obtained for
the case of gravity in \cite{V6}. For the (anti)selfdual case
these $C^2$ terms were shown to be complete.

The conclusion is that higher spin systems  are
in some sense simpler than the lower spin systems used in the
standard low energy theories. In fact,  the
 lower spin systems like Einstein, Yang-Mills, and others are
unlikely to be subsystems of the full higher spin gauge
theories, which can be obtained by a consistent truncation.
Indeed, lower spin fields form sources for
higher spin fields via higher spin currents analogous to those
considered in the section \ref{Higher Spin Currents}.
To single out the lower spin subsystems one has to implement
some limiting procedure, a kind of low-energy expansion. A most
natural possibility to achieve this is via  spontaneous
breaking of the higher spin symmetries down to some lower spin
symmetries. The symmetry breaking parameter equal to
 the inverse higher spin mass scale should then be identified
with  an expansion parameter reminiscent of   $\ga^\prime$
in superstring theory. Of course, some of the nice properties
of the original theory with unbroken higher spin gauge symmetries
may be lost in the resulting low-energy lower spin system.

\section{Nonlinear Higher-Spin Equations}
\label{Nonlinear Higher-Spin Equations}

In this section we resolve the problem of reconstruction of the d=3
and d=4 higher spin free differential algebra in all orders by
formulating some closed consistent system in an appropriately
extended space. The resulting formulation, based on certain
non-commutative Yang-Mills fields, is interesting on its own right
and, in particular, in the context of the recent developments in
the super\-string theory \cite{ch,SW}. The content of this section
is based on the papers \cite{Pr,more,Eq,PV,Bu}.

\subsection{Doubling of Spinor Variables and Non-Commutative Gauge Theory}
\label{Doubling of Spinor Variables and Non-Commutative Gauge Theory}

The key element of the construction consists of the
doubling of auxiliary Majorana spinor variables $Y_\nu$
in the higher spin 1-forms and 0-forms
\be
w(Y;Q|x)\longrightarrow W(Z;Y;Q|x)\,,\qquad
C(Y;Q|x)\longrightarrow B(Z;Y;Q|x)
\ee
and formulating equations which determine the dependence on
the additional variables $Z_\nu$ in terms of ``initial data"
\be
\label{inda}
w(Y;Q|x)=W(0;Y;Q|x)\,,\qquad
C(Y;Q|x)= B(0;Y;Q|x).
\ee
The variables $Q$ denote some discrete (Clifford)
variables which are different for $d=3$ and $d=4$ cases
and will be specified later.

To this end we introduce a new compensator-\-type spinor field
$S_\nu (Z;Y;Q|x)$ which carries only pure gauge degrees of freedom
and plays a role of a covariant differential along the
additional $Z_\nu$ directions. It is convenient to introduce
anticommuting $Z-$differentials $dZ^\nu dZ^\mu=-dZ^\mu dZ^\nu$
to interpret $S_\nu (Z;Y;Q|x)$ as $Z$ 1-forms,
\be
S=dZ^\nu S_\nu \,.
\ee

The full system of equations in d=3 and d=4 has the following form
\be
\label{dW}
dW=W*W\,,\qquad
\ee
\be
\label{dB}
dB=W*B-B*W\,,\qquad
\ee
\be
\label{dS}
dS=W*S-S*W\,,
\ee
\be
\label{SB}
S*B=B*S\,,
\ee
\be
\label{SS}
S*S= dZ^\nu dZ^\mu \,(-iC_{\nu\mu}+4 R_{\nu\mu}(B;c))\,,
\ee
where $C_{\nu\mu}$ is the charge conjugation matrix
and $R_{\nu\mu}(B;c) $ is certain star-product function
of the field $B$ and some central elements $c$ of the algebra.
The function $R_{\nu\mu}(B;c) $ encodes all information
about the higher spin dynamics and will be specified later.

In the analysis of the higher spin dynamics, a typical vacuum
solution for the field $S$ is
\be
S_0 = dZ^\nu Z_\nu \,.
\ee
{}From (\ref{z,f}) it follows then
that
\be
\label{S0f}
[S_0, f]_* =-2i\partial f\,,
\ee
where
\be
\label{partZ}
\partial =dZ^\nu \frac{\partial}{\partial Z^\nu }\,.
\ee
Interpreting the deviation of the full field $S$ from
the vacuum value $S_0$ as a $Z-$component of the gauge field
$W$,
\be
S=S_0 +2i dZ^\nu W_\nu\,,
\ee
one rewrites the equations (\ref{dW}), (\ref{dS}) and (\ref{SS}) as
\be
\label{calr}
{\cal R} =
 dZ^\nu dZ^\mu \,R_{\nu\mu}(B;c)
\ee
and the equations (\ref{dB}) and (\ref{SB}) as
\be
\label{cald}
{\cal D} B =0\,.
\ee
Here the generalized curvatures and covariant derivative
are defined by the relations
\be
\label{defcalr}
{\cal R} = (d+\partial ) (dx^\un W_\un +dZ^\nu W_\nu )
-(dx^\un W_\un +dZ^\nu W_\nu )\wedge
(dx^\un W_\un +dZ^\nu W_\nu )\,,
\ee
\be
\label{defcald}
{\cal D} ( A) = (d+\partial ) A-
(dx^\un W_\un +dZ^\nu W_\nu ) * A +
A* (dx^\un W_\un +dZ^\nu W_\nu ) \,.
\ee
(It is assumed that $dx^\un dZ^\nu =-dZ^\nu dx^\un$.)
We see that the function $R_{\nu\mu}$ in (\ref{SS})
identifies with the $ZZ$ components of the generalized
curvatures, while $xx$ and $xZ$ components of the
curvature vanish. The equation (\ref{cald})  means
that the curvature $R_{\nu\mu}$ is covariantly constant.

The equations (\ref{calr}) and (\ref{cald}) are consistent.
The Bianchi identities for (\ref{calr}) are satisfied as a
consequence of (\ref{cald}). The Bianchi identities for (\ref{cald})
are compatible with the equation (\ref{calr}) reducing to the
star-commutators of some functions of $B$ and central elements $c$
which vanish due to the simple fact that $F(B;c)*G(B;c)$=$G(B;c)*F(B;c)$
for any $F$ and $G$ since $B$ commutes to itself and central elements.

A related statement is that the equations (\ref{calr}) and (\ref{cald})
are gauge invariant under the transformations
\be
\label{gxz}
\delta (dx^\un W_\un +dZ^\nu W_\nu ) = (d+\partial) \gvep +
\gvep * (dx^\un W_\un +dZ^\nu W_\nu ) - (dx^\un W_\un +dZ^\nu W_\nu )*
\gvep\,,
\ee
\be
\label{deltaB}
    \delta B=[\gvep ,B]_* \,,
\ee
with an arbitrary gauge parameter $\gvep (Z;Y;Q|x)$.
In terms of the original variables the gauge transformations
(\ref{gxz}) have a form
\be
\label{deltaW}
      \delta W=d\gvep+[\gvep , W ]_* \,,
\ee
\be
\label{delta S}
      \delta S=[\gvep , S ]_*  \,.
\ee
Note that the gauge transformations for $Z-$components  $W_\nu$
of the gauge field  acquire
the inhomogeneous term due to the vacuum expectation value $S_0$
of $S$.

The consistency of the system of
equations (\ref{dW})-(\ref{SS}) guarantees that it admits a
perturbative solution as a system of differential equations
with respect to $Z_\nu$. This proves that all fields can be
expressed modulo gauge transformations in terms of the
``initial data" (\ref{inda})  identified with the physical higher
spin fields. Inserting thus obtained
expressions into (\ref{dW}) and (\ref{dB}) one finds
some (nonlinear) corrections to $Z-$independent parts of the
higher spin curvatures and covariant derivatives. (One has to take
into account that $(f*g )(0;Y)$ is generically different from
zero if $f(0;Y)=0$ and/or $g(0;Y)=0$ because $Z-$dependent
terms in $f$ and/or  $g$ contribute to the $Z-$independent
part of their star-product).
By construction, the resulting system of equations will be consistent as
a space-time free differential algebra, thus solving the problem.
Concrete details of this procedure for d=4 and d=3 are explained in
the sections \ref{Nonlinear Equations in d=4} and
\ref{Nonlinear Equations in d=3}  following to the
original papers \cite{Pr,more,Eq,PV}.

Since all components of curvatures and covariant derivatives
along space-time directions vanish, this allows one to
solve  all those equations, which contain space-time
derivatives, in the pure gauge form analogous to (\ref{pg})
and (\ref{gs})
\be
\label{PG}
W = -g^{-1}(Z;Y;Q|x)* d g(Z;Y;Q|x)\,,
\ee
\be
\label{GB}
B (Z;Y;Q|x) = g^{-1}(Z;Y;Q|x)* b (Z;Y;Q)* g(Z;Y;Q|x) \,,
\ee
\be
\label{GS}
S (Z;Y;Q|x) = g^{-1}(Z;Y;Q|x)*s (Z;Y;Q)* g(Z;Y;Q|x)
\ee
with some invertible $g(Z;Y;Q|x)$ and arbitrary $x-$independent
functions\\ $b (Z;Y;Q)$ and $s (Z;Y;Q)$. Due to the gauge invariance
of the whole system one is left only with the equations (\ref{SB})
and (\ref{SS}) for $b (Z;Y;Q)$ and $s(Z;Y;Q)$. These encode in
a coordinate independent way all information about the dynamics
of massless fields of all spins. From the perspective of the equations
(\ref{calr}) and (\ref{cald}) these equations imply some
algebraic (via particular form of $R_{\nu\mu}$) and differential
($dZ^\nu {\cal D}_\nu B =0$) constraints
on the $ZZ$ non-commutative star-product field strength built from
the potential $dZ^\nu W_\nu (Z;Y;Q)$.

The global symmetry of the system is identified with the
subalgebra of the local symmetry
(\ref{deltaB})-(\ref{delta S}) that leaves invariant
a vacuum solution $W_0$, $B_0$ and $S_0$ of the field equations.
{}From (\ref{deltaB})-(\ref{delta S}) it follows that the global
symmetry parameters satisfy
\be
\label{gl}
      d\gvep^{gl} =[W_0 , \gvep ]_* \,,\qquad
      [\gvep^{gl} ,B_0 ]_* =0\,,\qquad
      [\gvep^{gl} , S_0 ]_* =0  \,.
\ee
The first equation is consistent because $W_0$ satisfies (\ref{dW})
and can be solved as
\be
\gvep^{gl} (Z;Y;Q|x) = g^{-1}_0 (Z;Y;Q|x)*
\gvep^{gl}_0 (Z;Y;Q)* g_0(Z;Y;Q|x) \,.
\ee
The second one will be satisfied trivially since
we will only consider vacuum solutions with $B_0 = const$.
The third implies that $\gvep^{gl}_0$ belongs
to the centralizer of $s_0$.
We therefore conclude that the global symmetry algebra
is isomorphic to the centralizer of $S_0$. (Note that
$s_0=S_0 (x_0 )$ if $g_0 (Z;Y;Q|x_0 ) = 1$;
the centralizers of $S_0 (x)$ for different $x$
are pairwise isomorphic due to (\ref{GS}).)

In our approach, non-commutative gauge fields appear
in the auxiliary spinor space associated with the coordinates
$Z^\nu$. The dynamics of the higher spin gauge fields
is formulated entirely in terms of the corresponding non-commutative
gauge curvatures. For the first sight it is very different from
the non-commutative Yang-Mills model considered recently in
\cite{SW} in the context of the description of the new phase of
string theory, in which star-product is defined directly in terms
of the original space-time coordinates $x^\un$.
However, the difference may be not that significant
taking into account the relationships like (\ref{dxdydy}) and
(\ref{dxdyy}) between space-time and spinor derivatives,
which are themselves consequences of the equations (\ref{dW})
and (\ref{dB}) as will become clear in the sections
 \ref{Nonlinear Equations in d=4} and
\ref{Nonlinear Equations in d=3}. {}From this perspective, the
situation with the higher spin equations is reminiscent of
the Fedosov quantization approach developed in \cite{fed}
to solve the problem of quantization of symplectic structures.
In this approach the complicated problem of quantization of some (base)
manifold (coordinates $x^\un$) is reduced to a simpler problem of
quantization in the fibre endowed with the Weyl star-product structure
(analog of coordinates $Z^n$). An important
difference between the Fedosov's approach and the structures
underlying the higher spin equations is that the former is
based on the vector fiber coordinates $Z^n$,
while the higher spin dynamics prefers spinor coordinates
$Z^\nu$ (see also \cite{cast} for a discussion of the parallels
between the higher spin gauge theory and Fedosov quantization).

A few comments are now in order.

As explained in the sections \ref{Nonlinear Equations in d=4} and
\ref{Nonlinear Equations in d=3}, the discrete variables $Q$
sometimes do not commute with the spinor elements $dZ^\nu$ and
$Z^\nu$. One therefore has to be careful with the naive
interpretation of $\frac{\partial}{\partial Z^\nu}$ and $dZ^\nu$
taking into account some additional sign factors when necessary.

Sometimes, less trivial vacuum solutions
with $S_0\neq dZ^\nu Z_\nu$ turn out to be relevant. In that case,
$[S_0 , A]_* \neq \partial A$, and the interpretation
in terms of the usual noncommutative Yang-Mills theory becomes
less straightforward. Moreover, as shown in the
section \ref{Nonlinear Equations in d=3}, the same equations may
admit different vacuum solutions with essentially different
vacuum fields $S_0$. {}From that perspective it is practically more
convenient to treat  $S*S$ as the fundamental object
generalizing the non-commutative Yang-Mills strength.

The fact that the equations of motion (\ref{dW})-(\ref{SS})
are formulated in terms of fields $W$, $B$ and $S_\nu$ taking
values in the associative star-product algebra  $A$
allows one to extend the construction to the case with inner
symmetries by endowing all fields with the matrix indices, i.e.
by the extension $A\rightarrow A\otimes Mat_n$. The analysis of
possible reductions of the full system \cite{Pr} is parallel to
that of the section \ref{Extended Higher Spin Algebras}. The
perturbative analysis will work equally well provided that non-zero
vacuum fields $W_0{}^i{}_j$, $B_0{}^i{}_j$ and $S_0^\nu{}^i{}_j$ take
values in the singlet subalgebra $A\otimes I$ spanned by the
$n\times n$ unit matrices.

Consistency of the system (\ref{dW})-(\ref{SS}) is
not sufficient alone to fix a form of the
curvature $R_{\nu\mu}$. Some additional arguments taking into
account that the system should describe appropriate
relativistic dynamics have to be added. We will come back to
this point in the section \ref{Interaction Ambiguity}.

\subsection{Nonlinear Equations in d=4}
\label{Nonlinear Equations in d=4}

In the d=4 case higher spin dynamics is described
by the field variables\hfil\\  $ W(Z;Y;{\cal K}|x) $,
$ B(Z;Y;{\cal K}|x) $ and $ S(Z;Y;{\cal K}|x) $ depending on the
space-time coordinates $ x^{\un} $ $ (\un = 0 \div 3) $, spinor variables $
Z^\nu = (z^{\alpha }, \bar{z}^{\dot{\alpha}}), $ $ Y_\nu
=(y_{\alpha},\bar{y}_{\dot{\alpha}}) $ and two Klein operators $ {\cal
K}=(k,\bar{k}) $.  $ S $ is a 1-form with respect to the auxiliary
anticommuting
spinorial differentials $ dz^\alpha $ and $ d\bar{z}^{\dot{\alpha}} ,$
\be
S=s+{\bar{s}}\,,\qquad s=dz^\alpha  s_\alpha (Z;Y;{\cal K}|x)\,,\qquad \bar{s}
=d\bar{z}^{\dot{\alpha}} \bar{s}_{\dot{\alpha}} (Z;Y;{\cal K}|x)\,,
\ee
\be
dz^\alpha dz^\beta=-dz^\beta dz^\alpha\,,\quad
d\bar{z}^{\dot{\alpha}}\, d\bar{z}^{\dot{\beta}}= -d\bar{z}^{\dot{\beta}}\,
d\bar{z}^{\dot{\alpha}}\,,\quad  dz^\alpha d\bar{z}^{\dot{\beta}}
=-d\bar{z}^{\dot{\beta}} dz^\alpha\,.
\ee
By definition, the Klein operators $ k $ and $ \bar{k} $ anticommute with all
undotted and dotted spinors, respectively,
\be
\label{tildek}
kf(Z;\!Y;\!dZ;{\cal K})\!=\!f(\tilde{Z}\!;\tilde{Y}\!;d\tilde{Z};
{\cal K})k ,\quad \bar{k}f(Z\!;Y\!;dZ;{\cal K})\!=\!
f(-\tilde{Z}\!;-\tilde{Y}\!;-d\tilde{Z};{\cal K})\bar{k}
\ee
with
\be
\tilde{U}_\nu=(-u_\alpha,\,\bar{u}_{\dot{\alpha}})\,\,\,\,\mbox{for}
\,\,\,\,\,
U_\nu=(u_\alpha,\,\bar{u}_{\dot{\alpha}})\,.
\ee
In addition, it is required that
\be
k^2=\bar{k}^2=1\,,\qquad [k,\bar{k}]=0\,,
\qquad [k,dx^\un ]=[\bar{k},dx^\un ]=0;
\ee
\be
\{dx^\un ,dz^\alpha \}=0\,,\qquad\{dx^\un ,d\bar{z}^{\dot{\alpha}}\}=0\,.
\ee
In accordance with (\ref{tildek}), we assume in this section
that the Klein operators
$k$ ($\bar{k}$) anticommute with the differentials $dz^\ga$
$(d\bar{z}^\dgb )$.

The field variables $ W$, $ B $ and $ S $ obey the (anti-)hermiticity
(reality) conditions
\be
\label{reality}
W^\dagger =-W,\,\,\,\, B^\dagger =B,\,\,\,\, S^\dagger =-S
\ee
with the involution $ \dagger $
(i.e., $ (f\circ g)^\dagger =g^\dagger \circ  f^\dagger , $
 $(f^\dagger )^\dagger =f,$ $(\lambda f+\mu g)^\dagger
=\bar{\lambda} f^\dagger +\bar{\mu} g^\dagger ) $
 defined by the relations
\be
(z_\alpha )^\dagger =-\bar{z}_{\dot{\alpha}},\,\,\,
(y_\alpha )^\dagger =\bar{y}_{\dot{\alpha}},\,\,\,
(dz_\alpha )^{\dagger} =-d\bar{z}_{\dot{\alpha}}, \,\,\,
(dx^\nu )^\dagger =dx^\nu ,\,\,\, k^\dagger =\bar{k}\, .
\ee

Following to \cite{more} we fix the curvature on the right hand side
of (\ref{SS}) in the form
\be
\label{Rab}
dZ^\nu dZ^\mu \, R_{\nu\mu}(B)=- \frac{i}{4} \left (
dz_\alpha dz^\alpha \,F(B)*\kappa\,\,+ d\bar{z}_{\dot{\alpha}}\,
d\bar{z}^{\dot{\alpha}}\,\bar{F}(B)*\bar{\kappa}\right )\,.
\ee
Here $ F(B) $ is an arbitrary star-product function of the
0-form $ B $
\be
\label{F}
F(B)= \sum^{\infty}_{n=0} \frac{1}{n!} f_n
\,\,\underbrace{B*...*B}_{n} \,\,,
\ee
with some complex coefficients $f_n$. $ \bar{F}(B) $ is its complex
conjugate,
\be
\label{barF}
\bar{F}(B)= \sum^{\infty}_{n=0} \frac{1}{n!}\bar{f}_n
\,\,\underbrace{B*...*B}_{n}\,\, .
\ee
The operators
$ \kappa $ and $ \bar{\kappa} $ have the form
\be
\label{[kb]}
\kappa =k\ups,\qquad \bar{\kappa}
=\bar{k}\bu \,,
\ee
where $\ups$ and $\bu$ are the inner Klein operators (\ref{kk4}).

{}From (\ref{[uf]}) and (\ref{tildek}) it follows that
\be
\label{[bkb]}
\kappa *f(Z;Y;{\cal K};dZ)\!=\!f(Z;Y;{\cal K};d\tilde{Z})*\kappa ,\,\,
\bar{\kappa}
*f(Z;Y;{\cal K};dZ)\!=\!f(Z;Y;{\cal K};-d\tilde{Z})*\bar{\kappa},
\ee
i.e., $ \kappa (\bar{\kappa}) $ commutes with all quantities except for $
dz^\alpha (d\bar{z}^{\dot{\alpha}}) $.
As a result, the equation (\ref{SS}) with the
curvature tensor (\ref{Rab}) acquires a form of two mutually commuting
deformed oscillator relations with $F(B)$ and $\bar{F}(B)$ playing a
role of $\nu$ and $\bar{\nu}$, respectively (note that $B$ commutes
with $S$ and is covariantly constant according to (\ref{SB}) and
(\ref{dB})).

In order to prove that the system
(\ref{dW})-(\ref{SS})  is consistent, one has to use the
fact that the quantities
\be
c_1 =dz_\alpha dz^\alpha,\,\,\,\,\,\, \bar{c}_1
=d\bar{z}_{\dot{\alpha}}d\bar{z}^{\dot{\alpha}}
\ee
and
\be
c_2 =dz_\alpha dz^\alpha \kappa ,\,\,\,\,\,\, \bar{c}_1
=d\bar{z}_{\dot{\alpha}}d\bar{z}^{\dot{\alpha}} \bar{\kappa}
\ee
behave as central elements commuting with all elements of the
algebra. As for $
c_1 $ and $ \bar{c}_1, $ this is trivial.
For $ c_2 $ and $\bar{c}_2 , $ this is
 also the case, but now one has to take into account
that spinorial indices take
just two values. Indeed, due to the factors of $\kappa$ $(\kappa)$
, $ c_2 \,\,(\bar{c}_2) $
 anticommutes with all odd functions of
$ dz_\alpha \,\,(d\bar{z}_{\dot{\alpha}})  $.
However, when spinorial indices take just two values, all these
 potentially dangerous terms vanish
since $ (dz)^3\equiv 0, $ and therefore
$ c_2 $ and $ \bar{c}_2 $  commute with
everything. This property encodes the fact
that the equations under investigation make sense for
two-component spinors thus restricting our consideration to
the four-dimensional dynamics.

The equations (\ref{dW})-(\ref{SS}) are general
coordinate invariant (in the $x-$space sense)
and invariant under the higher spin gauge
transformations (\ref{deltaB})-(\ref{delta S}).

Now, we are in a position to analyze the equations
(\ref{dW})-(\ref{SS})   in the linearized
approximation. Let us assume that $ F(0)=0 $ (i.e. $ f_0=0 $ in
(\ref{F})). For this case a vacuum solution can be chosen in
the form
\be
B_0 =0\,,
\ee
\be
\label{S04}
S_0 =dz^\alpha z_\alpha +d\bar{z}^{\dot{\alpha}} \bar{z}_{\dot{\alpha}}
\ee
and
\be
\label{W04}
W_0 =\frac{1}{4i}[\omega_0{}^{\alpha \beta } (x)y_\alpha y_\beta
+\bar{\omega}_0{}^{\dot{\alpha}\dot{\beta}}(x) \bar{y}_{\dot{\alpha}}
\bar{y}_{\dot{\beta}}
+2\lambda h_0{}^{\alpha \dot{\beta}}(x)y_\alpha \bar{y}_{\dot{\beta}}]
\ee
with the $AdS_4$ gravitational vacuum fields discussed
in the section \ref{AdS Vacua}.
This ansatz solves the equations (\ref{dW})-(\ref{SS}).

In the lowest order, we obtain from  (\ref{SB})
 and (\ref{z,f}) that $ B $ is $ Z- $ independent
\be
B(Z;Y;{\cal K}|x)=C(Y;{\cal K}|x) + \mbox{higher order terms}.
\ee
Inserting this into (\ref{SS}) one arrives at the following
differential restrictions on the first-order part $ S_1 $ of $ S $ :
\be
\label{partS}
\partial S_1 \!=\frac{1}{2}[dz_\alpha dz^\alpha f_1
C(-z,\bar{y};{\cal K})k\exp{iz_\ga y^\ga }
\!+ d\bar{z}_{\dot{\alpha}} d\bar{z}^{\dot{\alpha}}
\bar{f}_1 C(y,-\bar{z};{\cal K}) \bar{k} \exp{i\bar{z}_\dga\bar{y}^\dga }]{},
\ee
where we have made use of
(\ref{uf}), (\ref{[kb]}) and (\ref{[bkb]}).
Using the simple fact that, for two-component spinors,
the general solution of the
equation
\be
 \frac{\partial}{\partial z^\alpha }f^\alpha(z)
=g(z)
\ee
is
\be  f_\alpha (z) =
  \frac{\partial}{\partial z^\alpha}\varepsilon(z)
+   \int_0^1dt\,t\,z_\alpha\,g(tz) \,,
\ee
where $ \varepsilon(z) $ is an arbitrary function, one deduces from
(\ref{partS}) that
\bee
\label{S1}
S_1 = \partial \varepsilon_1\!\!\!\! &+&\!\!\!\!
\int^1_0dt\,t[dz^\alpha z_\alpha f_1
C(-tz,\bar{y};{\cal K})k\exp{itz_\ga y^\ga }\nn\\ \!\!\!\! &+&\!\!\!\!
 d\bar{z}^{\dot{\alpha}}
\bar{z}_{\dot{\alpha}} \bar{f}_1 C(y,-t\bar{z};{\cal K}) \bar{k} \exp{it
\bar{z}_\dgb \bar{y}^\dgb }]{}\,.
\eee

The ambiguity in the function $ \varepsilon_1=\varepsilon_1 (Z;Y;{\cal K}) $
manifests invariance under the gauge transformations
(\ref{gxz}) in the $Z-$sector.
It is convenient to fix a gauge by requiring $ \partial
\varepsilon_1=0 $ in (\ref{S1}). This gauge fixing is not complete as
it does not fix the gauge transformations with
\be
\label{WE}
\varepsilon_1 (Z;Y)=\xi_1 (Y) +
\mbox{higher order terms} \,.
\ee

Thus, the field $ S $ is entirely
expressed in terms of the 0-form $ B $, disappearing as an independent
dynamical variable. In this sense, $ S $ can be thought of as a sort of
a pure gauge compensator field. This result is not surprising
since it means that the noncommutative gauge connection $S$ is
reconstructed in terms of the noncommutative curvature $F(B)$ uniquely
modulo gauge transformations.

Now, let us analyze the  equation (\ref{dS}).
In the first order, one gets
\be
\label{partw}
2i\partial W_1 =dS_1 -[W_0 ,S_1 ]_*\,\,.
\ee
Using the fact that generic solution of the equation
\be
 \frac{\partial}{\partial z^\alpha } \varphi(z)=\chi_\alpha (z)
\ee
has the form
\be
 \varphi (z)=  const +\int^1_0 dt\,z^\alpha \, \chi_\alpha (tz)
\ee
provided that
 $  \frac{\partial}{\partial z^\alpha } \chi^\alpha (z)
 \equiv 0 $ and $ \alpha =1,2,$ one finds from (\ref{partw})
\bee
\label{W1}
W_1(Z;Y;{\cal  K})=w (Y;{\cal K})
+\frac{i}{2} \int^1_0 dt\,\,
\ls&{}&\ls \Big ( z^\alpha
[W_0,s_{1\,\alpha}]_* (tz,\bar{z};Y;{\cal K})\nn\\
\ls &&\ls +\bar{z}^{\dot{\alpha}}
[W_0,\bar{s}_{1\,\dot{\alpha}} ]_*(z,t\bar{z};Y;{\cal K} )\Big)
\eee
(the terms with $ z^\alpha ds_{1\,\alpha} +\bar{z}^{\dot{\alpha}}
 d\bar{s}_{1\,\dot{\alpha}} $ vanish due to the formula
(\ref{S1}) with $ \partial \varepsilon_1 =0 $
because $ z^\alpha z_\alpha \equiv \bar{z}^{\dot{\alpha}}
 \bar{z}_{\dot{\alpha}} \equiv 0 $).
Let us stress that, as shown in the section
\ref{Doubling of Spinor Variables and Non-Commutative Gauge Theory},
(\ref{partw}) is consistent as
the system of differential equations with respect to $
\frac{\partial}{\partial z} $, $  \frac{\partial}{\partial
\bar{z}}$ and $\frac{\partial}{\partial x} $.
As a  result, it is enough to analyze
the equations (\ref{dW}) and (\ref{dB})
at $ Z^\nu =0 $. For other values of $ Z^\nu , $
 (\ref{dW}) and (\ref{dB}) will hold
automatically provided that the equations (\ref{dS})-(\ref{SS})
are solved.

Thus, to derive dynamical higher spin equations,
one has to insert (\ref{W1}) into (\ref{dW}) and (\ref{dB})
interpreting $ w (Y;{\cal K}|x) $ and $ C(Y;{\cal K}|x) $  as generating
functions for the higher spin fields. Note that the gauge
transformations (\ref{WE}) preserving the
linearized gauge condition $ \partial
\varepsilon_1 =0 $ identify with the gauge transformations
for $ w (Y;{\cal K}|x) $, the higher spin gauge transformations.

The straightforward analysis of
(\ref{dW}) at $ Z=0 $, based on (\ref{star2}), (\ref{W04}) and
(\ref{S1}) is elementary. The final result is
\bee
\label{dgo}
dw (Y;{\cal K}) &=&w *\wedge\, w(Y;{\cal K})\nn\\
 &+&\frac{i\lambda^2 }{8}\Big [f_1 h_\alpha {}^{\dot{\beta}}\wedge h^{\alpha
\dot{\delta}} k\frac{\partial}{\partial \bar{y}^{\dot{\beta}}}
\frac{\partial}{\partial \bar{y}^{\dot{\delta}}} (C(0,\bar{y};k,\bar{k})-
C(0,\bar{y};-k,-\bar{k}))\nn\\
&+&\bar{f}_1 h^\alpha
{}_{\dot{\beta}}\wedge h^{\gamma \dot{\beta}}\bar{k}\frac{\partial}{\partial
y^\alpha} \frac{\partial}{\partial y^\gamma} (C(y,0;k,\bar{k})-
C(y,0;-k,-\bar{k}))\nn\\
 &-&f_1 h_\alpha {}^{\dot{\beta}}\wedge
h^{\alpha \dot{\delta}} k\,\bar{y}_{\dot{\beta}} \bar{y}_{\dot{\delta}}
(C(0,\bar{y};k,\bar{k})+ C(0,\bar{y};-k,-\bar{k}))\nn\\
&-&\bar{f}_1 h^\alpha {}_{\dot{\beta}} \wedge h^{\gamma \dot{\beta}}\bar{k}\,
y_\alpha  y_\gamma (C(y,0;k,\bar{k})+ C(y,0;-k,-\bar{k}))\Big ]\nn\\
&+& \mbox{higher order terms}\,.
\eee
Note that all terms with the Lorentz connection
$\go_{0\,\alpha \beta} $ and $ \bar{\go}_{0\,\dot{\alpha} \dot{\beta}} $
in the $ C- $ dependent part of this formula cancel. This is a particular
manifestation of the general property discussed in the section
\ref{Free Differential Algebras and Unfolded Formulation} that
the higher spin equations are
covariant with respect to the local Lorentz transformations.

The equation (\ref{dB}) reduces to
\be
\label{dc}
dC=W_0*C-C*W_0 + \mbox{higher order terms}\,.
\ee
The equations (\ref{dgo}) and (\ref{dc}) thus
describe all dynamical information contained in the linearized
system (\ref{dW})-(\ref{SS}).

To make contact with the section \ref{$AdS_4$}
let us expand the fields as
\be
w(Y\!;\!{\cal K}|x\!)=\!\!\!\!\sum_{A,B = 0,1}
\!\!\!\!w^{AB} (Y|x)k^A \bar{k}^B ,\quad
C(Y\!;{\cal K}|x\!)=\!\!\!\!\sum_{A,B = 0,1}\!\!\!\!\lambda^{-2}
C^{AB} (Y|x)k^A \bar{k}^B.
\ee
Inserting these expressions back into
(\ref{dgo}) and (\ref{dc}) one finds that the equations for the fields
$ w^{AA}$ and $ C^{A\,\,1-A} $ reduce exactly to the Central
On-Mass-Shell Theorem (\ref{R_1}) and (\ref{DC}).

The fields
$ w^{A\,\,1-A} $ and $ C^{AA} $  are
auxiliary and do not describe nontrivial
degrees of freedom as was first shown in
\cite{aux}. Another way to see this is to observe that
the set of free equations for $C^{AA}$ decomposes into an infinite
set of finite subsets of equations for homogeneous polynomials in
spinor variables. Because $C=C^{AA}(x_0 )$ plays a role
of initial data in the pure gauge solution of the
covariant constancy conditions we conclude
that each of the subsystems contained in $C^{AA}$
describes at most a finite number of degrees of freedom.
Moreover, because the $AdS_4$ algebra
$o(2,3)$ is noncompact, it does not admit
finite-dimensional unitary representations and, therefore,
these degrees of freedom should not appear in a unitary
theory (ruled out by appropriate boundary conditions).
As a result, $C^{AA}$ describes some topological fields
which carry no degrees of freedom in the unitary case.
Let us note that the equations for $C^{A1-A}$ decompose into
infinite subsystems each identified with the equations
for some dynamical fields.

The auxiliary fields
can be truncated away from the full system by the discrete symmetry
\be
\tau (W(Z\!;Y\!;k,\bar{k}|x)\!)\! =\! W(Z\!;Y\!;-k,-\bar{k}|x\!),\quad
\tau (S(Z\!;Y\!;k,\bar{k}|x)\!)\! =\! S(Z\!;Y\!;-k,-\bar{k}|x\!),
\ee
\be
\tau (B(Z;Y;k,\bar{k}|x)) = -B(Z;Y;-k,-\bar{k}|x)
\ee
which takes place at least if $F(B)$ is odd. We therefore do
not discuss the auxiliary fields here in more detail
(the only comment is that the appropriate sector of the equations
(\ref{dgo}) and (\ref{dc}) just reproduces the analog of the
Central-On-Mass-Shell theorem for the sector of auxiliary fields).

Let us note that the operators $k$ and $\bar{k}$ that appear
explicitly in the equations via (\ref{Rab})
flip a type of the representation linking
gauge fields (1-forms) in the adjoint
representation of the higher spin algebra to the
0-forms $C$ in the twisted representation and vice versa.
We see that every massless spin appears twice
due to the Klein operator $ k\bar{k} $ giving rise to the labels $ A $
for the higher spin gauge fields $ w^{AA} $.
It is impossible to get rid of
$ k\bar{k} $ by any field redefinition. (In the purely bosonic case
however the operator $ k\bar{k} $ belongs to the center of
the algebra and therefore can be replaced by a constant.) Note that
in the recent paper \cite{bls} it was shown that some particle
models formulated in terms of twistor variables exhibit analogous
spectra of spins of massless particles. It is an interesting problem
to establish some direct connection between these superparticle
models and higher spin algebras.

Taking into account higher-order terms one  can in principle
reconstruct all higher-order corrections to the higher spin equations.
Thus the system (\ref{dW})-(\ref{SS}) indeed
solves the problem of reconstruction of the nonlinear
higher spin free differential algebra discussed in the section
\ref{Free Differential Algebras and Unfolded Formulation}.
Note that the nonlinear terms in the expansion (\ref{F})
contribute only to the nonlinear corrections in the resulting
free differential algebra and therefore describe some
ambiguity in the higher spin interactions.

\subsection{Nonlinear Equations in d=3}
\label{Nonlinear Equations in d=3}

The full nonlinear $d=2+1$ system is formulated \cite{PV} in terms of
the functions $W(z;y;\psi_{1,2},k,\rho | x)$,
$B(z;y;\psi_{1,2},k,\rho | x)$, and $S_\gal(z;y;\psi_{1,2},k,\rho | x)$
that depend on the space-time coordinates $x^\un$ $(\un=0,1,2)$, auxiliary
spinors $z_\gal$, $y_\gal$ $(\gal=1,2)$,
$[y_\gal,y_\beta]=[z_\gal,z_\beta]=[z_\gal,y_\beta]=0$, a pair of Clifford
elements $\{\psi_i,\psi_j\}=2\delta_{ij}$ $(i=1,2)$ that commute to
all other generating elements, and another pair of Clifford-type elements $k$
and $\rho$ which have the following properties
\be
\label{Klein} k^2=1\,,\: \rho^2 =1 \,,\: k\rho +\rho k =0\,,\:
   ky_\gal=-y_\gal k\,,\: kz_\gal=-z_\gal k\,, \:  \rho y_\gal=y_\gal \rho
   \,,\:
   \rho z_\gal=z_\gal \rho\,.
\ee

The field variables obey the (anti-)hermiticity
(reality) conditions
\be
\label{reality3}
W^\dagger =-W,\,\,\,\, B^\dagger =B,\,\,\,\, S^\dagger_\ga =-S_\ga
\ee
with the involution $ \dagger $ defined by the relations
\be
(z_\alpha )^\dagger =-{z}_{{\alpha}},\quad
(y_\alpha )^\dagger ={y}_{{\alpha}},\quad
(\psi_i )^\dagger ={\psi_i},\quad
k^\dagger =k ,\quad
\rho^\dagger =\rho ,\quad
(dx^\un )^\dagger =dx^\un \,.
\ee

To define the $d=3$ equations of motion we have to fix a form
of the curvature $R_{\nu\mu}$ in (\ref{SS}). In d=3, spinor indices
take two values and therefore $R_{\ga\gb}$ is proportional to
the d=3 charge conjugation matrix $\epsilon_{\ga\gb}$. The
appropriate choice is
\be
\label{R3}
R_{\ga\gb} =- \frac{i}{4} \epsilon_{\ga\gb} B * \kappa\,,
\ee
where
\be
\label{kappa}
\kappa = k\ups \equiv k\exp iz_\ga y^\ga\,.
\ee

With the aid of the involutive automorphism $\rho\to -\rho$,
$S_\gal\rightarrow -S_\gal$
         the system (\ref{dW})-(\ref{SS}) can be truncated
to the one with the
fields $W$ and $B$ independent of $\rho$ and
$S_\gal$ linear in $\rho$,
\be
\label{tru}
    W(\!z,y;\psi_{1,2},k,\!\rho | x\!)\!=\!W(\!z,y;\psi_{1,2},k | x\!),\quad
    B(\!z,y;\psi_{1,2},k,\!\rho | x\!)\!=\!
    B(\!z,y;\psi_{1,2},k | x\!),
\ee
\be
\label{truS}
    S_\gal(z,y;\psi_{1,2},k,\rho | x)=
      \rho s_\gal(z,y;\psi_{1,2},k | x)\,.
\ee
It is  this reduced system that describes higher spin
interactions of matter fields in d=3 \cite{PV}.
For this system one finds taking into account
(\ref{[UF]}), (\ref{Klein}), and (\ref{kappa}) that
\be
\label{propK}
    \kappa*W=W*\kappa\,,\qquad \kappa*B=B*\kappa\,,\qquad
\ee
and
\be
\label{propK1}
    \kappa*S_{\ga}=-S_{\ga}*\kappa \,.
\ee
The additional minus sign in
(\ref{propK1}) is due to the factor of $\rho$ in (\ref{truS}).
We observe that for d=3
the relations (\ref{SS}) have a form of the deformed
oscillator relations (\ref{modosc}) with $B$ playing a role of
the central element $\nu$.

Let us note that in this section we assume that the differentials
$dz^\ga$ commute with all variables except for
themselves and the space-time
differentials $dx^\un$ to which they anticommute. An alternative
formulation applied to the d=4 analysis in the section
\ref{Nonlinear Equations in d=4} is to require
$dz^\ga$ to anticommute with the Klein operator $k$.
Clearly, this is equivalent to the substitution $dz^\ga \to \rho dz^\ga$.

Note that the parameter $\gvep=\gvep(z,y;\psi_{1,2},k | x)$
of the higher spin gauge transformations (\ref{deltaB})-(\ref{delta S})
is independent of
$\rho$ and therefore commutes with $\kappa$. Thus, the element $\kappa$,
that appears explicitly in the constraints (\ref{SS}),
belongs to the center of the gauge algebra.

To elucidate the dynamical content of the system
(\ref{dW})-(\ref{SS}), one first of all has to
find an appropriate vacuum solution.
We consider  vacuum solutions with
\be
\label{B_0}
   B_0=\nu\,,
\ee
where $\nu$ is some constant independent of the space-time coordinates
and auxiliary variables. As a result,
the equations (\ref{dB}) and
(\ref{SB}) hold trivially and
the vacuum fields $W_0$ and $S_{0\ga}$ have to satisfy
\be
\label{dW_0}
   dW_0=W_0 *\wedge W_0\,,
\ee
\be
\label{WS_0}
   dS_{0\ga} =W_0*S_{0\ga}-S_{0\ga}*W_0\,,
\ee
\be
\label{SS_0}
   [S_{0\ga},\,S_{0\gb}]_*=-2i\gep_{\ga\gb}(1+\nu \kappa )  \,.
\ee

For $\nu=0$, the standard choice is $S_{0\ga}=\rho z_\ga$.
For general $\nu$, a class of solutions of the equations (\ref{SS_0})
is found in \cite{PV}. Here we reproduce the following three
most important solutions
\be
\label{S_0}
   S_{0\ga}^{\pm}=\rho \left(z_\ga+\nu (z_\ga\pm y_\ga)
    \int_0^1dtt e^{it(zy)} k \right) \,,
\ee
and
\bee
\label{S sym}
   S^{sym}_{0\ga} (z,y)\!\!\!\!\! & = &\!\!\!\!\! \rho z_\ga -
\rho  {\nu\over 8}
     \int^1_{-1} ds (1-s)
  \!\!  \left[ e^{ \frac{i}{2} (s+1) (zy) }(y_\ga + z_\ga)\!  * \!
     \Phi \left({1\over 2}, 2;  -\kappa \, \mbox{ln} \,|s|^{\nu} \right)
        \right.  \nonumber \\
   \!\!\!\!\! & &\!\!\!\!\! \left. {} + e^{ \frac{i}{2} (s+1) (zy) }
(y_\ga - z_\ga) *
     \Phi \left({1\over 2}, 2;  \kappa \, \mbox{ln} |s|^{\nu}\right)
        \right]  * \kappa \, ,
\eee
where $\Phi(a,c;x)$ is the degenerate hypergeometric function
\be
\Phi (a,c;x ) = 1+\frac{ax}{c1!}+\frac{a(a+1)x^2}{c(c+1)2!}+ \ldots \,.
\ee
The ambiguity in the solutions of the equation (\ref{SS_0})
takes its origin in the gauge transformation (\ref{delta S}).
All three
solutions $S^\pm_{0\ga}$ and $S^{sym}_{0\ga}$ belong to the same gauge
equivalence class.  It is not hard to see that $S^\pm_{0\ga}$ solve
(\ref{SS_0}).  To prove that $S^{sym}_{0\ga}$ solves (\ref{SS_0})
is more tricky \cite{PV}.

$S^\pm_{0\ga}$ and $S^{sym}_{0\ga}$ have the properties
\be
   S^\pm_{0\ga} (z,y;k,\rho) = - \bar{S}^\mp_{0\ga} (-z,y;k,\rho) \,,\qquad
   S^\pm_{0\ga} (z,y;k,\rho) = i S^\mp_{0\ga} (-iz,iy;k,\rho) \,,
\ee
and
\be
\label{pr Ss}
   S^{sym}_{0\ga} (z,y;k,\rho) = - \bar{S}^{sym}_{0\ga} (-z,y;k,\rho) ,\quad
   S^{sym}_{0\ga} (z,y;k,\rho) = i S^{sym}_{0\ga} (-iz,iy;k,\rho) \,.
\ee
In fact, the solution $S^{sym}_{0\ga}$ is fixed by the properties
(\ref{pr Ss}). It is particularly useful for the analysis of
truncations of the system being invariant under
the discrete symmetries and reality conditions.
When  the analysis is independent of the
particular form of a vacuum solution the symbol $S_{0\ga}$
will be used for any one of them.

Now, let us turn to the equation (\ref{WS_0}). Since $dS_{0\ga}=0$, we get
\be
\label{Wt}
   [W_0, S_{0\ga}]_*=0 \,.
\ee
Thus, $W_0$ belongs to the subalgebra $A_S\subset A$ spanned by
elements
which commute to $S_{0\ga}$, i.e. $A_S$ is the centralizer of $S_{0\ga}$.
For the case of $\nu = 0$,
 $A_S$ is the subalgebra of functions independent of $z$.
To find  $A_S$ for general $\nu$ we construct
generating elements $\ty_\ga$ commuting with $S^\pm_{0\ga}$
(\ref{S_0}) and $S^{sym}_{0\ga}$ (\ref{S sym}). The final result is
\cite{PV}
\be
\label{tilde y}
   \ty^\pm_\ga (z,y) = y_\ga+\nu (z_\ga \pm y_\ga)\int_0^1dt(t-1)
   e^{it(zy)}k \,,
\ee
\bee
\ls\ls\label{y sym}
  \ty^{sym}_\ga (z,y) \ls &=\ls &  y_\ga + k \; {\nu\over 8} \;
     \int^1_{-1} ds (1-s)
     \exp\left\{\frac{i}{2}(s+1)(zy)\right\} \nn\\
\ls\ls\ls&{\times}&\ls \left[ (y_\ga + z_\ga)
     \Phi \left({1\over 2}, 2;  - k  \mbox{ln} |s|^{\nu} \right)
        \right.
      \left. {} \!- (y_\ga - z_\ga)
     \Phi \left({1\over 2},  2;   k  \mbox{ln} |s|^{\nu}\right)
        \right] .
\eee
Remarkably, $\ty_\ga$ again obey the commutation relations of
the form (\ref{modosc}) but now with the Klein operator $k$
instead of $\kappa$
\be
\label{y com}
   [\ty_\ga,\ty_\gb]_*=
      2i\gep_{\ga\gb}(1+\nu k)\,,\quad
      \ty_\ga k=-k \ty_\ga \,.
\ee
It is elementary to check that $[\ty^\pm_\ga\,,S^\pm_{0\gb}]_* =0$. The
fact that $[\ty^{sym}_\ga\,,S^{sym}_{0\gb}]_* =0$ is less trivial
\cite{PV}.
Note that, as a by-product,  the deformed oscillator
algebra is realized in terms of the embedding into the tensor
product of two Weyl algebras equipped with
the generating element $k$.

Since $k$, $\psi_1$ and $\psi_2$ commute with $S_{0\ga}$,
the subalgebra $A_S$ is spanned by
the power series of $\ty_\ga$, $\psi_1$, $\psi_2$ and $k$,
i.e. its generic element has the form
\be
\label{A}
    f (z,y;\psi_{1,2},k)= \sum_{B,C,D=0}^1\sum_{n=0}^\infty \frac1{n!}\;
    f^{BCD}_{\ga_1\ldots\ga_n}\; k^B \psi_1^C\psi_2^D \;
    \ty^{\ga_1}*\ldots*\ty^{\ga_n}\,,
\ee
where $f^{BCD}_{\ga_1\ldots\ga_n}$ are totally symmetric multispinors
(i.e. we choose the Weyl ordering).
According to (\ref{Wt}), $W_0$ has a form (\ref{A}).

Because the commutation relations (\ref{y com}) have a form of the
deformed oscillator algebra (\ref{modosc}), we recover the
d3 higher spin algebra discussed in the section
\ref{Higher-Spin Symmetries} for an arbitrary value of $\nu$
as the stability algebra of the chosen
vacuum solution characterized by the parameter $\nu$.
One can now choose $AdS_3$ solution
of the vacuum equations (\ref{dW}) in the form (\ref{w})
identifying
 $\psi$ e.g. with $\psi_1$,
\be
\label{wpsi1}
w_0 =
\frac{1}{8i}
\left ( \go^{\ga\gb} (x)\{\ty_\ga (\nu ),\ty_\gb (\nu )\}
+\lambda h^{\ga\gb} (x)\psi_1 \{\ty_\ga (\nu ),\ty_\gb (\nu )\} \right )\,.
\ee
This completes the construction of the background solution.
Let us emphasize that the form of the constraint (\ref{SS}) leads
in a rather non-trivial way
to the AdS background geometry via realization of the vacuum
centralizer $A_S$ in terms of the deformed oscillators $\ty_\ga$.

Once a vacuum solution is known, one can study the system
(\ref{dW})-(\ref{SS})
perturbatively expanding the fields as
\be
\label{pert}
   B=B_0+B_1+\ldots \,,\qquad
   S_\gal=S_{0\gal}+S_{1\gal}+\ldots \,,\qquad
   W=W_0+W_1+\ldots \,.
\ee
Substitution of these expansions
into (\ref{dW})-(\ref{SS}) gives in the lowest order
\be
\label{WW_1}
   D_0\,W_1=0\,,
\ee
\be
\label{WB_1}
   D_0\,C=0\,,
\ee
\be
\label{WS_1}
   D_0\, S_{1\gal}=[W_1\,,\,S_{0\gal}]_* \,,
\ee
\be
\label{SS_1}
   [S_{0\gal}\,,\,S_{1\beta}]_*-[S_{0\beta}\,,\,S_{1\gal}]_*=
      -2i\epsilon_{\gal\beta}\,C*\kappa \,,
\ee
\be
\label{SB_1}
   [S_{0\gal}\,,\,C]_*=0 \,,
\ee
where we denote $C=B_1$ and $D_0$ is the background covariant
derivative.

The system (\ref{WW_1})-(\ref{SB_1}) is analyzed
as follows. {}From (\ref{SB_1}), one concludes that
$C$ has a form analogous to (\ref{A}),
i.e. $C=C (\hat{y};\psi_{1,2},k | x)$.
Expanding $C$ as
\be
\label{dynaux}
   C=C^{aux}(\hat{y};\psi_1,k | x)
    +C^{dyn}(\hat{y};\psi_1 ,k | x)\psi_2 \,,
\ee
with
\be
    C^{dyn}(\hat{y};\psi_1 ,k | x)= \sum_{A,B = 0,1}
      \sum^\infty_{n=0}
    \frac{1}{n!}  C^{dyn}{}^{A\,B\gal_1\ldots\gal_n}(x)(k)^A(\psi_1 )^B
    \ty_{\gal_1}\ldots \ty_{\gal_n}\,,
\ee
and
\be
    C^{aux}(\hat{y};\psi_1 ,k | x)=\sum_{A,B = 0,1}
      \sum^\infty_{n=0}
    \frac{1}{n!}  C^{aux}{}^{A\,B\gal_1\ldots\gal_n}(x)(k)^A(\psi_1 )^B
    \ty_{\gal_1}\ldots \ty_{\gal_n}\,,
\ee
one observes that the covariant derivative $D_0$ acts differently
on $C^{aux}$ and $C^{dyn}$ because the factor $\psi_1$ that
appears in the vacuum solution (\ref{wpsi1}) anticommutes to
$\psi_2$. As a result, the equation
(\ref{WB_1}) in the sector of $C^{dyn}$ turns out to be
equivalent to the equation (\ref{cald0}).
{}From the analysis of the section \ref{$AdS_3$} we
conclude that $C^{dyn}$  describes four spin 0 and four spin 1/2
matter fields with the masses given in (\ref{M}) and (\ref{M f}).
The overall doubling for each mass is due to $\psi_1$ while
the doubling with the mass splitting in the  boson sector is due
to $k$ (cf  (\ref{pm})). Altogether, matter fields
 form a d3 massive hypermultiplet with respect to the $N=2$
supersymmetry algebra $osp(2,2)\oplus osp(2,2)$
discussed in the section \ref{Higher-Spin Symmetries}.

In the sector of $C^{aux}$, the covariant derivative acts
as in the adjoint representation of the deformed
oscillator algebra,
\be
D_0 C^{aux} = d C^{aux} - w_0 * C^{aux}  + C^{aux}* w_0 \,.
\ee
An important difference between the adjoint
representation and twisted representation is that the
set of equations for $C^{aux}$ decomposes into an infinite
set of finite subsets of equations for homogeneous polynomials,
i.e. for $ C^{aux}{}^{A\,B\gal_1\ldots\gal_n}(x)$ with any fixed $n$.
This is a simple consequence of (\ref{q2com}).
(The equations for $C^{dyn}$ contain a finite set
of infinite subsystems, each describing a matter field.)
Because $C=C^{aux}(\hat{y};\psi_1,k | x_0 )$ plays a role
of initial data in the pure gauge solution of the
covariant constancy conditions we conclude \cite{Eq,BPV}
that each of the subsystems contained in $C^{aux}$
describes at most a finite number of degrees of freedom.
Moreover, because the $AdS_3$ algebra
$o(2,2)$ is noncompact, these degrees of freedom
cannot appear in a unitary theory, ruled out by appropriate
boundary conditions. As a result, $C^{aux}$ describes
some topological fields which carry no degrees of freedom
in the unitary case.

The next step consists of resolution of the constraints (\ref{SS_1})
to reconstruct the auxiliary field $S_{1\gal}$ as a linear
functional of $C$, $S_{1\gal}=S_{1\gal}(C)$, up to a gauge ambiguity.
Then, (\ref{WS_1}) allows one to express a part of degrees of freedom
in $W_1$ via $C$, while the rest modes, which belong to the kernel
of the mapping $[S_{0\gal}\,, \dots]_*$, remain free.
These free modes are again arbitrary functions of
$\hat{y}_\gal$, i.e.
 $  W_1 = \omega (\hat{y};\psi_{1,2},k | x) + \Delta W_1 (C) $,
where $\omega (\hat{y};\psi_{1,2},k | x)$ corresponds to the
higher spin gauge fields. The dynamical equations
on them are imposed by eq.~(\ref{WW_1}) after
(\ref{WS_1}) is solved.
Eq.~(\ref{WW_1}) describes the $C$-dependent first-order corrections
to the higher spin strengths for $\omega$, which are argued
in the section \ref{Integrating Flow} to vanish \cite{PV}.
As a result one arrives at the d3 Central On-Mass-Shell Theorem
(\ref{d3l}).

One proceeds analogously in the highest orders. The equations
(\ref{WW_1})-(\ref{SB_1}) are consistent and admit a perturbative
solution in powers of $C$ and $\nu$.

\subsection{Interaction Ambiguity}
\label{Interaction Ambiguity}

Historically, we arrived at the equations for d=4
higher spin fields via perturbative analysis of the
higher spin free differential
algebra. In the end, the system of equations self-organized
into the compact form (\ref{dW})-(\ref{SS}) and (\ref{Rab}).
In this section we focus on some general
features underlying this construction which hopefully will
be useful for the analysis of higher spin systems in different
dimensions and have already been applied for
the d=3 \cite{Eq,PV} and d=2 \cite{d2} systems.

In principle, there are two ways for possible
generalizations of the equations (\ref{dW})-(\ref{SS}):
a generalization of the equations within the same set
of fields $W,$ $B$ and $S^\nu$ or  modifications by using
larger sets of fields. The latter problem, being quite
interesting in the context of the off-mass-shell
(Lagrangian) formulation of the higher spin theories, is not
yet enough clarified and is not discussed here.

Within the set of the variables $W,$ $B$ and $S^\nu$ the most
interesting question is why $R_{\nu\mu}$ has the particular
form (\ref{Rab}).
One can generalize (\ref{SS}) to
\be
\label{SSgen}
S*S \!=\!-i[dz_\alpha \,dz^\alpha (G(B)+F(B)*\kappa )
\!+\!d\bar{z}_{\dot{\alpha}}d\bar{z}^{\dot{\alpha}} (\bar{G} (B)
\!+\!\bar{F} (B)*\bar{\kappa})+dz^\ga d\bar{z}^\dgb H_{\ga\dgb} (B)]
\ee
with arbitrary complex functions $ G(B) $ and $ F(B) $
and Hermitian
$H_{\ga\dgb}(B)$
\be
\bar{H}_{\ga\dgb} = H_{\dgb\ga}\,
\ee
being some star-product expansions analogous to (\ref{F}).
This form of $R_{\nu\mu}$ is consistent with $x-$ and $Z-$ Bianchi
identities provided that the rest equations
(\ref{dW})-(\ref{SB}) are true. The term with $H_{\ga\dgb}$
was not considered in \cite{more} since it
looks weird breaking down Lorentz invariance explicitly (because
it is impossible to construct a vector from a scalar field $B$
without introducing some exterior vector). Note that since $\kappa$
and $\bar{\kappa}$ anticommute with $dz^\ga$ and $d\bar{z}^\dga$,
they can only appear in the $dzdz$ and $d\bar{z}d\bar{z}$
sectors, respectively (otherwise the system becomes inconsistent).

The ambiguity in $ G(B) $ is artificial within the perturbative analysis
in powers of $B$ provided that $ G(0)\neq 0 $.
Indeed, one can get rid of the dependence on $ G $ by means of
the following field redefinition
which does not affect the equations (\ref{dW})-(\ref{SB})
\be
\label{SSB}
S\rightarrow S^\prime =[G(B)]^{-\frac{1}{2}}dz^\alpha s_\alpha
+[\bar{G}(B)]^{-\frac{1}{2}}d\bar{z}^{\dot{\alpha}} \bar{s}_{\dot{\alpha}}\,.
\ee
We therefore set $G=1$.

One can also use the ambiguity in the invertible field redefinitions
\be
\label{BfB}
B\rightarrow B^\prime =f(B)\,.
\ee
In particular, if $ F(B) $ is some real function, $ F(B)=\bar{F}(B), $
one can choose $ f(B) $ to coincide with the inverse of $ F(B) $,
thus reducing the problem to the case with $ F(B)=\bar{F}(B) =B. $
However, generally, $ F(B) $ is some complex function of $ B $ ,
while $ f(B) $ is real, $ \bar{f}(B)=f(B), $ because
$ B $ itself is real in accordance with the reality conditions
(\ref{reality}). Assuming that
$ F(B) $  starts with the linear term $ f_1 B $
one can fix this freedom by requiring that
 $ F(B)*\bar{F}(B)=B*B $
 and, therefore
\be
\label{FB}
F(B)=B\exp[i\varphi (B)]\,\,,\,\,\,\, \bar{F}(B)=B\exp[-i\varphi (B)]\,,
\ee
where $ \varphi (B) $ is an arbitrary real function of $ B. $
Note that one cannot use field redefinitions mixing
$s_\ga$ and $\bar{s}_\dga$ because of the properties of
the Klein operators $\kappa$ and $\bar{\kappa}$. Therefore, we are left
with the formula (\ref{SSgen}) with $G=1$,
$F(B)$ of the form (\ref{FB}) and arbitrary $H_{\ga\dgb}(B)$.

Let us now discuss the question of Lorentz invariance of
the higher spin equations. The gauge transformations
(\ref{delta S}) act on the spinor variables $Z^\nu$ but not on
the exterior index $\nu$ of $S^\nu$. Moreover,
there is no local symmetry at all rotating this
index\footnote{In that respect, there is some difference between
non-commutative gauge fields $S_\nu$ and
base-space differential forms $dx^\un W_\un$ possessing
diffeomorphisms rotating the form index $\un$.}.
The generators of the Lorentz symmetry acting on the spinor
variables are
\be
   L^{tot}_{\ga\gb}={i\over 4}(\{z_{\ga}, z_{\gb}\}_*-
     \{y_{\ga}, y_{\gb}\}_*) \,,\qquad
   \bar{L}^{tot}_{\dga\dgb}={i\over 4}(\{\bar{z}_{\dga}, \bar{z}_{\dgb}\}_*-
     \{\bar{y}_{\dga}, \bar{y}_{\dgb}\}_*) \,.
\ee
Actually, by (\ref{y,f}) and (\ref{z,f}),
the infinitesimal local Lorentz transformations
\be
\label{loc L}
   \gd f=
[\eta^{\ga\gb}L^{tot}_{\ga\gb}\,,\; f]_*
+[\bar{\eta}^{\dga\dgb} \bar{L}^{tot}_{\dga\dgb}\,,\; f]_*
\ee
with the parameter $\eta^{\ga\gb}(x)$
 rotate properly the spinor generating elements,
\be
   \gd z_{\ga} = 2\eta_\ga{}^\gb z_{\gb} \,,\quad
   \gd y_{\ga} = 2\eta_\ga{}^\gb y_{\gb} \,,\quad
   \gd \bar{z}_{\dga} = 2\bar{\eta}_\dga{}^\dgb \bar{z}_{\dgb} \,,\quad
   \gd \bar{y}_{\dga} = 2\bar{\eta}_\dga{}^\dgb \bar{y}_{\dgb} \,.
\ee
Although the system (\ref{dW})-(\ref{SS}) is invariant
under the local Lorentz transformations (\ref{loc L})
(not acting on the outer spinor index in $S_\nu$),
this symmetry is spontaneously broken due to the constraint
(\ref{SS}) because the r.h.s. of (\ref{SS}) has a
non-vanishing vacuum value and therefore $S_{\nu}$ itself must have a
non-vanishing vacuum expectation value as is manifested by
the vacuum solution (\ref{S04}).

As explained in the section
\ref{Doubling of Spinor Variables and Non-Commutative Gauge Theory}
the global symmetry identifies with the centralizer of the vacuum
value $S_0$ (\ref{S04}) isomorphic to the star-product algebra of
$Z-$independent functions. It contains the Lorentz subalgebra spanned
by the generators bilinear in
$y^\ga$ and $\bar{y}^\dga$
\be
\label{l4glob}
l^{gl}_{\ga\gb}=-{i\over 4}\{y_{\ga}, y_{\gb}\}_* \,,\qquad
\bar{l}^{gl}_{\dga\dgb}=-{i\over 4}\{\bar{y}_{\dga}, \bar{y}_{\dgb}\}_* \,.
\ee
By its definition, the full global symmetry acts properly in all
orders in interactions. In particular this is true for the
AdS subalgebra spanned by the bilinears in
$y^\ga$ and $\bar{y}^\dga$ and its Lorentz part (\ref{l4glob}).
So, we conclude that the higher spin field equations with
arbitrary $G$, $F$ and $H_{\ga\dgb}$ possess global Lorentz
symmetry. This happens because this global symmetry leaves
invariant $S_0$ and does not act on the indices carried
by $H_{\ga\dgb}$.

The situation with local symmetries is different because, as is
known from the example of
 supergravity \cite{PVN}, their form is deformed by
curvature-dependent terms compared to the global symmetry
algebra. In particular,  AdS translations acquire some
curvature-dependent corrections
which transform them into diffeomorphisms. The local Lorentz
symmetry remains undeformed however, with the Lorentz
connection entering only via the usual Lorentz covariant derivatives.
The local Lorentz symmetry guarantees the equivalence between the
Cartan (frame) and the Riemanian formulations of gravity, providing
the meaningful interpretation of spinors and tensors. It is therefore
reasonable to require the higher spin equations to have standard
(undeformed)
local Lorentz symmetry. Surprisingly, this simple requirement is
highly restrictive and to large extend fixes a form of the higher spin
equations.

The question is whether there exists a local Lorentz
symmetry which rotates properly spinor indices of the dynamical
fields identified with the ``initial data" (\ref{inda})
without a contradiction with the constraints
(\ref{SB}) and (\ref{SS}) in all orders of perturbations. The answer
is that this is indeed true provided that $H_{\ga\dgb} = 0$.

Indeed, for the curvature $R_{\nu\mu}$ (\ref{Rab}) with $H_{\ga\dgb} =0$,
the constraints (\ref{SS}), (\ref{SB})
have a form of the deformed oscillator algebra (\ref{modosc})
in the dotted and undotted sectors. As a result,
the elements
\be
   M_{\ga\gb}={i\over 4} \{s_\ga, s_\gb\}_*
\ee
obey the Lorentz commutation relations and
rotate properly $s_{\ga}$,
\be
\label{M_}
   [M_{\ga\gb}\,,\;s_{\gamma}]_*=\gep_{\ga\gga} s_{\gb}+
       \gep_{\gb\gga} s_{\ga} \,,\qquad
   [M_{\ga\gb}\,,\;\bar{s}_{\dga}]_*= 0\,.
\ee
Analogously,
\be
   \bar{M}_{\dga\dgb}={i\over 4} \{\bar{s}_\dga, \bar{s}_\dgb\}_*
\ee
obey the Lorentz commutation relations and
rotate properly $\bar{s}_{\dga}$,
\be
   [\bar{M}_{\dga\dgb}\,,\;\bar{s}_{\dot{\gamma}}]_*
=\gep_{\dga\dot{\gga}}
       \bar{s}_{\dgb}+ \gep_{\dgb\dot{\gamma}} \bar{s}_{\dga} \,,\qquad
   [\bar{M}_{\dga\dgb}\,,\;{s}_{\ga}]_*= 0\,.
\ee

Now we require that the gauge fixings used to
reconstruct $S_\nu$ in terms
of $B$ (i.e. of the type we have used
to gauge away $\gvep_1$ in (\ref{S1})) do not contain any exterior
objects transforming as nontrivial representations of the Lorentz
group. In other words $S_\nu (Z;Y)$ is required to be reconstructed
in terms of $C(Y)=B(0;Y)$ only by using $Z_\nu$ and $Y_\nu$.
{}From the analysis of the section (\ref{Nonlinear Equations in d=4})
it is clear that this can be achieved in all orders in interactions.

Let us define
\be
\label{l}
   l_{\ga\gb}=L^{tot}_{\ga\gb} - M_{\ga\gb} \,,\qquad
   \bar{l}_{\dga\dgb}=\bar{L}^{tot}_{\dga\dgb} - \bar{M}_{\dga\dgb} \,
\ee
with $l_{\ga\gb}$ and $\bar{l}_{\dga\dgb}$
to be identified with the generators of the local Lorentz transformations
acting on the physical fields.
Taking into account (\ref{SB}), we obtain
\be
\label{del B}
   \gd B=[\eta^{\ga\gb} l_{\ga\gb}\,,\; B]_*=
           \eta^{\ga\gb} [L^{tot}_{\ga\gb}\,,\; B]_* \,,
\ee
i.e. $l_{\ga\gb}$ rotates properly the field $B$.

For the gauge fields $W$ we have
\be
\label{del W}
   \gd W = D(\eta^{\ga\gb} l_{\ga\gb})=
       (d\eta^{\ga\gb}) l_{\ga\gb} +
       \eta^{\ga\gb} [L^{tot}_{\ga\gb}\,,\; W]_* \,.
\ee
Here $D(f)=df-[W,f]_*$ and therefore
$D(\eta^{\ga\gb})=d(\eta^{\ga\gb})$, since $\eta^{\ga\gb}(x)$
is proportional to the unit element of the star-product algebra.
Also, $D(l_{\ga\gb}) =
[L^{tot}_{\ga\gb}, W]_*$ because $d L^{tot}_{\ga\gb} = 0$ and
$D M_{\ga\gb} = 0$
(cf. eq.(\ref{dS})). {}From (\ref{del W}) one concludes that the gauge
field for a true local Lorentz symmetry is
\be
\label{Wl} W_L = \go_L^{\ga\gb}
l_{\ga\gb} \,,
\ee
while the other gauge fields are rotated properly under the
local Lorentz transformations. (The analysis in the sector of
dotted spinors is analogous.)

By assumption, the auxiliary field $S_{\nu}$ is expressed
via $B$ by the constraint (\ref{SS}) in a Lorentz covariant way.
We therefore have
\be
\label{}
   [\eta^{\ga\gb} L^{tot}_{\ga\gb}\,,\; s_\gga(B)]_* =
      \eta^{\ga\gb}(\gep_{\ga\gga} s_{\gb} (B)
      + \gep_{\gb\gga} s_{\ga} (B) )
      + \frac{\gd s_\gga (B)}{\gd B} \gd B \,,
\ee
\be
[\eta^{\ga\gb} L^{tot}_{\ga\gb}\,,\; \bar{s}_{\dot{\gga}}(B) ]_* =
       \frac{\gd \bar{s}_{\dot{\gga}}(B) }{\gd B} \gd B \,,
\ee
where $ \gd B = \eta^{\ga\gb} [L^{tot}_{\ga\gb}, B]_* $.
Making use of (\ref{M_}), we find
\be
\label{del S}
   \gd S_\nu = [\eta^{\ga\gb} l_{\ga\gb}\,,\; S_\nu]_*=
       \frac{\gd S_\nu}{\gd B}\;\gd B  \,.
\ee
As a result, the local Lorentz rotations generated by $l_{\ga\gb}$
and $\bar{l}_{\dga\dgb}$
do not act on the index $\nu$ of $S_\nu$, acting only on the physical
fields $B$. This is just the desired result that the transformation
law respects the solution of the constraints for $S$ in terms of $B$.

Thus, the fact that the
 constraints (\ref{SS}), (\ref{SB}) have a form of the deformed
oscillator algebra (\ref{modosc}) guarantees
that the local Lorentz symmetry remains unbroken. This property
restricts a form of the curvature $R_{\nu\mu}$ ruling out the term with
$H_{\ga\dgb}$. We conclude that, modulo perturbative field redefinitions,
the most general form of the curvature $R_{\nu\mu}$ compatible with
local Lorentz invariance is (\ref{Rab}) with $F(B)$ of the form (\ref{FB}).

Beyond the perturbative analysis,
the most general form of the equation (\ref{SS})
compatible with local Lorentz invariance is
\be
\label{nonper}
S*S =-i[dz_\alpha \,dz^\alpha (G(B)+F(B)*\kappa )
+d\bar{z}_{\dot{\alpha}}d\bar{z}^{\dot{\alpha}} (\bar{G} (B)
+\bar{F} (B)*\bar{\kappa})]\,.
\ee
In principle, the theory may admit phases with different
behavior of the functions $G$ and $F$. One interesting possibility
is $G(0)=0$. The straightforward interpretation in terms of
AdS higher spin fields is not obvious for that case but it may
correspond to the $w_\infty$ limit of
the underlying algebras \cite{Aq} having therefore some relevance
to conformal models. Similar statement concerns the strong coupling
limit $F(B)\to \infty$ equivalent to $G(B)\to 0$ by a rescaling
of $S_\nu$.

The meaning of the ambiguity in one real function
$\varphi (B)$ which affects higher spin interactions is not yet
clear.  For general $\varphi$, some discrete symmetries of the
system turn out to be lost. For more details on this issue
we refer the reader to the original paper \cite{more}.
Note that even the simplest choice $\varphi (B)=0$ is interesting
enough leading to consistent nontrivial higher spin dynamics.

The analysis of the d=3 case is parallel to d=4.
The constraint (\ref{SS}) associated with $R_{\ga\gb}$ (\ref{R3})
again has a form of the deformed oscillator algebra (\ref{modosc})
thus guaranteeing local Lorentz invariance. There is no ambiguity
in the function $\varphi (B)$ because the fields are real and,
perturbatively, all nonlinearities in $R_{\ga\gb}$ can be
compensated by field redefinitions analogous to (\ref{SSB}) and
(\ref{BfB}). Beyond the perturbative analysis,
the most general form of the equation (\ref{SS}) is
\be
S*S =-i dz_\alpha \,dz^\alpha (G(B)+F(B)*\kappa )\,.
\ee

The d=3 equations are very close to the $d=4$ self-dual higher spin
equations introduced in \cite{more}. Indeed, it is the Minkowski
signature $(+---)$ that forces
the left and right sectors to be conjugated to each other. For the
signatures $(----)$ and $(++--)$ these sectors
are independent. For that reason, for the cases admitting selfduality,
the functions $F(B)$ and $\bar{F}(B)$ are no longer
conjugated to each other but become independent real functions.
One can therefore set
\be
\label{SD}
\bar{F}(B) =0\,,\qquad F(B) = B
\ee
or vice versa. (For $\bar{F}(B) =0$ one can achieve $F(B) = B$ by the
field redefinition (\ref{BfB}).)

Note that in the standard selfduality
equations a half of (higher spin) Weyl tensors (field strengths in the
spin 1 case) vanishes. In the self-dual higher spin equations
of the form (\ref{SD}) the situation is different because all higher spin
Weyl tensors are still contained in the generating function
$B(0;Y;k,\bar{k}|x)$. However a half of the Weyl tensors
decouples from the higher spin curvatures
(i.e. only another half of them survive in (\ref{R_1})).
It is interesting to clarify a relationship of this
form of the self-dual higher spin equations with that proposed
in \cite{chog}.

\section{Integrating Flow}
\label{Integrating Flow}

An interesting property of the d=3 equations (\ref{dW})-(\ref{SS})
is \cite{PV} that they admit a flow which
expresses solutions of the full system in terms of
free fields. Since we use the perturbation expansion
in powers of the physical fields identified
with the deviation $C(Y;Q|x)=B(0;Y;Q|x)$ from its vacuum value
$\nu$, let us introduce a formal perturbation expansion
parameter $\eta$ as follows
\be
\label{fr}
   B(\eta)=\nu+\eta {\cal B}(\eta) \,.
\ee
Simultaneously, the rest of the  fields acquire a formal
dependence on $\eta$, i.e. $W=W(\eta)$ and $S_{\gal}=S_{\gal}(\eta)$.
The d=3 system takes a form
\be
\label{WWe}
   dW=W*\wedge W  \,,\quad d{\cal B}=W*{\cal B}-{\cal B}*W \,,\quad
   dS_\gal=W*S_\gal-S_\gal*W  \,,
\ee
\be
\label{SSe}
   S_\gal*S^\gal=-2i(1+\nu \kappa +\eta {\cal B}*\kappa ) \,,\quad
   S_\gal*{\cal B}={\cal B}*S_\gal \,.
\ee
The parameter $\eta$ drops out all the equations except for the
non-commutative curvature in (\ref{SSe}).
As a result, the expansion in powers of $\eta$ is equivalent to
the expansion in powers of the curvature fluctuations.

Now, one observes that for the limiting case $\eta=0$ the system
(\ref{WWe}), (\ref{SSe}) reduces to the free one. Indeed, setting
\be
\label{indat}
   {\cal B}|_{\eta = 0}=B_1\equiv
   C\,,\qquad W|_{\eta = 0} =W_0\equiv w\,, \quad
   S_\gal|_{\eta = 0}=S_{0\gal} \,,
\ee
we see that at $\eta=0$ the system (\ref{WWe}),
(\ref{SSe}) solves in terms of $S_\gal =S_{0\gal}$ with the gauge field
$W = w(\hat{y}(\nu);\psi_{1,2},k | x)$
satisfying the vacuum equations $dw = w *\wedge w$,
and
${\cal B} = C(\hat{y}(\nu);\psi_{1,2},k | x)$
satisfying the free field equations  (\ref{WB_1}).
This is similar to contractions of Lie algebras.
For all values of $\eta \neq 0$, the systems of equations
(\ref{WWe}), (\ref{SSe}) are pairwise equivalent since the field
redefinition (\ref{fr}) is non-degenerate. On the other hand,
although the field redefinition (\ref{fr}) degenerates at $\eta =0$,
eqs.~(\ref{WWe}), (\ref{SSe}) still make sense for $\eta=0$,
describing the free field dynamics.

Remarkably, the two inequivalent systems turn out to be related
by the following flow with respect to $\eta$:
\be
\label{ps 1}
   \frac{\partial W}{\partial\eta}=
     (1-\mu)\;{\cal B}*\frac{\partial W}{\partial\nu}
     +\mu\; \frac{\partial W}{\partial\nu}*{\cal B} \,,
\ee
\be
\label{ps 2}
  \frac{\partial {\cal B}}{\partial\eta}=
   (1-\mu)\;{\cal B}*\frac{\partial {\cal B}}{\partial\nu}
   +\mu\; \frac{\partial {\cal B}}{\partial\nu}*{\cal B}   \,,
\ee
\be
\label{ps 3}
   \frac{\partial S_\ga}{\partial\eta}=
    (1-\mu)\; {\cal B}*\frac{\partial S_\ga}{\partial\nu}
    +\mu\; \frac{\partial S_\ga}{\partial\nu} * {\cal B} \,,
\ee
where $ \mu $ is an arbitrary parameter.
By applying $\frac{\partial}{\partial\eta}$ to the both sides
of eqs.(\ref{WWe}), (\ref{SSe}) one concludes that for any $ \mu $
the system (\ref{ps 1})-(\ref{ps 3}) is compatible with
(\ref{WWe}), (\ref{SSe}). Therefore, solving
(\ref{ps 1})-(\ref{ps 3}) with the initial data (\ref{indat})
satisfying the free equations of motion
we can express solutions of the full nonlinear system at $\eta=1$
via solutions of the free system at $\eta =0$. Note that
all fields acquire nontrivial dependence on the parameter $\nu$
via the deformed oscillators (\ref{tilde y}) or (\ref{y sym}).

This approach is
very efficient at least perturbatively and allows one to derive
the corresponding nonlinear field redefinitions order by order.
In particular, using  this flow, one easily proves \cite{PV}
that the d3 higher spin gauge field strengths do not admit
nontrivial sources linear in fields. This is of course expected
result because, in accordance with the d3 Central On-Mass-Shell
Theorem (\ref{d3l}), d3 higher spin fields do not admit Weyl tensors.
Nevertheless, even at the linearized level it is a
complicated technical problem to find a form of an appropriate
field redefinition for arbitrary $\nu$ without using the flow
(\ref{ps 1})-(\ref{ps 3}).

The flows (\ref{ps 1})-(\ref{ps 3}) at different $\mu$ develop within
the same gauge equivalence class.
To see that any variation of $\mu$ is induced
by some gauge transformation one has to
find such a gauge parameter $\gvep$ that
\be
\label{mu ev}
   {\partial W\over \partial\mu}  = D\gvep \,,\quad
   {\partial {\cal B}\over \partial\mu}  = [\gvep\,,\,{\cal B}]_* \,,\quad
   {\partial S_\ga\over \partial\mu}  = [\gvep\,,\,S_\ga]_* \,,\quad
\gvep |_{ \eta =0} =0\,,
\ee
where $D\gvep = d\gvep - [W\,,\,\gvep]_*$. The compatibility
condition of (\ref{mu ev})
 with (\ref{ps 1})-(\ref{ps 3}) is satisfied if
\be
\label{eq e}
   \frac{\partial \gvep}{\partial\eta} =
   \frac{\partial {\cal B}}{\partial\nu} +
   (1-\mu)\; {\cal B} * \frac{\partial \gvep}{\partial\nu}
   +\mu\; \frac{\partial\gvep}{\partial\nu} * {\cal B}  \,,
\ee
which condition just fixes the $\eta$-dependence of $\gvep$.
Thus, one is free to choose any value of $\mu$.

One has to be careful in making statements
on the locality of the mapping induced by the flow
(\ref{ps 1})-(\ref{ps 3}). Indeed, although it does not contain
explicitly space-time derivatives, it contains them implicitly
via highest components $C_{\ga(n)}$ of the generating function
$C(\hat{y})$ which are identified with the highest derivatives
of the matter fields according to (\ref{Cn}).
For example, the equation (\ref{ps 2}) at $\mu=0$ in the zero order
in $\eta$ reads
\be
\label{ps 22}
   \frac{\partial}{\partial\eta}{\cal B}_1(z,y)=
        C(\ty)*\frac{\partial C(\ty)}{\partial\nu} \,.
\ee
Because of nonlocality of the star-product,
for each fixed rank multispinorial component of the
left hand side of this formula there appears, in general,
an infinite series involving bilinear combinations of the components
$C_{\ga(n)}$ with all $n$ on the right hand side of (\ref{ps 22}).
Therefore, in accordance with (\ref{3dxdydy}),
the right hand side of (\ref{ps 22}) effectively involves
space-time derivatives of all orders, i.e. the transformation laws
(\ref{ps 1})-(\ref{ps 3}) can effectively describe some nonlocal
transformation. This means that we cannot treat
the system (\ref{WWe}), (\ref{SSe}) as locally equivalent to the free
system. Instead we can only claim that there
exists a nonlocal mapping between the free and nonlinear system.
This mapping is reminiscent of the Nicolai mapping in supersymmetric
models \cite{Nic}.

At the linearized level, however, the transformations
induced by the integrating flow (\ref{ps 1})-(\ref{ps 3}) are
local for the following simple reason. In this case, all field
redefinitions are linear in the matter fields $C$ and
only the  zero-order (vacuum) part $W_0$ of the
 higher spin gauge fields $W$ contributes.
Since the background gravitational 1-forms (\ref{w}) are bilinear
in the auxiliary variables $\hat{y}_\ga$, the transformations for
physical fields, induced by (\ref{ps 1}), contain at most two
derivatives in $\hat{y}_\ga$. In accordance with (\ref{3dxdydy}),
this is equivalent to the statement that
the linearized field transformations contain only a finite number of
space-time derivatives and therefore are local.
Thus, the fact that the equations of motion for higher spin
fields do not acquire sources linear in the matter fields is the
well-defined local statement. This is not expected to be true for
the second-order analysis of bilinear
higher spin currents constructed from the matter fields.

In fact,                                the 2-forms
$J (C)$ dual to currents  of an arbitrary
spin constructed from massless matter fields
in $AdS_3$ are shown in \cite{PV2} to be exact
\be
J(C) = {\cal D} U(C)
\ee
in the class of infinite expansions in powers of
derivatives (i.e. with $U(C)$ depending on all components
of $C(Y)$),
thus explaining how a nonlocal field redefinition
induced by the flow (\ref{ps 1})-(\ref{ps 3}) can compensate
higher spin current interactions of matter fields
\be
R_1\equiv {\cal D} w = J(C)\,.
\ee
Let us stress that this phenomenon has no analog in the
flat space.

The existence of the integrating flow
(\ref{ps 1}) takes its origin in the simple fact that
$B$ behaves like a constant in the system (\ref{dW})-(\ref{SS}):
it commutes to $S_\gal$ and satisfies the covariant constancy condition.
Knowledge of the vacuum solution with $B=\nu $ can therefore be used to
reconstruct the full dependence on $B$. Indeed, the meaning of (\ref{ps 1}) is
that a derivative with respect to $\eta{\cal B}$ is the same as that with
respect to $\nu$.
Since the parameter $\eta$ can be interpreted as the coupling constant,
the idea to integrate the higher spin equations by integrating
a flow with respect to $\eta$ has much in common
with the coupling constant evolution method developed
in \cite{Kir} in application to quantum mechanics.

One can proceed analogously in the d=4 case by the substitution
\be
F(B ) = \nu +\eta{\cal F}({\cal B})\,,\qquad
\bar{F}( B) = \bar{\nu} +\bar{\eta}\bar{{\cal F}}({\cal B})\,
\ee
in (\ref{Rab}) or, equivalently, reintroducing nonzero
$f_0 =\nu$ and $\bar{f}_0 = \bar{\nu}$.
This leads to the
two flows commuting with the system (\ref{dW}) and to each other,
\be
\label{ps 14}
   \frac{\partial X}{\partial\eta}=
     (1-\mu)\;{{\cal F}({\cal B})}*\frac{\partial X}{\partial\nu}
     +\mu\; \frac{\partial X}{\partial\nu}*{{\cal F}({\cal B})} \,,
\ee
\be
\label{ps 34}
   \frac{\partial X}{\partial\bar{\eta}}=
     (1-\bar{\mu})\;{\bar{{\cal F}}({\cal B})}*\frac{\partial
     X}{\partial\bar{\nu}} +\bar{\mu}\; \frac{\partial
X}{\partial\bar{\nu}}*{\bar{{\cal F}}({\cal B})} \,
\ee
for $X=W$, $S$ or $B$.

For the first sight the existence of the d=4 flows
is paradoxical because it establishes a connection between the
full nonlinear problem  and the
free system with only vacuum fields in the sector of gauge fields
$W$. That was fine  for the Chern-Simons
d=3 higher spin dynamics but sounds surprisingly for the d=4 case.
Indeed, let us discuss the example of Einstein gravity.
Einstein equations (\ref{e1}), (\ref{e2}) can be rewritten as
\begin{equation}
R_{\alpha \dot{\beta}} =0,\quad \!
R_{\alpha_1\alpha_2}=\eta h^{\gamma_1 \dot{\delta}} \wedge
h^{\gamma_2}{}_{\dot{\delta}} C_{\alpha_1 \alpha_2 \gamma_1
\gamma_2}\,,\quad
\bar{R}_{\dot{\beta}_1 \dot{\beta}_2}=
\bar{\eta} h^{\eta\dot{\delta}_1}\wedge
h_\eta{}^{\dot{\delta}_2} \bar{C}_{\dot{\beta}_1 \dot{\beta}_2 \dot{\delta}_1
\dot{\delta}_2}\,,
\ee
where the parameter $\eta$ is introduced in a way it comes
from the system (\ref{dW})-(\ref{SS}).
In the limit $\eta = 0$, Einstein equations therefore reduce
to the vacuum equations of the AdS space. The dynamical equations of
the massless spin 2 field reappear as equations on the Weyl tensor
described by the covariant constancy
equation (\ref{c2}) and (\ref{bc2}), (equivalently by (\ref{dB})).
To understand how this free vacuum system is related to the
nonlinear one with $\eta \neq 0$, one has to take into account
the difference between the d=3 and d=4 problems discussed in the
end of the section \ref{AdS Vacua}.

For d=3, the ansatz (\ref{w}) describes the vacuum         solution
with the same $AdS_3$ gravitational fields $\go^{\ga\gb} (x)$ and
$h^{\ga\gb} (x)$ for arbitrary $\nu$.
Therefore, the differentiation with respect to $\nu$ in the d=3
case when using the flow to iterate solutions will only act
on the deformed oscillators (\ref{tilde y}) or  (\ref{y sym})
but not on the space-time dependent coefficients in (\ref{w})
identified with the background gravitational fields. As a result,
implementation of the integration flow in d=3 has a form of some
(may be effectively nonlocal) field redefinition.

The d=4 ansatz (\ref{anz}) only describes AdS geometry for the case
of $\nu =\bar{\nu}= 0$.
For $\nu \neq 0$ it does not solve the vacuum equations. Therefore,
in the d=4 case one first of all has to find a true
vacuum solution $w_0 (\nu ,\bar{\nu} )$ of the problem with
$\nu \neq 0$ and $\bar{\nu} \neq 0$, such that
$w_0 (0 ,0 )$ reduces to the $AdS_4$ solution. {}From the fact that
(\ref{anz}) is inconsistent away from $\nu =\bar{\nu}= 0$
it follows that the dependence on the space-time
coordinates $x^\un $ and spinor coordinates $Z_\mu$ and $Y_\mu$
is mixed nontrivially in such a vacuum solution. In particular,
$\frac{\partial W_0 (x) }{\partial \nu}\vert_{\nu =0}$
does not express directly via the $AdS_4$ gravitational fields
identified with $ W_0 (x) \vert_{\nu =0}$ as was the case for d=3.
As a result, the d=4 flow describes a
change of variables explicitly containing some functions of
space-time coordinates (via
$\frac{\partial W_0 (x) }{\partial \nu}$), which depend on a
particular gauge choice and cannot be directly
expressed via the background gravitational fields.
Therefore, in d=4, the integration flow generates
not a field redefinition but some change of variables
containing an explicit dependence on the space-time coordinates.
The d=4 field transform induced by the flow is nonlocal even at
the linearized level. The conclusion therefore is that the
flows (\ref{ps 14}), (\ref{ps 34}) indeed allow one to solve the
nonlinear system in terms of the free system via some expansion
of the form
\be
\label{nc}
W = W_0 + \eta \ga_1 (x) C +\eta^2 \ga_2 (x) C^2 +\ldots
\ee
In particular, such an expansion reconstructs the metric tensor
in terms of the curvature tensor and in that sense is
analogous to the normal coordinate expansion. In d=3 the
coefficients $\ga_n (x)$ express locally via the metric tensor.

Thus the integration flow in d=4 higher spin system
provides a systematic way for the derivation of the coefficients
of the expansions like (\ref{nc}).
The resulting procedure is pure algebraic at any given order in
$\eta$ (equivalently $C$). In particular,  it can be used \cite{BV}
to reconstruct the potentials
$w_1$ corresponding to solutions of the free field equations
described in the section \ref{``Plane Waves" in $AdS_4$}.

Note that the fact that some system can be integrated order
by order with the help of a nonlocal change of variables is
rather trivial. What is special about the higher spin systems is
that such changes of variables are described in a systematic
and constructive way by a simple flow with respect to an
additional evolution parameter.

\section{Conclusions}

Higher spin gauge theories are based on the
infinite-dimensional higher spin symmetries. Their role
 in the higher spin theories is as fundamental
the role of the supersymmetry algebra
discovered by Golfand and Likhtman \cite{GL}
for supersymmetric theories. The
higher spin symmetries are realized by the algebras of
oscillators carrying spinorial representations of the space-time
symmetries. These star-product algebras are nonlocal in the
auxiliary spinor spaces
in the usual quantum-mechanical sense typical for the
Moyal product. One point illustrated in this
contribution is that the dynamical higher spin field
equations transform this nonlocality into
nonlocality in the space-time coordinates, i.e.
the quantum mechanical nonlocality
of the higher spin algebras may imply
some space-time nonlocality of the higher spin gauge theories
at the interaction level.
The same time the higher spin gauge theories remain local
at the linearized level.

Another important implication of the star-product origin of
the higher spin algebras is that the space-time symmetries
are simple (semisimple for d=3)
and therefore correspond to AdS geometry rather than to the flat one.
The space-time symmetries are realized in terms of bilinears
in spinor oscillators according to the well-known isomorphisms
$o(2,2)\sim sp(2;R)\oplus sp(2;R)$ and $o(3,2)\sim sp(4;R)$. This
phenomenon has two consequences. On the one hand, it explains why the
theory
is local at the linearized level. The reason is that bilinears
in the non-commuting auxiliary coordinates can lead to at most two
derivatives in the star-products. On the other hand, the fact that
higher spin models require AdS geometry is closely
related to their potential nonlocality at the interaction level
because it allows expansions with arbitrary high space-time
derivatives, in which the coefficients carry  appropriate
(positive or negative) powers of the cosmological constant
fixed by counting of dimensions.
As a result, higher spin symmetries link together such seemingly
distinct concepts as AdS geometry, space-time nonlocality
of interactions and quantum mechanical nonlocality of
the star-products in auxiliary spinor spaces.
Taking into account the recent developments related to the
role of the AdS space \cite{adsconf} and star-product
\cite{mmoyal,ch,SW} in the string theory
this triple looks too natural to be just a coincidence.
Another consequence of the star-product origin of the
higher spin symmetries is that higher spin theories are
 based on the  associative structure rather than
on the Lie-algebraic one. As a result, higher spin gauge theories with
non-Abelian symmetries classify in a way analogous to the Chan-Paton
symmetries in oriented and non-oriented strings.

The algebraic structures underlying
dynamical systems associated with infinite sets of higher spin gauge
fields turn out to be interesting and deep. The full geometric
understanding is still lacking however. The parallelism with the
Fedosov quantization is very suggestive in that respect,
although higher spin gauge theories involve spinor oscillators
instead of the vector ones used in \cite{fed}. The analysis of
this issue may be useful to shed light on a
relationship with the non-commutative
Yang-Mills regime \cite{SW} in string theory.

\section*{Acknowledgments}

The author is grateful
to R.Metsaev and E.Sezgin for stimulating discussions and to M.~Sivakumar
for drawing my attention to missed terms in Eq.~(29) of the original version
of this paper. This research was supported in part by
INTAS, Grant No.96-0308 and by the RFBR Grant No.99-02-16207.

\section*{Appendix. Notation}
Underlined Latin indices are used for differential forms and vector fields
in d-dimensional space-time with coordinates $x^\un$,
\be
\um\,,\un \,,\ldots = 0,\ldots ,d-1\,,\qquad
\partial_\un = \frac{\partial}{\partial x^\un}\,,\qquad
d=dx^\un \partial_\un \,.
\ee
Indices from the middle of the Latin Alphabet denote fiber vectors,
\be
m\,,n \,,\ldots = 0,\ldots ,d-1\,,\qquad
 \eta^{nn} = (1,-1,\ldots -1 )\,.
\ee
In the flat space, base and fiber indices are sometimes identified.

The indices  $i$ and $j$ are often used for inner symmetries.

Letters from the middle of the Greek alphabet are reserved for
spinors in d-dimensions
\be
\mu\,,\nu \,,\ldots = 1,\ldots ,2^{[d/2]}\,
\ee
but sometimes are also used for symplectic indices in the star-product.
Dirac gamma matrices
$\gamma^n{}_\nu{}^\mu$ satisfy
\be
\{\gamma^n \,,\gamma ^m \} = 2\eta^{nm}\,.
\ee
Two-component spinorial indices are denoted by the Greek indices
from the beginning of the Alphabet
\be
\ga ,\gb \ldots = 1,2 \,,\quad
\dga ,\dgb \ldots = 1,2 \,,\quad
\epsilon_{\alpha\beta}=-\epsilon_{\beta\alpha}\,,\quad
\epsilon_{12}=\epsilon^{12}=1\,,\quad
\epsilon_{\dga\dgb}=\epsilon_{\ga\gb} ,
\ee
\be
A^\ga =\epsilon^{\alpha\beta}A_\gb\,,\qquad
A_\ga =A^\gb \epsilon_{\beta\alpha}\,,\qquad
A^\dga =\epsilon^{\dga\dgb}A_\dgb\,,\qquad
A_\dga =A^\dgb \epsilon_{\dgb\dga}\,.
\ee


\section*{References}



\begin{thebibliography}{77}

\bibitem{GL} Yu.A. Golfand and E.P. Likhtman,  {\it JETP Lett.}
{\bf 130}, 452 (1971);\\
in ``Problems of Theoretical Physics", Memorial Volume to Igor E.Tamm,
``Nauka" (Moscow, 1972), p. 37.

\bibitem{M} E. Witten, String Theory Dynamics in Various Dimensions,
    {\it Nucl. Phys.\/} B {\bf 443}, 85 (1995), hep-th/9503124.

\bibitem{mmoyal} A. Connes, M.R. Douglas and A. Schwarz,
``Noncommutative
  Geometry and Matrix Theory: Compactification on Tori", hep-th/9711162;\\
  M. Douglas and C. Hull, ``D-Branes and Noncommutative Torus",
  hep-th/9711165,
 M. Berkooz, ``Nonlocal Field Theories and the
  Noncommutative Torus", hep-th/9802069;\\
    N. Nekrasov and
A. Schwarz, ``Instantons on noncommutative ${\bf R^4}$
  and (2,0) superconformal six dimensional theory", hep-th/9802068.

\bibitem{ch} Chong-Sun Chu, Pei-Ming Ho,  {\it Nucl. Phys.\/} B {\bf 550}, 151
(1999), hep-th/9812219; hep-th/9906192.

\bibitem{SW} N. Seiberg and E. Witten, String Theory and Non-Commutative
Geometry, hep-th/9908142.

\bibitem{adsconf} J. Maldacena,
    ``The Large $N$ Limit of Superconformal Field Theories and Supergravity",
      hep-th/9711200;\\ S. Ferrara and C. Fronsdal,
``Conformal Maxwell Theory as a Singleton Field Theory on $ADS_5$, IIB
Branes
and Duality", hep-th/9712239;\\
 M. Gunaydin and D. Minic, ``Singletons, Doubletons and $M$-theory",
      hep-th/9802047;\\ S.S. Gubser, I.R. Klebanov, and A.M. Polyakov,
     ``Gauge Theory Correlators from Non-Critical String Theory",
      hep-th/9802109;\\
 E. Witten, ``Anti De Sitter Space and Holography",
      hep-th/9802150.

\bibitem{cm} S. Coleman and J. Mandula,
{\it Phys. Rev.} D {\bf 159}, 1251 (1967);\\
 R. Haag, J. Lopuszanski and M. Sohnius,
    {\it Nucl. Phys.\/} B {\bf 88}, 257 (1975);\\
J. Niederle, No-Go Theorems on the Connection of External and
Internal Symmetries, Proceedings of the International Seminar on
High Energy Physics and Quantum Field Theory, Protvino, July 1982.

\bibitem{diff} C. Aragone and S. Deser,
        {\it Phys. Lett.} B {\bf 86}, 161 (1979);\\
    F.A. Berends, J.W. van Holten, P. van Niewenhuizen and B. de Wit,
        {\it J. Phys.} A {\bf 13}, 1643 (1980).

\bibitem{WF} B. de Wit and D.Z. Freedman,
{\it Phys. Rev.} D {\bf 21}, 358 (1980).

\bibitem{pos} A.K. Bengtsson, I. Bengtsson and L. Brink, {\it Nucl. Phys.\/}
     B {\bf 227}31, 41 (1983);\\
   F.A. Berends, G.J. Burgers and H. van Dam, {\it Z. Phys.\/}
     C {\bf 24}, 247 (1984);
      {\it Nucl. Phys.\/} B {\bf 260}, 295 (1985);\\
    A.K.H. Bengtsson and I. Bengtsson, {\it Class. Quant. Grav.\/}
      {\bf 3}, 927 (1986);\\
  A.K.H. Bengtsson, {\it Class. Quant. Grav.\/} {\bf 5}, 437 (1988);\\
     E.S. Fradkin and R.R. Metsaev,
    {\it Class. Quant. Grav.\/} {\bf 5}, L89 (1991);\\
    R.R. Metsaev, {\it Mod. Phys. Lett.\/} A {\bf 6}, 359, 2411 (1991);
    {\it Mod. Phys. Lett.\/} A {\bf 8}, 2413 (1993);
     {\it Class. Quant. Grav.\/} {\bf 10}, L39 (1993);
     {\it Phys.~Lett.\/} B {\bf 309}, 39 (1993).

\bibitem{FV1} E.S. Fradkin and M.A. Vasiliev, {\it Phys. Lett.\/}
    B {\bf 189}, 89 (1987); {\it Nucl. Phys.\/} B {\bf 291}, 141 (1987).

\bibitem{con} M.A. Vasiliev, {\it Phys. Lett.} B {\bf 243}, 378 (1990).

\bibitem{Pr}M.A. Vasiliev, {\it Class. Quant. Grav.} {\bf 8}, 1387 (1991).

\bibitem{more} M.A. Vasiliev, {\it Phys. Lett.} B {\bf 285}, 225 (1992).

\bibitem{Witstr} E. Witten, Contribution to Strings 98.

\bibitem{FVpr} E.S. Fradkin and M.A. Vasiliev, ``Model of Supergravity
with Minimal Electromagnetic Interaction", Preprint FIAN, N197 (1976).

\bibitem{DF} D.Z. Freedman and A. Das, {\it Nucl. Phys.\/} B {\bf 120},
221 (1977).

\bibitem{rev} M.A. Vasiliev,
Higher-Spin Gauge Theories in Four, Three and Two Dimensions,
{\it Int. J. Mod. Phys.} D {\bf 5}, 763 (1996).

\bibitem{V2}M.A. Vasiliev, {\it Fortschr. Phys.\/} {\bf 35}, 741 (1987).

\bibitem{Fr}   C.~Fronsdal, {\it Phys. Rev.\/} {\bf D18} (1978) 3624; {\bf D20}
              (1979) 848;\\
              J.~Fang and C.~Fronsdal, {\it Phys. Rev.\/} {\bf
              D18} (1978) 3630; {\bf D22} (1980) 1361.

\bibitem{cur}
   F.A. Berends, G.J. Burgers and H. van Dam,
      {\it Nucl.~Phys.\/} B {\bf 271}, 429 (1986)\,;\\
 D. Anselmi, Theory of Higher Spin Tensor Currents and Central Charges,
 {\tt hep-th/9808004}.

\bibitem{curan} D. Anselmi,
Higher-Spin Current Multiplets in Operator Product
 Expansions, {\tt hep-th/9906167}.

\bibitem{LV}   V.E. Lopatin and M.A. Vasiliev,
 {\it Mod. Phys. Lett.} A {\bf 3}, 257 (1988).

\bibitem{V}   M.A. Vasiliev, {\it Nucl.~Phys.} B {\bf 301}, 26 (1988).

\bibitem{CHC}   M.A. Vasiliev, in preparation.

\bibitem{PV2}    S.F. Prokushkin and M.A. Vasiliev, Currents of
Arbitrary Spin in $AdS_3$,
{\it Phys. Lett.} B (1999) (in press), {\tt hep-th/9906149};
Cohomology of Arbitrary Spin Currents in $AdS_3$,
Theor. Math. Phys. (in press), {\tt hep-th/9907020}.

\bibitem{hsa4} E.S. Fradkin and M.A. Vasiliev, {\it Dokl. Acad. Nauk.}
   {\bf 29}, 1100 (1986); {\it Ann. of Phys.} {\bf 177}, 63 (1987).

\bibitem{OP1} M.A. Vasiliev, {\it Fortschr. Phys.} {\bf 36}, 33 (1988).

\bibitem{KV1} S.E. Konstein and M.A. Vasiliev, {\it Nucl. Phys.\/}
B  {\bf 331}, 475 (1990) 475.

\bibitem{bl} M.P. Blencowe, {\it Class. Quantum Grav.} {\bf 6}, 443 (1989).

\bibitem{BBS} E. Bergshoeff, M. Blencowe and K. Stelle, {\it Comm. Math.
Phys.}
{\bf 128}, 213 (1990).

\bibitem{Aq} M.A. Vasiliev, {\it JETP Lett.} {\bf 50}, 374 (1989);
{\it Int. J. Mod. Phys.} A {\bf 6}, 1115 (1991).

\bibitem{Eq} M.A. Vasiliev, {\it Mod. Phys. Lett.} A {\bf 7}, 3689 (1992).

\bibitem{PV} S.F. Prokushkin and M.A. Vasiliev,
 {\it Nucl. Phys.} B {\bf 545}, 385 (1999); {\tt hep-th/9806236}.

\bibitem{BF} B.L. Feigin, {\it Uspehi Mat. Nauk} {\bf 43}, 169 (1988).

\bibitem{H} M. Bordemann, J. Hoppe and P. Schaller,
{\it Phys. Lett.} {\bf 232}, 199 (1989).

\bibitem{FLU} E.S. Fradkin and V.Ya. Linetsky, {\it Mod. Phys. Lett.}
A {\bf 5}, 1967 (1990).

\bibitem{wig} E. Wigner, {\it Phys. Rev.} D {\bf 77}, 711 (1950).

\bibitem{des} L.M. Yang, {\it Phys. Rev.} D {\bf 84}, 788 (1951);\\
D.G. Boulware and S. Deser, {\it Il Nouvo Cimento}, {\bf XXX}, 231 (1963);\\
N. Mukunda, E.C.G. Sudarshan, J.K. Sharma and C.L. Mehta, {\it J. Math.
Phys.}
{\bf 21}, 2386 (1980).

\bibitem{BWV} E. Bergshoeff,
B. de Wit, and M.A. Vasiliev, {\it Nucl. Phys.} B {\bf 366}, 315 (1991).

\bibitem{HOPPE} J. Hoppe, MIT Ph.D Thesis (1982); Quantum theory of a
relativistic surface, workshop on Constraints theory and relativistic
dynamics (Florence, 1986), eds. G. Longhi and L. Lusanna (World Scientific,
Singapore 1987) p. 267.

\bibitem{BPV}
   A.V. Barabanschikov, S.F. Prokushkin, and M.A. Vasiliev,
   {\it Rus. Theor. Math. Phys.} {\bf 110}, 295 (1997),
   {\tt hep-th/9609034}.

\bibitem{d3gr} A. Achucarro and P.K. Townsend, {\it Phys. Lett.} B {\bf 180},
89 (1986);\\
 E. Witten, {\it Nucl. Phys.} B {\bf 311}, 46 (1989).

\bibitem{FVA} E.S. Fradkin and M.A. Vasiliev,
{\it Mod. Phys. Lett.} A {\bf 3}, 2983 (1988).

\bibitem{VA}   M.A. Vasiliev, {\it Nucl. Phys.} B {\bf 307}, 319 (1988).

\bibitem{Ann}   M.A. Vasiliev, {\it Ann. Phys.} (N.Y.) {\bf 190}, 59 (1989).

\bibitem{KV}  S.E. Konstein and M.A. Vasiliev, {\it Nucl. Phys.\/}
B {\bf 312}, 402 (1989).

\bibitem{kirill} A.A. Kirillov, ``Elements of Representation Theory",
Second Edition. ``Nauka" (Moscow, 1978).

\bibitem{SS}  E. Sezgin and P. Sundell,
``Higher Spin $N=8$ Supergravity",
  hep-th/9805125, ``Higher Spin $N=8$ Supergravity in $AdS_4$",
  hep-th/9903020.

\bibitem{sym}  F.A. Berezin, {\it Mat. Sbornik} {\bf 86}, 578 (1971); see also
F.A. Berezin ``The method of Second Quantization'', Nauka (Moscow, 1986)
and
  references therein;\\ F.A. Berezin and M.S. Marinov, {\it Ann. of Phys.}
  {\bf 104}, 336 (1977);\\ F. Bayen, M. Flato, C. Fronsdal, A. Lichnerowicz
and
 D. Sternheimer, {\it Ann. of Phys.} {\bf 110}, 61, 111 (1978).

\bibitem{BS} F.A.Berezin and M.A. Shubin, ``Schr\"odinger Equation",
Moscow University Press (Moscow, 1983).

\bibitem{moyal} J.E. Moyal, {\it Proc. Cambridge Phil. Soc.} {\bf 45},
99 (1949).

\bibitem{BV} K. Bolotin and M.A. Vasiliev, in preparation.

\bibitem{un} M.A. Vasiliev, {\it Class. Quant. Grav.} {\bf 11}, 649 (1994).


\bibitem{PR}
R. Penrose and W. Rindler, {\it Spinors and space-time\/},
Cambridge Univ. Press (Cambridge, 1984), Vol. 1.

\bibitem{FDA} P. van Nieuwenhuizen, in {\it ``Group Theoretical Methos
in Physics''}, Proceedings, (Eds. M. Serdaroglu and E.In\"onu),
(Istanbul,1982) p. 228;
Lecture Notes in Physics, Springer-Verlag, (New York/Berlin, 1983), Vol.
180.\\
R. D'Auria, P. Fre, P.K. Townsend, and P. van Nieuwenhuizen,
 {\it Ann. Phys.} (N.Y.) {\bf 155}, 423 (1984).

\bibitem{V6}  M.A. Vasiliev, {\it Nucl. Phys.\/} B {\bf 324}, 503 (1989).

\bibitem{aux}  M.A. Vasiliev, {\it Nucl. Phys.\/} B {\bf 307}, 319 (1988).

\bibitem{Bu}  M.A. Vasiliev, {\it Nucl. Phys. Proc. Suppl.\/} B {\bf 324}, 241
(1997).

\bibitem{fed} B. Fedosov, {\it J. Diff. Geometry} {\bf 40}, 213 (1994).

\bibitem{cast} C. Castro, W Geometry from Fedosov's
deformation quantization, hep-th/9802023.

\bibitem{PVN} P. van Nieuwenhuizen, {\it Phys. Rep.\/} {\bf 68}, 189 (1981).

\bibitem{bls}
I. Bandos, J. Lukierski and D. Sorokin, Superparticle Models with
Tensorial Central Charges, hep-th/9904109.

\bibitem{d2}  M.A. Vasiliev, {\it Phys.Lett.\/} B {\bf 363}, 51 (1995).

\bibitem{chog}  Ch. Devchand and V. Ogievetsky {\it Nucl. Phys.\/} B
{\bf 481}, 188 (1996),  hep-th/9606027.

\bibitem{Nic} H. Nicolai, {\it Nucl. Phys.} B {\bf 176}, 419 (1980).

\bibitem{Kir} D.A. Kirzhnitz,
{\it JETP} {\bf 49}, 1544 (1965);\\
in {\it ``Problems of Theoretical Physics''}, Igor E. Tamm Memorial Volume,
``Nauka" (Moscow, 1972), p. 74.

\end{thebibliography}
\end{document}